\documentclass[10pt,a4paper,twocolumn]{article}
\usepackage[margin=19mm,columnsep=6mm]{geometry}
\usepackage[T1]{fontenc}
\usepackage{amsmath,amssymb,array,booktabs,float,graphicx,tikz}
\usepackage{newtxtext,newtxmath}
\usepackage[authoryear,round]{natbib}
\usepackage[hidelinks]{hyperref}
\usetikzlibrary{arrows.meta,calc,decorations.pathmorphing,positioning}
\graphicspath{{figures/}{./}}
\DeclareMathOperator*{\minimize}{minimize}
\newcommand{\vect}[1]{\boldsymbol{#1}}
\begin{document}
\title{Gradient-based optimization of buoyancy distributions for lazy-wave risers}
\author{Nicolò Pollini \quad Kazem Jaber\\[3pt]
\small Department of Structural Engineering and Construction Management\\
\small Faculty of Civil and Environmental Engineering\\
\small Technion--Israel Institute of Technology, Haifa, Israel\\
\small \href{mailto:nicolo@technion.ac.il}{nicolo@technion.ac.il}}
\date{}
\twocolumn[\begin{@twocolumnfalse}
\maketitle
\begin{abstract}
This paper presents a novel gradient-based approach for optimizing the buoyancy distribution of lazy-wave offshore risers. The static response is modeled with a total potential energy formulation for a geometrically nonlinear, inextensible beam on an elastic seabed. The total potential energy is discretized by finite differences and the resulting nonlinear equilibrium system is solved by a two stage continuation procedure, in which the line length and then the buoyancy load are increased incrementally, with Newton--Raphson iterations and an Armijo line search. The design variables are the uplift ratios at the nodes where buoyancy modules may be installed. The optimization minimizes a quadratic measure of buoyancy demand subject to constraints on the maximum curvature and effective tension. The gradients of the constraints are computed by means of a discrete adjoint sensitivity analysis. Three design parametrizations are compared: a continuous formulation with independent nodal uplift ratios, a grouped formulation with fixed module assignments, and a formulation inspired by discrete material optimization that simultaneously selects the uplift ratios of a small catalog of module types and their placement along the riser. The latter yields practical designs with a limited number of module sizes. The framework provides a flexible and computationally efficient tool for preliminary design of lazy-wave configurations.
\end{abstract}\vspace{1em}
\end{@twocolumnfalse}]
\section{Introduction}
\label{sec:intro}

Flexible pipelines and power cables are essential components of offshore energy systems. They transport hydrocarbons, electrical power, and communication signals between floating structures and seabed facilities. Although their operational purposes differ, both belong to the class of subsea slender structures and may undergo large geometric deformations while interacting with environmental loads and the seabed \citep{bai2014subsea,Bai2021,guo2013offshore}. Despite these different applications, they share many structural characteristics and can generally be analyzed using similar approaches during preliminary analysis and design. 
The ability of these systems to accommodate large deformations and seabed interaction depends strongly on their structural configuration. The free-hanging catenary is among the simplest and most widely used configurations \citep{trapper2019feasible,ALBINO2021103660}. It requires no distributed buoyancy and is comparatively straightforward to install. Under submerged self weight, the line assumes a catenary profile between the floating structure and the seabed. The line is connected to the floating structure at the hang-off point (HOP) and first contacts the seabed at the touchdown point (TDP). This arrangement is particularly attractive when environmental conditions and motions induced by the floating platform are moderate \citep{CERIK2024118810}.

The expansion of offshore energy developments into deeper water creates important design and operational challenges. As water depth increases, the longer suspended span generally produces greater effective tension at the HOP because of the accumulated submerged weight. Moreover, motions of the supporting floating platform generate dynamic loads in the attached pipeline or cable. The combination of high static tension and platform induced excitation may increase fatigue demand and stress concentrations, particularly near the touchdown region \citep{BORGES20141,Guo2014,jmse12112000}.

Lazy-wave configurations have therefore been investigated as an alternative to conventional free hanging catenaries \citep{Yue2011,Hu2019,FelisitaEtAl2016}. A lazy wave is formed by attaching buoyancy modules along a selected portion of the line. These modules are buoyant volumes that may be made from lightweight materials such as syntactic foam or constructed as thin walled air cans. They are distributed along a prescribed section of the line \citep{Bai2005,wang2014numerical}. The resulting upward force creates an elevated arch and divides the system into an upper section, a buoyancy section, and a lower section \citep{RUAN201473}. A schematic representation is provided in Figure~\ref{fig:lazywave}. 
By supporting part of the submerged weight, the buoyancy modules alter both the equilibrium geometry and the tension distribution. The compliant wave shaped profile can also accommodate platform motion through geometric changes in the buoyancy section. Consequently, important response quantities may occur not only at the HOP and TDP but also near the transitions into and out of the buoyancy section. The design must therefore balance competing requirements, such as buoyancy demand, effective tension, curvature, and geometric feasibility.

Previous studies have examined the analysis and design optimization of lazy-wave systems.  Their practical optimization is challenging from a computational point of view because each design evaluation requires the solution of a geometrically nonlinear equilibrium problem. 
Several authors studied the design optimization of offshore lazy-wave configurations by means of gradient-free optimization algorithms, such as, for example, genetic algorithms or particle swarm optimization algorithms \citep{dePina2011tailoring,Ai2019performance,Rentschler2019design,Wang2024stepwise,Zhao2025optimization,Schnepf2025efficient,Hou2026configuration}. 

In practice, design optimization requires repeated structural analyses and design updates until prescribed stopping criteria are met. The structural analysis tool employed within the optimization loop must therefore be computationally efficient while remaining sufficiently accurate and reliable, particularly during the preliminary and conceptual design phases. Commercial offshore analysis packages, such as OrcaFlex, RIFLEX, OFFPIPE, and Cable3D, as well as general purpose finite element software, such as Abaqus, provide advanced modeling capabilities and high fidelity structural analyses. However, their use as inner solvers in optimization loops may be computationally expensive and impractical during the early design stages, when numerous alternative configurations and design possibilities must be explored and assessed. This limitation is not necessarily overcome by replacing graphical user interface (GUI) workflows with scripting interfaces, since the computational cost associated with the underlying numerical models and solution procedures remains. Moreover, commercial software may effectively operate as a black box, limiting access to the underlying formulations and, more importantly, to consistent derivatives of the structural response with respect to the design variables. Consequently, optimization may have to rely on computationally demanding gradient-free algorithms or finite difference approximations of the sensitivities, which, respectively, are computationally expensive and may introduce numerical inaccuracies.

As a result, throughout the design process, engineers often rely on simplified two dimensional static models to reduce computational costs and facilitate repeated structural analyses. Although these models neglect certain physical phenomena and may provide lower accuracy than more sophisticated numerical approaches, they can adequately capture the dominant effects governing the overall riser configuration. Consequently, they represent a computationally efficient and practical alternative, particularly during the preliminary and intermediate stages of the design process \citep{Lenci2005simple}. 
Recent studies have proposed simplified approaches to reduce the modeling complexity and the computational cost of determining equilibrium configurations for lazy-wave profiles \citep{RUAN201473,wang2014numerical,Wang2015nonlinear,Wang2015abandonment,Wang2015transfer}. 
In a recent contribution, \citet{trapper2020feasible} presented a novel numerical approach for the static structural analysis of steel lazy-wave risers. The method models the entire riser, including its suspended, buoyancy, and seabed interaction, as a single continuous structure governed by geometrically nonlinear beam theory. The formulation accounts for large deformations, bending stiffness, distributed loading, and riser–seabed interaction through a total potential energy minimization principle. The equilibrium equations are obtained by directly differentiating the discretized energy functional and are solved using an incremental and iterative Newton–Raphson procedure. This approach avoids the need to formulate separate governing equations and enforce compatibility conditions between the different riser sections, providing a relatively simple and computationally efficient framework for structural analysis. The work discussed by \citet{trapper2020feasible} extends the approach initially presented in \citet{trapper2019feasible}, and subsequently extended to the case of inelastic seabed behavior \citep{trapper2020staticanalysis}, long free spanning offshore pipelines \citep{trapper}, and pipe lay on rough seafloor \citep{trapper2022roughseafloor}.

In this paper, we present a novel gradient-based approach for the design optimization of lazy-wave riser configurations. The proposed framework builds upon the flexible and computationally efficient structural analysis method developed by \citet{trapper2020feasible}, which is significantly further extended by incorporating analytical design sensitivities derived using the adjoint variable method \citep{Michaleris1994tangent,pollini2018adjoint}. The approach aims to optimize the spatial distribution and sizing of buoyancy modules while satisfying constraints on structural response quantities, such as effective tension and curvature. Different design parameterizations are proposed and investigated, ranging from a continuous nodal based distribution of buoyancy to more practical configurations characterized by a limited number of available module sizes, whose dimensions and assignment along the riser are simultaneously optimized. The latter formulation is inspired by optimization techniques traditionally employed in structural topology optimization \citep{Bendsoe2004topology,sigmund2013topology} and discrete material optimization \citep{Stegmann2005discrete,Hvejsel2011material}, enabling the treatment of practical design requirements within a gradient-based framework. By combining an efficient structural analysis procedure with accurate analytical sensitivity information, the proposed approach provides a computationally efficient methodology for exploring and optimizing lazy-wave riser configurations. Several numerical examples are presented to assess the performance of the proposed formulations and to investigate the influence of different design parameterizations and structural constraints on the optimized configurations. Although the numerical examples presented in this work focus on offshore risers, the proposed approach is also suitable for the design optimization of lazy-wave power cables for floating offshore wind turbines, once their specific structural properties and loading conditions are considered \citep{Rentschler2019design,CERIK2024118810,Liu2025deeplearning,Ahmad2023optimisation,Abrahamsen2024reference,PerezRua2024exact,Holcombe2026surrogate}.

The remainder of this paper is organized as follows. Sec.~\ref{sec:problem} introduces the engineering problem and formulates the design optimization problem. Sec.~\ref{sec:response} presents the nonlinear structural analysis formulation, its numerical implementation, and validation against reference results. Sec.~\ref{sec:sensitivity} develops the adjoint sensitivity analysis approach and verifies the analytical gradients against finite difference approximations. Sec.~\ref{sec:examples} presents and discusses the numerical optimization results obtained using different buoyancy design parameterizations and structural constraints. Finally, Sec.~\ref{sec:end} concludes the paper with a summary of the main findings and final remarks.

\section{Problem description}
\label{sec:problem}

This section introduces the engineering problem considered in this study and defines the corresponding design optimization setting. The physical configuration, loading conditions, and principal modeling assumptions are presented first, followed by the definition of the buoyancy design variables, objective function, structural constraints, and equilibrium equations governing the optimization problem.

\subsection{Engineering problem}

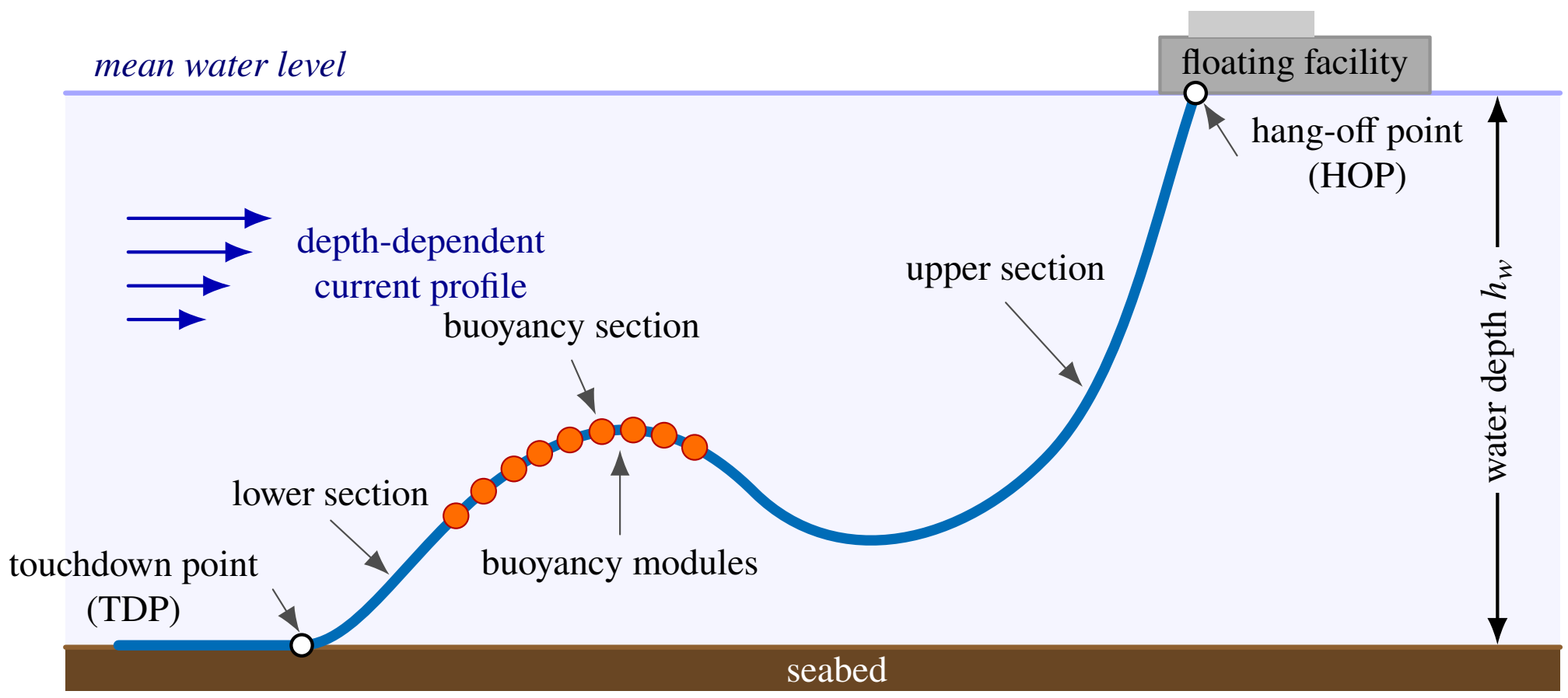
\begin{figure*}[htbp!]
    \centering
    \begin{tikzpicture}[
        x=0.82cm,
        y=0.82cm,
        >=Latex,
        line cap=round,
        line join=round,
        label text/.style={font=\small,align=center},
        callout/.style={-{Latex[length=2.2mm]},thin,draw=black!70},
        current/.style={-{Latex[length=2.4mm]},blue!70!black,thick}
    ]
        \fill[blue!4] (0,0.45) rectangle (15.5,6.2);
        \draw[blue!35,very thick] (0,6.2) -- (15.5,6.2);
        \node[anchor=south west,font=\small\itshape,text=blue!45!black]
            at (0.15,6.22) {mean water level};

        \fill[brown!55!black] (0,0) rectangle (15.5,0.45);
        \draw[brown!75!black,very thick] (0,0.45) -- (15.5,0.45);
        \node[font=\small,text=white] at (8.0,0.22) {seabed};

        \fill[gray!65] (11.35,6.2) rectangle (14.15,6.78);
        \draw[gray!85,thick] (11.35,6.2) rectangle (14.15,6.78);
        \fill[gray!40] (11.65,6.78) rectangle (12.95,7.05);
        \node[font=\small] at (12.75,6.49) {floating facility};

        \draw[cyan!45!blue,line width=2.5pt]
            (0.55,0.47) -- (2.45,0.47)
            .. controls (3.10,0.48) and (3.65,1.65) .. (4.65,2.30)
            coordinate[pos=0.78] (module1)
            coordinate[pos=0.89] (module2)
            coordinate[pos=1] (module3)
            .. controls (5.55,2.90) and (6.35,2.85) .. (7.15,2.05)
            coordinate[pos=0.10] (module4)
            coordinate[pos=0.22] (module5)
            coordinate[pos=0.35] (module6)
            coordinate[pos=0.48] (module7)
            coordinate[pos=0.61] (module8)
            coordinate[pos=0.74] (module9)
            .. controls (8.00,1.22) and (9.30,1.50) .. (10.20,2.45)
            .. controls (11.00,3.30) and (11.30,4.95) .. (11.72,6.20);

        \foreach \p in {1,...,9} {
            \fill[orange!85!red] (module\p) circle (0.13);
            \draw[red!70!black,thin] (module\p) circle (0.13);
        }

        \fill[white] (2.45,0.47) circle (0.11);
        \draw[black,thick] (2.45,0.47) circle (0.11);
        \fill[white] (11.72,6.20) circle (0.11);
        \draw[black,thick] (11.72,6.20) circle (0.11);

        \draw[current] (0.65,4.90) -- (2.15,4.90);
        \draw[current] (0.65,4.55) -- (1.95,4.55);
        \draw[current] (0.65,4.20) -- (1.72,4.20);
        \draw[current] (0.65,3.85) -- (1.47,3.85);
        \node[label text,anchor=west,text=blue!55!black] at (2.25,4.40)
            {depth-dependent\\current profile};

        \node[label text] (lower) at (2.75,2.05) {lower section};
        \draw[callout] (lower.south) -- (3.36,1.20);

        \node[label text] (buoy) at (5.25,3.75) {buoyancy section};
        \draw[callout] (buoy.south) -- (5.50,2.86);

        \node[label text] (upper) at (9.75,4.35) {upper section};
        \draw[callout] (upper.south) -- (10.55,3.15);

        \node[label text,anchor=east] (tdp) at (2.15,1.05) {touchdown point\\(TDP)};
        \draw[callout] (tdp.east) -- (2.43,0.60);

        \node[label text,anchor=west] (hop) at (12.15,5.55) {hang-off point\\(HOP)};
        \draw[callout] (hop.west) -- (11.80,6.10);

        \node[label text,anchor=north] (modules) at (5.75,1.62) {buoyancy modules};
        \draw[callout] (modules.north) -- (5.75,2.50);

        \draw[<->,thick] (14.85,0.47) -- (14.85,6.18);
        \node[label text,fill=blue!4,inner sep=2pt,rotate=90]
            at (14.85,3.33) {water depth $h_w$};
    \end{tikzpicture}
    \caption{Schematic of the lazy-wave configuration and the principal geometric features considered in the analysis.}
    \label{fig:lazywave}
\end{figure*}

This study considers lazy-wave offshore pipelines deployed between a floating facility and a frictionless elastic seabed at water depth \(h_w\), as illustrated in Figure~\ref{fig:lazywave}. Although the formulation and numerical examples focus on pipelines or offshore risers, the proposed analysis and optimization framework is also applicable to lazy-wave power cables for floating offshore wind turbines. Such applications require specific structural properties, loading conditions, and boundary conditions, but the underlying response, sensitivity, and optimization procedures remain unchanged. The buoyancy modules occupy a prescribed portion of the line and divide the configuration into three segments: an upper section adjacent to the hang-off point (HOP), a buoyancy section in which the modules are installed, and a lower section adjacent to the touchdown point (TDP).

The upper section goes from the HOP to the beginning of the buoyancy section. Its equilibrated configuration has a catenary profile and transfers the loads to the floating facility. The buoyancy section contains uniformly distributed modules whose net upward force exceeds the local submerged self weight, which causes the desired hogbend characterizing the so called lazy-wave profiles. The lower section forms the sagbend and approaches the seabed at the TDP. Beyond the TDP, the line is supported by a linear elastic seabed, where tangential seabed friction is neglected. The applied loads comprise the submerged self-weight of the line, the net upward force generated by the buoyancy modules, and steady hydrodynamic drag associated with a current profile that varies with depth. The combined action of these loads, together with seabed contact and the end conditions, determines the equilibrium configuration and its global response. Dynamic response is outside the scope of this work.

\subsection{Design optimization problem}

The design optimization problem determines the spatial distribution and magnitude of buoyancy along a prescribed section of the pipeline. In the numerical formulation, the buoyancy layout is represented by nodal uplift ratios that may vary over the active design nodes. Let
\begin{equation}
    \vect{r}=\begin{bmatrix}r_1&r_2&\cdots&r_{n_d}\end{bmatrix}^{\mathsf T}
    \label{eq:design_vector}
\end{equation}
collect the uplift ratios associated with the $n_d$ active design nodes. If the design variable $r_i$ is associated with node $i$, the net vertical nodal load at that node is
\begin{equation}
    w_{i}=w_g\Delta s\,(1-r_i),
    \label{eq:design_nodal_weight}
\end{equation}
where $w_g$ is the submerged structural weight per unit length. Thus, $r_i=0$ corresponds to the unbuoyed pipe, $r_i=1$ locally balances the submerged structural weight, and $r_i>1$ produces a net upward load. Outside the prescribed buoyancy section, $r=0$. In the calculations reported here the design bounds are
\begin{equation}
    \underline{r} \le r_i \le \overline{r} ,\qquad i=1,\ldots,n_d,
    \label{eq:design_bounds}
\end{equation}
where for example $\underline{r}=0$ and $\overline{r}=3$. 
The design optimization problem minimizes a quadratic measure of the buoyancy uplift ratios while satisfying prescribed structural response constraints. The objective is defined as
\begin{equation}
    f(\vect r)=\vect r^{\mathsf T}\vect r,
    \qquad
    \frac{\partial f}{\partial \vect r}=2\vect r.
    \label{eq:objective_buoyancy}
\end{equation}
Eq.~\eqref{eq:objective_buoyancy} defines the quadratic buoyancy objective and its explicit gradient. The quadratic objective was selected following preliminary numerical comparisons with the linear criterion $\sum_i r_i$, in which it led to smoother convergence to the final designs. It penalizes large local uplift ratios more strongly and, for a given total uplift, favors a more even distribution among the design nodes. The objective is therefore interpreted as a quadratic measure of buoyancy demand. 

Constraints are imposed on the maximum curvature and maximum effective tension along the riser. In their original form, these requirements are written as
\begin{align}
    \max_{i=2,\ldots,N+1}|\kappa_i|
    &\le \kappa_{\mathrm{lim}},
    \label{eq:max_curvature_constraint}\\
    \max_{i=1,\ldots,N+1}|T_i|
    &\le T_{\mathrm{lim}},
    \label{eq:max_tension_constraint}
\end{align}
where $\kappa_i$ and $T_i$ denote the discrete curvature and effective tension, respectively. The maximum operator is not differentiable. To obtain constraints suitable for gradient-based optimization, the maximum values are approximated using the following differentiable $p$-norm aggregates:
\begin{align}
    \kappa_p &=\left(\sum_{i=2}^{N+1}\kappa_i^p\right)^{1/p} \le \kappa_{\mathrm{lim}},
    \label{eq:kappa_p}\\
    T_p &=\left(\sum_{i=1}^{N+1}T_i^p\right)^{1/p} \le T_{\mathrm{lim}}.
    \label{eq:tension_p}
\end{align}
where $\kappa_{\mathrm{lim}}$ and $T_{\mathrm{lim}}$ are the prescribed upper bounds on curvature and effective tension, respectively. The exponent $p$ is chosen as a large even number, so that $\kappa_i^p=|\kappa_i|^p$ and $T_i^p=|T_i|^p$, and no absolute values are required in Eqs.~\eqref{eq:kappa_p} and~\eqref{eq:tension_p}.
As $p$ increases, the aggregates in Eqs.~\eqref{eq:kappa_p} and~\eqref{eq:tension_p} approximate with increasing accuracy Eqs.~\eqref{eq:max_curvature_constraint} and~\eqref{eq:max_tension_constraint}. 
Using these aggregates, the optimization problem is formulated as
\begin{equation}
    \begin{aligned}
        \minimize_{\vect r \in \mathbb{R}^{n_d}}\quad & \vect r^{\mathsf T}\vect r\\
        \text{subject to}\quad
        & \kappa_p\bigl(\vect\theta(\vect r),\vect r\bigr)\le \kappa_{\mathrm{lim}},\\
        & T_p\bigl(\vect\theta(\vect r),\vect r\bigr)\le T_{\mathrm{lim}},\\
        & \vect R\bigl(\vect\theta(\vect r),\vect r\bigr)=\vect 0,\\
        & \underline{\vect r} \le \vect r\le \overline{\vect r}.
    \end{aligned}
    \label{eq:optimization_problem}
\end{equation}
Here, $\underline{\vect r}$ and $\overline{\vect r}$ denote the lower and upper bounds on the design variables, respectively, with the inequalities understood componentwise. The vector $\vect R(\vect\theta(\vect r),\vect r)$ denotes the discrete structural equilibrium residual, which must vanish at an equilibrium configuration. The vector $\vect\theta$ contains the state variables describing the riser configuration and depends implicitly on the design vector $\vect r$, whose entries are the buoyancy uplift ratios. Consequently, each evaluation of the objective and structural constraints requires the solution of $\vect R(\vect\theta(\vect r),\vect r)=\vect 0$ to determine the equilibrium state $\vect\theta(\vect r)$.
The structural analysis used to compute these equilibrium configurations is presented in the following section. 

\section{Static structural response analysis}
\label{sec:response}

The static response analysis presented in this section follows the minimum energy approach developed by \citet{trapper2019feasible} for offshore pipeline installation and subsequently extended to steel lazy-wave risers by \citet{trapper2020feasible}. The suspended, buoyant, touchdown, and seabed supported portions are treated as a single geometrically nonlinear beam rather than as separate analytical segments. 
The analysis approach is presented in detail because it is tightly coupled to the sensitivity formulation developed subsequently in Sec.~\ref{sec:sensitivity}. In particular, the analytical gradients are constructed from the discrete equilibrium residual and its Jacobian, so a clear definition of the state variables, loads, contact model, discretization, and nonlinear solution procedure is required to understand in detail and possibly reproduce the proposed optimization approach. Thus, in what follows, the continuous formulation is first summarized and then specialized to the finite difference equations used in this work. The numerical implementation developed in this work relies on an Armijo line search to improve the robustness of the Newton--Raphson iterations. 

\subsection{Continuous formulation}
\label{subsec:contform}

\begin{figure*}[htbp!]
    \centering
    \begin{tikzpicture}[
        x=0.82cm,
        y=0.82cm,
        >=Latex,
        line cap=round,
        line join=round,
        axis/.style={-{Latex[length=2.4mm]},thick},
        load/.style={-{Latex[length=2.1mm]},thin},
        dim/.style={<->,thin},
        pipe/.style={blue!65!black,line width=2.2pt}
    ]
        \fill[blue!4] (0.25,0.75) rectangle (13.2,5.8);
        \draw[axis] (0.25,5.8) -- (14.2,5.8) node[above left] {$x$};
        \node[anchor=north west,font=\scriptsize,text=blue!50!black]
            at (0.45,5.74) {mean water level ($y=0$)};
        \draw[axis] (-1.05,5.8) -- (-1.05,-0.75) node[above right] {$y$};
        \draw[gray!55,dashed] (0.25,0.78) -- (0.25,5.8);
        \node[anchor=north,font=\small] at (0.25,0.38) {$x=0$};

        \fill[gray!18] (0.25,0.42) rectangle (10.1,0.75);
        \draw[gray!70,thick] (0.25,0.75) -- (10.1,0.75);
        \foreach \xx in {0.4,0.8,...,9.8}
            \draw[gray!55] (\xx,0.75) -- ++(0.22,-0.33);
        \node[anchor=west,font=\scriptsize] at (10.2,0.60) {seabed};

        \foreach \yy/\len in {4.65/1.55,4.30/1.32,3.95/1.10}
            \draw[blue!65!black,thick,decorate,
                  decoration={snake,amplitude=1.2pt,segment length=6pt},
                  -{Latex[length=2.0mm]}]
                (1.25,\yy) -- ++(\len,0);
        \node[anchor=west,text=blue!65!black] at (2.95,4.28) {$v_x(y)$};

        \draw[pipe]
            (0.25,0.72) -- (1.85,0.72)
            .. controls (2.55,0.70) and (2.80,0.38) .. (3.55,0.54)
            .. controls (4.35,0.72) and (4.55,2.45) .. (5.75,2.85)
            .. controls (6.80,3.22) and (7.45,1.65) .. (8.65,1.72)
            .. controls (10.05,1.80) and (11.25,4.20) .. (12.25,5.80);

        \fill[black] (0.20,0.67) rectangle (0.30,0.77);
        \node[anchor=south west,font=\small\bfseries] at (0.42,1.08) {FP};
        \draw[black,thick] (-0.48,0.30) -- (-0.48,1.14);
        \draw[black,thick] (-0.20,0.30) -- (-0.20,1.14);
        \foreach \yy in {0.34,0.50,...,1.10}
            \draw[black!65] (-0.48,\yy) -- ++(-0.20,0.12);
        \foreach \yy in {0.46,0.72,0.98} {
            \fill[white] (-0.34,\yy) circle (0.065);
            \draw[black,thin] (-0.34,\yy) circle (0.065);
        }
        \draw[black,thick] (-0.20,0.72) -- (0.25,0.72);

        \fill[white] (12.25,5.80) circle (0.10);
        \draw[black,thick] (12.25,5.80) circle (0.10);
        \node[anchor=north west,font=\small\bfseries] at (12.35,5.72) {HOP};
        \draw[load,red!70!black,very thick] (12.25,5.80) -- (13.65,5.80)
            node[midway,above] {$P_0$};

        \draw[load] (1.00,1.02) -- (1.90,1.02) node[midway,above] {$s$};
        \draw[dim] (-1.62,0.75) -- (-1.62,5.80);
        \node[fill=white,inner sep=1pt] at (-1.62,3.28) {$h_w$};

        \foreach \xx/\yy in {4.062/0.921,4.411/1.462,4.789/2.065,5.272/2.590,6.112/2.906}
            \draw[load,orange!85!red] (\xx,\yy) -- ++(0,0.62);
        \node[text=orange!80!red] at (5.30,3.55) {$w_b$};
        \foreach \xx/\yy in {6.746/2.679,7.413/2.194,8.050/1.821}
            \draw[load] (\xx,\yy) -- ++(0,-0.62);
        \node at (8.55,1.05) {$w_g$};

        \coordinate (A) at (10.600,3.190);
        \draw[gray!70,dashed] (A) -- ++(1.20,0);
        \draw[gray!70,dashed] ($(A)+(-0.35,-0.47)$) -- ($(A)+(0.65,0.88)$);
        \draw[->] ($(A)+(0.50,0)$) arc[start angle=0,end angle=53.5,radius=0.50];
        \node[anchor=west] at ($(A)+(0.66,0.48)$) {$\theta(s)$};
        \coordinate (B) at (9.466,2.021);
        \draw[load] (B) -- ++(0.82,0);
        \node[anchor=west] at ($(B)+(0.88,0)$) {$f_x$};
        \draw[load] (B) -- ++(0,-0.72);
        \node[anchor=north] at ($(B)+(0,-0.78)$) {$f_y$};

        \draw[dim] (2.95,0.75) -- (2.95,0.508);
        \node[anchor=south,font=\small]
            at (2.95,0.82) {$e(s)$};
        \foreach \xx/\yy in {2.138/0.693,2.415/0.634,2.681/0.562,2.952/0.508,3.243/0.498}
            \draw[decorate,decoration={coil,aspect=0.45,segment length=3.0pt,amplitude=2.0pt},thin]
                (\xx,\yy) -- (\xx,0.02);
        \node[anchor=east] at (1.92,0.28) {$k_s$};
        \draw[brown!65!black,thick] (1.88,0.02) -- (3.48,0.02);
        \foreach \xx/\yy in {1.98/-0.06,2.23/-0.16,2.48/-0.29,2.73/-0.39,2.98/-0.36,3.23/-0.22,3.43/-0.08}
            \draw[load,green!45!black] (\xx,\yy) -- (\xx,0.02);
        \draw[green!45!black,thin]
            (1.98,-0.06) .. controls (2.35,-0.24) and (2.55,-0.40) .. (2.73,-0.39)
            .. controls (3.02,-0.38) and (3.22,-0.19) .. (3.43,-0.08);
        \node[text=green!35!black] at (2.70,-0.58) {$R_s(s)$};
    \end{tikzpicture}
    \caption{Coordinate system, positive directions, and principal loads used in the structural formulation. The horizontal coordinate is referenced from the FP, whereas the vertical coordinate is referenced from the HOP and increases downward.}
    \label{fig:coordinate_system}
\end{figure*}
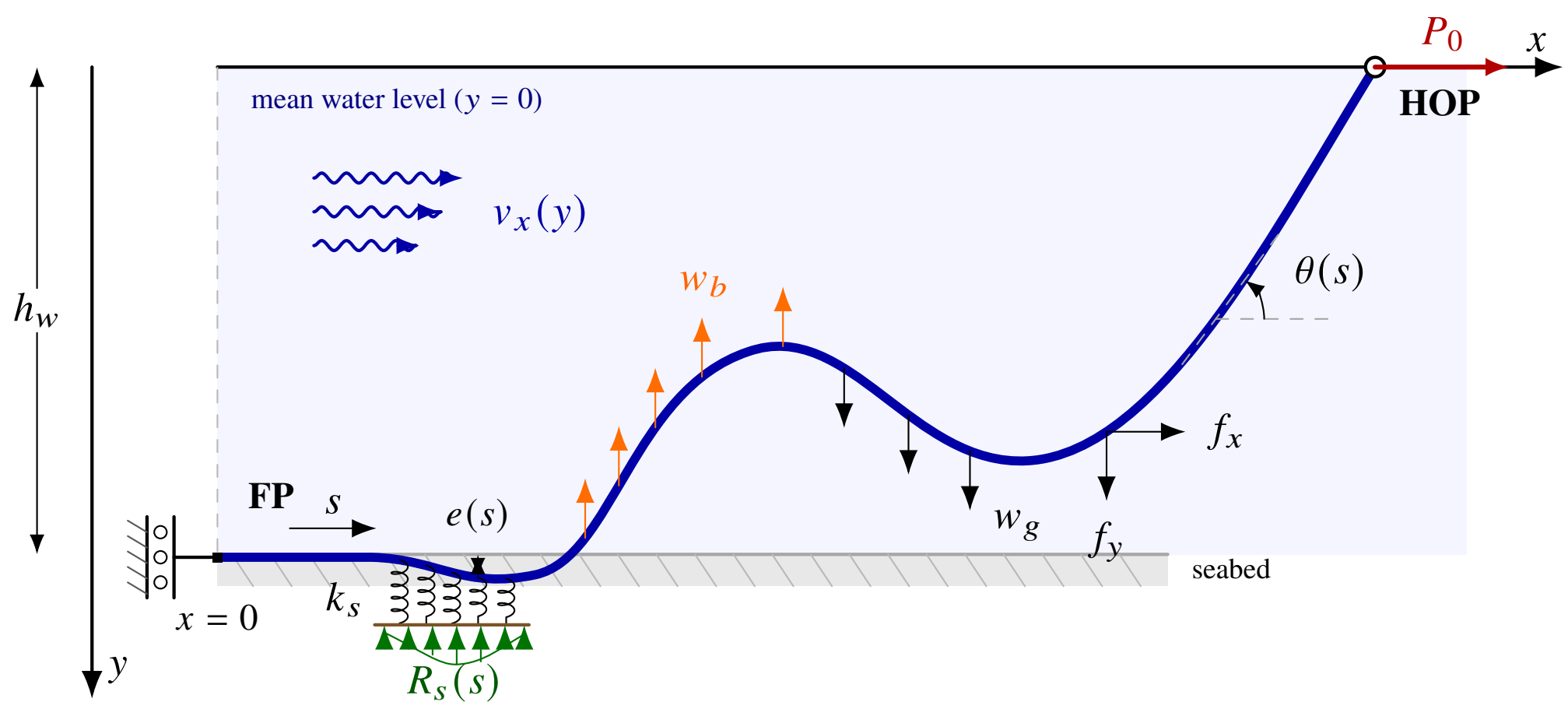

The line is modeled as an inextensible Euler-Bernoulli beam undergoing large planar rotations. The material response and seabed foundation are linear elastic, the seabed is horizontal and frictionless, and soil suction is neglected. 
The $x$-axis is horizontal and the $y$-axis is positive downward, with the seabed located at $y=h_w$. The $s\in[0,L]$ coordinate denotes the arc length measured from the fixed point (FP) at the laid end running toward the HOP. Let $\theta(s)$ denote the tangent angle. The coordinate system and principal sign conventions are illustrated in Figure~\ref{fig:coordinate_system}. 
The line coordinates are:
\begin{align}
    x(s) &= \int_0^s \cos\theta(\xi)\,\mathrm d\xi,
    \label{eq:x_reconstruction}\\
    y(s) &= -\int_s^L \sin\theta(\xi)\,\mathrm d\xi.
    \label{eq:y_reconstruction}
\end{align}
With this convention, $x(0)=0$ and $y(L)=0$ are satisfied automatically. At the FP the rotation is prescribed, whereas the HOP is modeled as a ball joint. The geometric and natural end conditions are therefore
\begin{equation}\label{eq:boundary_conditions}
\begin{split}
    &\theta(0)=2\pi, \\
    &y(L)=0, \\
    &M(L)=EI\,\theta'(L)=0.
\end{split}
\end{equation}
At the HOP, the floating facility applies a prescribed horizontal force $P_0$ to the riser. This external load is defined as positive when directed to the right, as illustrated in Figure~\ref{fig:coordinate_system}. It should be noted that $P_0$ is an input to the equilibrium problem, while the horizontal HOP position is obtained from the equilibrium configuration. The optimization therefore compares buoyancy layouts under the same prescribed horizontal force and water depth, with the corresponding horizontal span allowed to vary. This force controlled setting supports preliminary configuration design for a specified horizontal load transferred to the floating facility. 
The horizontal coordinates are referenced to the seabed point FP. Consequently, differences in the plotted HOP coordinates should not be interpreted as prescribed motions of the floating facility: the profiles can equivalently be translated to a common HOP location, so that changes in touchdown position are displayed relative to the facility. The horizontal distance between the HOP and FP is not prescribed.

The submerged structural weight per unit length is denoted by $w_g>0$. Introducing the nonnegative uplift-ratio field $r(s)$, the net distributed vertical load is written as
\begin{equation}
    w(r(s), s)=w_g\,[1-r(s)],
    \label{eq:generalized_weight}
\end{equation}
within the buoyancy section, while $r(s)=0$ elsewhere. The seabed penetration and reaction are
\begin{align}
    e(s) &= \max\bigl(0,y(s)-h_w\bigr),
    \label{eq:seabed_penetration}\\
    R_s(s) &= k_s e(s),
    \label{eq:seabed_reaction}
\end{align}
where $k_s$ is the Winkler stiffness per unit line length.

For the loads due to sea current, the depth dependent horizontal velocity is represented by a parabolic profile
\begin{equation}
    v_x(y)=v_0\left[1-\left(\frac{y}{h_w}\right)^2\right].
    \label{eq:current_profile}
\end{equation}
The normal and tangential Morison drag components per unit length are
\begin{align}
    f_n &=-\frac{1}{2}\rho_w C_nD
        \left|v_x(y)\sin\theta\right|v_x(y)\sin\theta,
    \label{eq:normal_drag}\\
    f_\tau &=\frac{1}{2}\rho_w C_\tau\pi D
        \left|v_x(y)\cos\theta\right|v_x(y)\cos\theta,
    \label{eq:tangential_drag}
\end{align}
with global components
\begin{equation}\label{eq:global_drag}
\begin{split}
    &f_x=-f_n\sin\theta+f_\tau\cos\theta, \\
    &f_y=f_n\cos\theta+f_\tau\sin\theta.
\end{split}
\end{equation}

The energy and external work terms are written as follows:
\begin{equation}
\begin{split}
    \Pi \left(\theta(s), r(s)\right)= &
        \frac{1}{2}\int_0^L EI\left(\frac{\mathrm d\theta}{\mathrm ds}\right)^2\mathrm ds
        \\ &+\frac{1}{2}\int_0^L k_s e(s)^2\,\mathrm ds \\
        &-\int_0^L w(r(s), s)y(s)\,\mathrm ds\\
        &-\int_0^L f_x(s)x(s)\,\mathrm ds \\
        &-\int_0^L f_y(s)y(s)\,\mathrm ds  \\
        &-P_0x(L). 
    \label{eq:total_potential}
\end{split}
\end{equation}
The first two terms are the bending and seabed elastic energies, while the remaining terms account for the external loads. In the absence of current, equilibrium follows from the stationarity of the total potential energy. When current is present, the hydrodynamic drag forces are treated as configuration dependent external loads through their virtual work. 
Their dependence on the local tangent angle and the depth dependent current velocity is fully included when differentiating the resulting equilibrium residual to construct the consistent Jacobian and the adjoint operator.

\subsection{Finite difference numerical approximation}

To compute the equilibrium configuration for a prescribed vector of uplift ratios $\vect r$, the arc-length domain $[0,L]$ is divided into $N$ uniform subintervals of length $\Delta s=L/N$, resulting in $N+1$ angular nodes. The coordinate relations in Eqs.~\eqref{eq:x_reconstruction}--\eqref{eq:y_reconstruction} are then approximated as
\begin{equation}
    x_i=\Delta s\sum_{j=1}^{i-1}\cos\theta_j,
    \qquad
    y_i=-\Delta s\sum_{j=i}^{N}\sin\theta_j.
    \label{eq:discrete_coordinates}
\end{equation}
The prescribed FP angle is $\theta_1=2\pi$. At the HOP, the zero-moment condition is imposed in the implementation as
\begin{equation}
    \theta_{N+1}=\theta_N,
    \label{eq:hop_discrete_bc}
\end{equation}
so that the independent equilibrium variables of the final configuration are
\begin{equation}
    \vect \theta=\begin{bmatrix}\theta_2&\theta_3&\cdots&\theta_N\end{bmatrix}^{\mathsf T}.
    \label{eq:reduced_state_vector}
\end{equation}
This explicit treatment of the HOP condition is also accounted for in the adjoint system in Sec.~\ref{sec:sensitivity}. 
The nodal seabed stiffness and net vertical load are
\begin{equation}\label{eq:nodal_loads}
\begin{split}
    &k_i^{\mathrm{SB}}=
    \begin{cases}
        k_s\Delta s, & y_i\ge h_w,\\
        0, & y_i<h_w,
    \end{cases}\\
    &w_i=w(s_i, \vect r)\Delta s.
\end{split}
\end{equation}
The nodal drag components include the tributary length $\Delta s$,
\begin{equation}
\begin{split}
    f_{n,i} &=-\frac{1}{2}\rho_w C_nD\Delta s
       |v_i\sin\theta_i|v_i\sin\theta_i,\\
    f_{\tau,i} &=\frac{1}{2}\rho_w C_\tau\pi D\Delta s
       |v_i\cos\theta_i|v_i\cos\theta_i,
\end{split}
\end{equation}
where $v_i=v_x(y_i)$, followed by the transformation in Eq.~\eqref{eq:global_drag}. It is convenient to introduce the cumulative quantities
\begin{align}
    K_i^{\mathrm{SB}}&=\sum_{k=1}^{i}k_k^{\mathrm{SB}},
    &W_i&=\sum_{k=1}^{i}w_k,\notag\\
    F_{x,i}&=\sum_{k=i}^{N+1}f_{x,k},
    &F_{y,i}&=\sum_{k=1}^{i}f_{y,k}.
    \label{eq:cumulative_quantities}
\end{align}

Using a left Riemann approximation, the equilibrium equations follow from the discrete virtual work balance. This reduces to stationarity of the discrete total potential energy with respect to the free angles:
\begin{equation}
    \frac{\partial \Pi (\vect \theta,  \vect r)}{\partial \theta_i} = 0 \text{ for } i=2, \dots, N.
\end{equation}
The equilibrium residual associated with an interior node $i=2,\ldots,N$ can be written in the following compact form:
\begin{equation}\label{eq:fd_residual}
\begin{split}
    &R_i=\frac{EI}{\Delta s}
       \left(-\theta_{i-1}+2\theta_i-\theta_{i+1}\right) \\
    &+\Delta s\cos\theta_i
      \sum_{j=1}^{N}K_{\min(i,j)}^{\mathrm{SB}}\Delta s\sin\theta_j
      \\ &+h_w K_i^{\mathrm{SB}}\Delta s\cos\theta_i \\
    &+F_{x,i}\Delta s\sin\theta_i
      +\left(W_i+F_{y,i}\right)\Delta s\cos\theta_i \\
    &  +P_0\Delta s\sin\theta_i=0.
\end{split}
\end{equation}
In matrix notation the equilibrium equations are written as follows: 
\begin{equation}
    \vect R(\vect\theta,\vect r)
    =\vect K_b\vect\theta+\vect F(\vect\theta,\vect r)=\vect 0,
    \label{eq:discrete_equilibrium}
\end{equation}
where $\vect K_b$ contains the constant bending contribution, with the familiar tridiagonal stencil $(-1,2,-1)EI/\Delta s$, and $\vect F$ contains the seabed, gravity, buoyancy, current, and HOP load terms. 
The Jacobian matrix is denoted by
\begin{equation}
    \vect J=\frac{\partial\vect R}{\partial\vect\theta}.
    \label{eq:tangent_matrix}
\end{equation}

\subsection{Incremental and iterative solution approach}

The equilibrium equations in Eq.~\eqref{eq:discrete_equilibrium} form a nonlinear algebraic system that is solved using the Newton-Raphson method. Because convergence to the fully buoyant configuration can be sensitive to the initial estimate, the solution is obtained through a two stage continuation procedure. In the first stage, a free hanging catenary provides the initial configuration, and the wet length of the riser is increased in a step wise manner by one subinterval $\Delta s$ at each step. This procedure mimics the progressive release of the riser into the water. Neglecting bending stiffness in the initial catenary estimate gives
\begin{equation}\label{eq:initial_catenary}
\begin{split}
    L_0&=\frac{P_0}{w_g}
    \sinh\left[\operatorname{arcosh}\left(\frac{w_g h_w}{P_0}+1\right)\right],\\
    \theta_i^{(0)}&=2\pi-\tan^{-1}\left[\frac{w_g}{P_0}(i-1)\Delta s\right].
\end{split}
\end{equation}
The configuration defined in Eq.~\eqref{eq:initial_catenary} is then used to start the computations in the first stage, where the length of the riser is increased progressively. 
The converged configuration at each length increment initializes the next increment while the active part of the seabed that produces reaction forces on the riser is updated. In the second stage, the total length is held fixed and the prescribed buoyancy load is ramped from zero to its full value. The current implementation uses 100 equal buoyancy increments.

At a generic nonlinear iteration $j$, let $\vect R$ and $\vect J$ denote the residual and Jacobian restricted to the current free angular variables. The Newton direction is obtained from
\begin{equation}
    \vect J^{(j)}\Delta\vect \theta^{(j)}=-\vect R^{(j)}.
    \label{eq:newton_step}
\end{equation}
To improve convergence of the Newton--Raphson iterations in both stages, the software implementation developed uses a merit function based on the residual:
\begin{equation}
    \phi(\vect \theta)=\frac{1}{2}\|\vect R(\vect \theta)\|_2^2,
    \qquad
    \nabla\phi=\vect J^{\mathsf T}\vect R.
    \label{eq:merit_function}
\end{equation}
Starting with $\alpha=1$, a trial state
\begin{equation}
    \vect \theta_{\mathrm{tr}}=\vect \theta^{(j)}+\alpha\Delta\vect \theta^{(j)}
    \label{eq:newton_trial}
\end{equation}
is accepted when the Armijo sufficient decrease condition is satisfied \citep{nocedal2006numerical}:
\begin{equation}
    \phi(\vect \theta_{\mathrm{tr}})
    \le
    \phi(\vect \theta^{(j)})
    +c_A\alpha\,\nabla\phi(\vect \theta^{(j)})^{\mathsf T}\Delta\vect \theta^{(j)}.
    \label{eq:armijo_condition}
\end{equation}
Otherwise, $\alpha$ is reduced according to $\alpha\leftarrow\rho_A\alpha$ and the residual is reevaluated, including the current contact set and configuration dependent loads. The implementation uses
\begin{equation}
    c_A=10^{-4},\qquad \rho_A=0.5,
    \label{eq:armijo_parameters}
\end{equation}
with the full Newton step always attempted first. The backtracking search is used in both stages of the structural analysis approach, that is, in the length continuation and in the 100 step buoyancy continuation. In both stages, convergence is based on
\begin{equation}
    \|\Delta\vect \theta\|_\infty < \epsilon,
    \label{eq:newton_tolerance}
\end{equation}
where, for example $\epsilon = 10^{-6}$.

For clarity, the implemented two-stage structural analysis approach for offshore risers may be summarized as follows:
\begin{enumerate}
    \item Construct the initial free-hanging catenary from Eq.~\eqref{eq:initial_catenary};
    \item Increase the line length one subinterval at a time, update seabed contact, and solve each equilibrium state by Newton-Raphson with Armijo backtracking;
    \item Hold the final length fixed, apply the buoyancy uplift effects in 100 increments, and solve each increment by Newton-Raphson with Armijo backtracking while updating the contact set;
    \item Recover geometry and internal force quantities from the converged layout (defined in terms of the angular field $\vect \theta$).
\end{enumerate}


After equilibrium is obtained, curvature, bending moment, and shear are evaluated using
\begin{equation}
\begin{split}
    &\kappa_i=\frac{\theta_i-\theta_{i-1}}{\Delta s},
    \quad
    M_i=EI\kappa_i,
    \quad
    i=2,\ldots,N+1,\\
    &S_i=-\frac{M_i-M_{i-1}}{\Delta s},
    \quad
    i=3,\ldots,N+1.
\end{split}
\label{eq:response_recovery}
\end{equation}
Here, following the indexing of \citet{trapper2020feasible}, $\kappa_i$ and $M_i$ are associated with the segment between nodes $i-1$ and $i$, so that the HOP condition in Eq.~\eqref{eq:hop_discrete_bc} gives $\kappa_{N+1}=M_{N+1}=0$. The shear force $S_i$ is obtained from the bending moments of the two adjacent segments. In particular, $S_{N+1}$ is the shear force at the HOP.
The axial tension is recovered from cumulative equilibrium of the horizontal and vertical nodal loads and the HOP reaction. The corresponding expressions for the nodal tension, the vertical reaction at the HOP, and the tension at the HOP are given in Eqs.~(47)--(49) of \citet{trapper2020feasible}.  

\subsection{Assessment of the nonlinear structural analysis solver}

\begin{figure}[htbp!]
    \centering
    \includegraphics[width=0.95\linewidth]{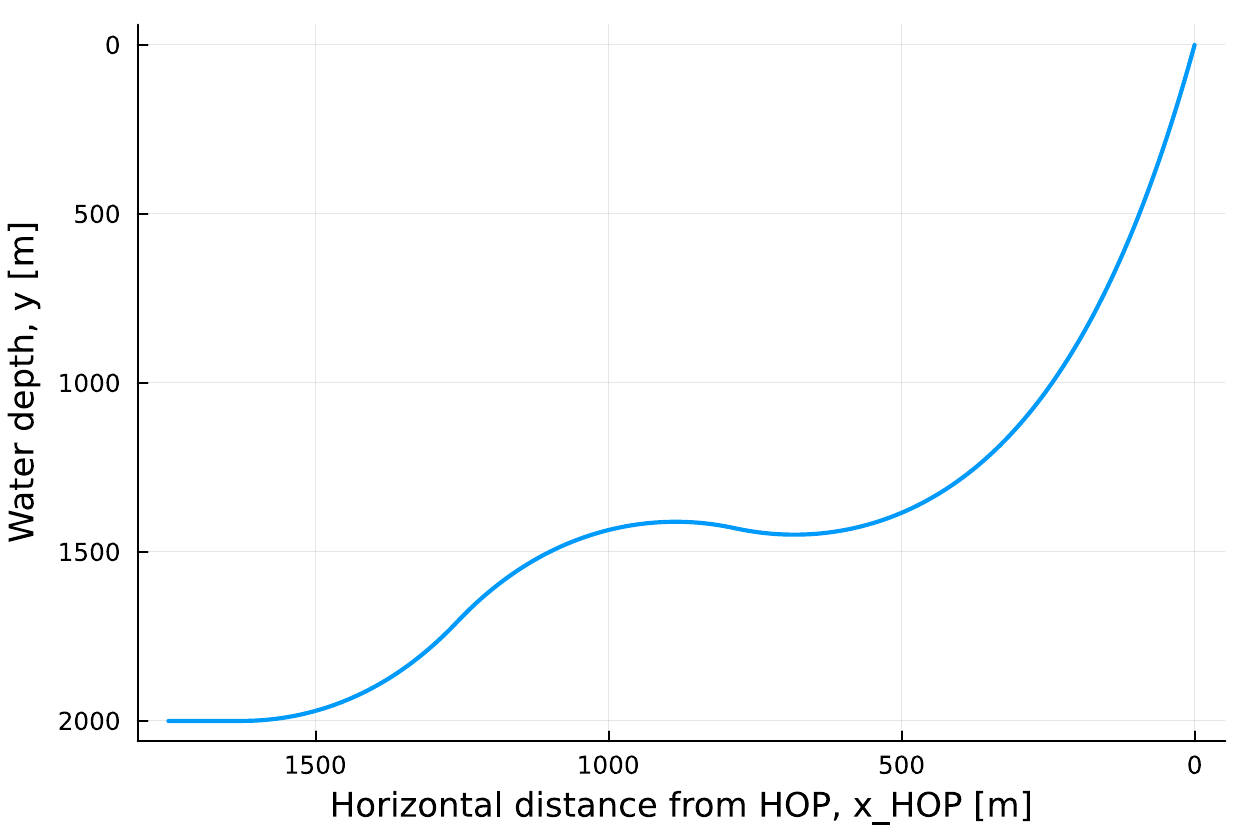}
    \caption{Computed equilibrium configuration for the baseline case.}
    \label{fig:trapper_configuration}
\end{figure}

\begin{figure}[htbp!]
    \centering
    \includegraphics[width=0.95\linewidth]{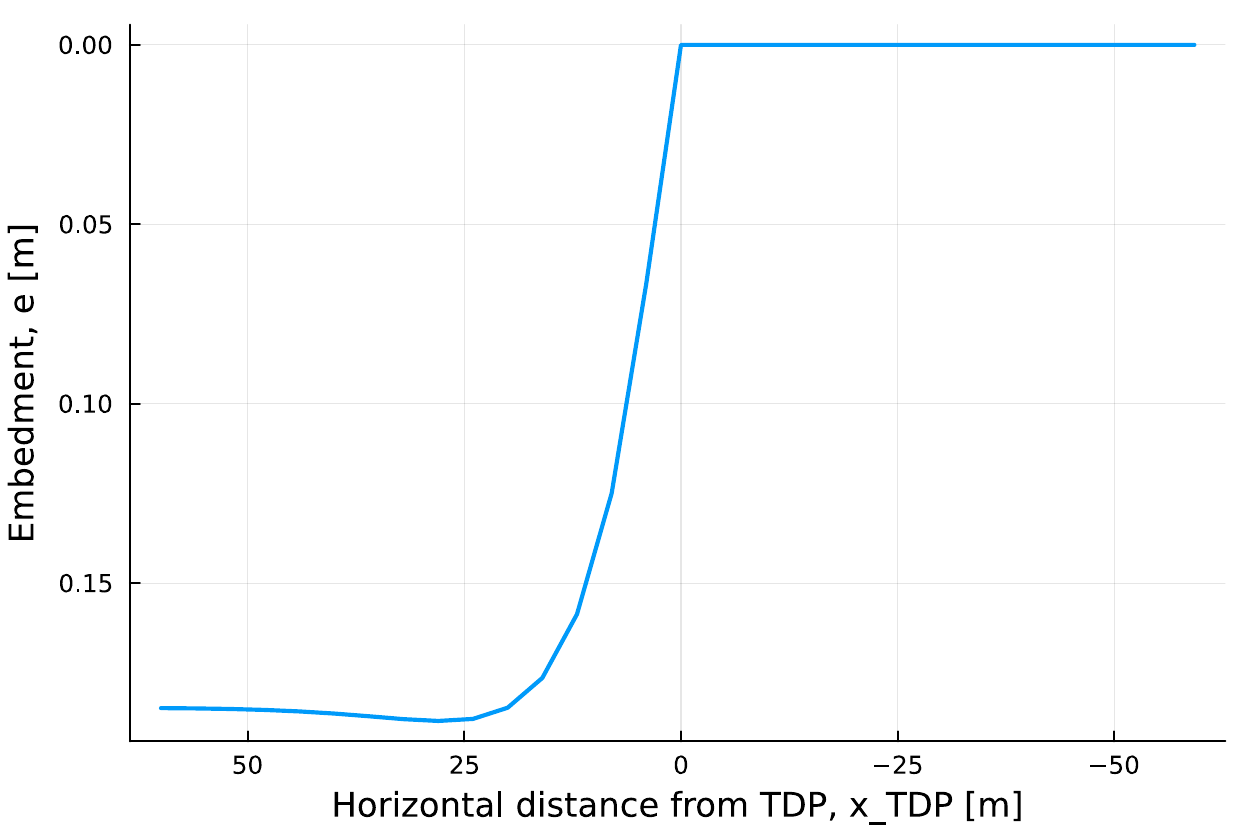}
    \caption{Computed seabed embedment distribution for the baseline case. The maximum value is compared with the reference result in Table~\ref{tab:trapper_validation}.}
    \label{fig:trapper_embedment}
\end{figure}

\begin{figure}[htbp!]
    \centering
    \includegraphics[width=0.95\linewidth]{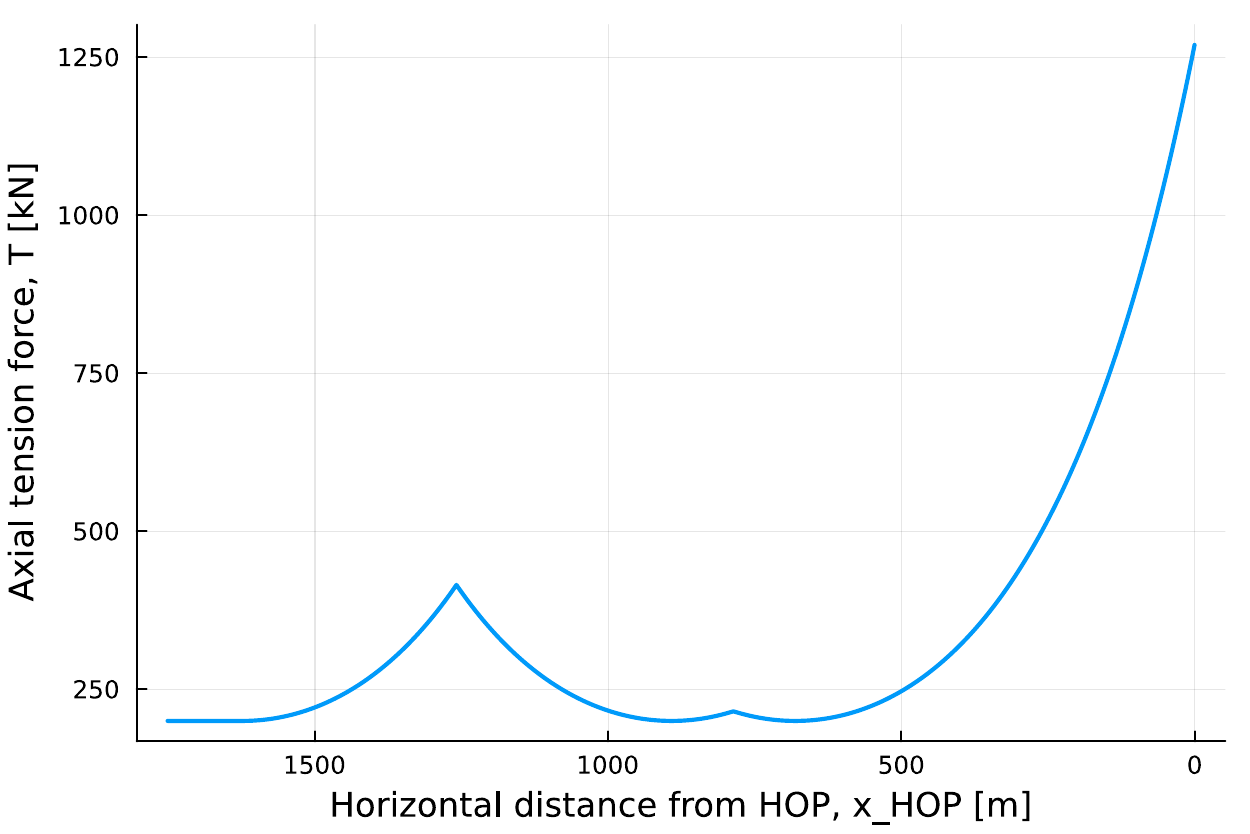}
    \caption{Computed axial-tension distribution for the baseline case. The maximum value is compared with the reference result in Table~\ref{tab:trapper_validation}.}
    \label{fig:trapper_tension}
\end{figure}

\begin{figure}[htbp!]
    \centering
    \includegraphics[width=0.95\linewidth]{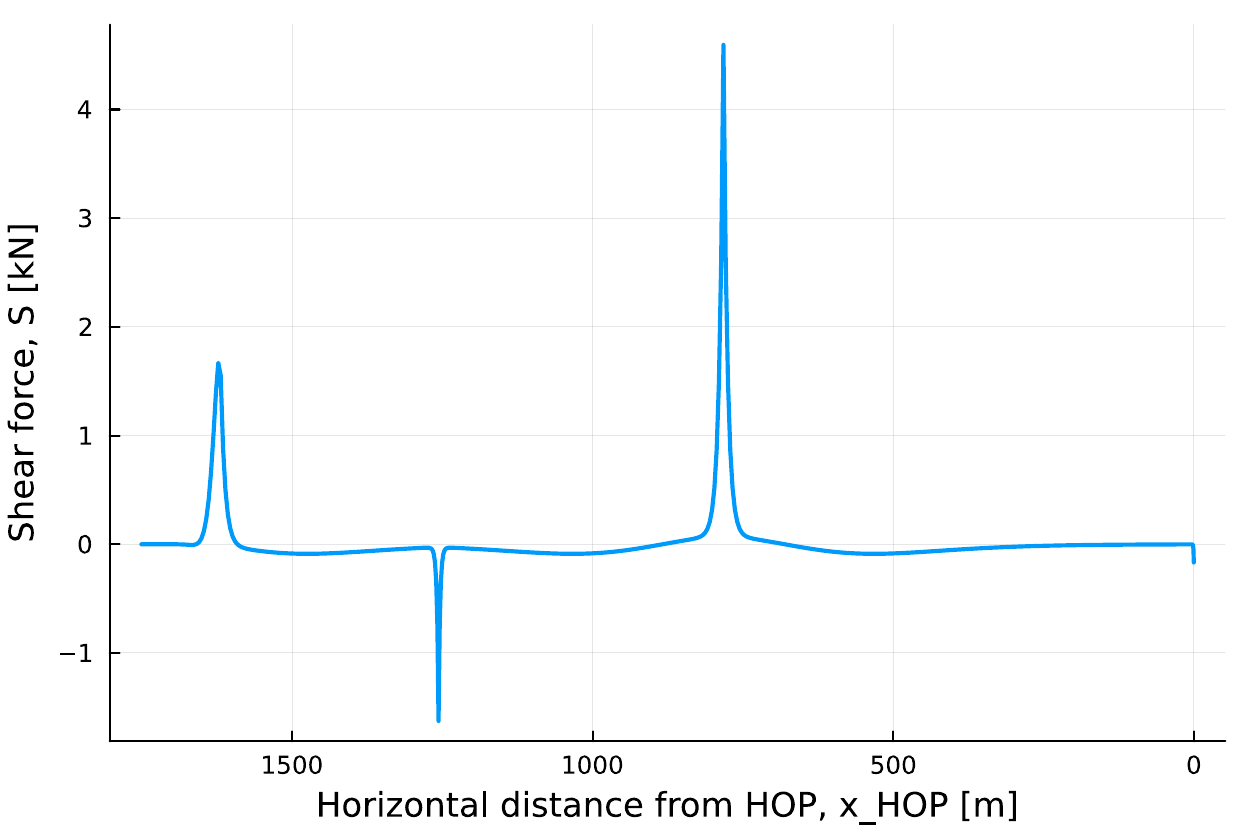}
    \caption{Computed shear-force distribution for the baseline case. Table~\ref{tab:trapper_validation} compares the peak at the lower buoyancy boundary with the reference range.}
    \label{fig:trapper_shear}
\end{figure}

\begin{figure}[htbp!]
    \centering
    \includegraphics[width=0.95\linewidth]{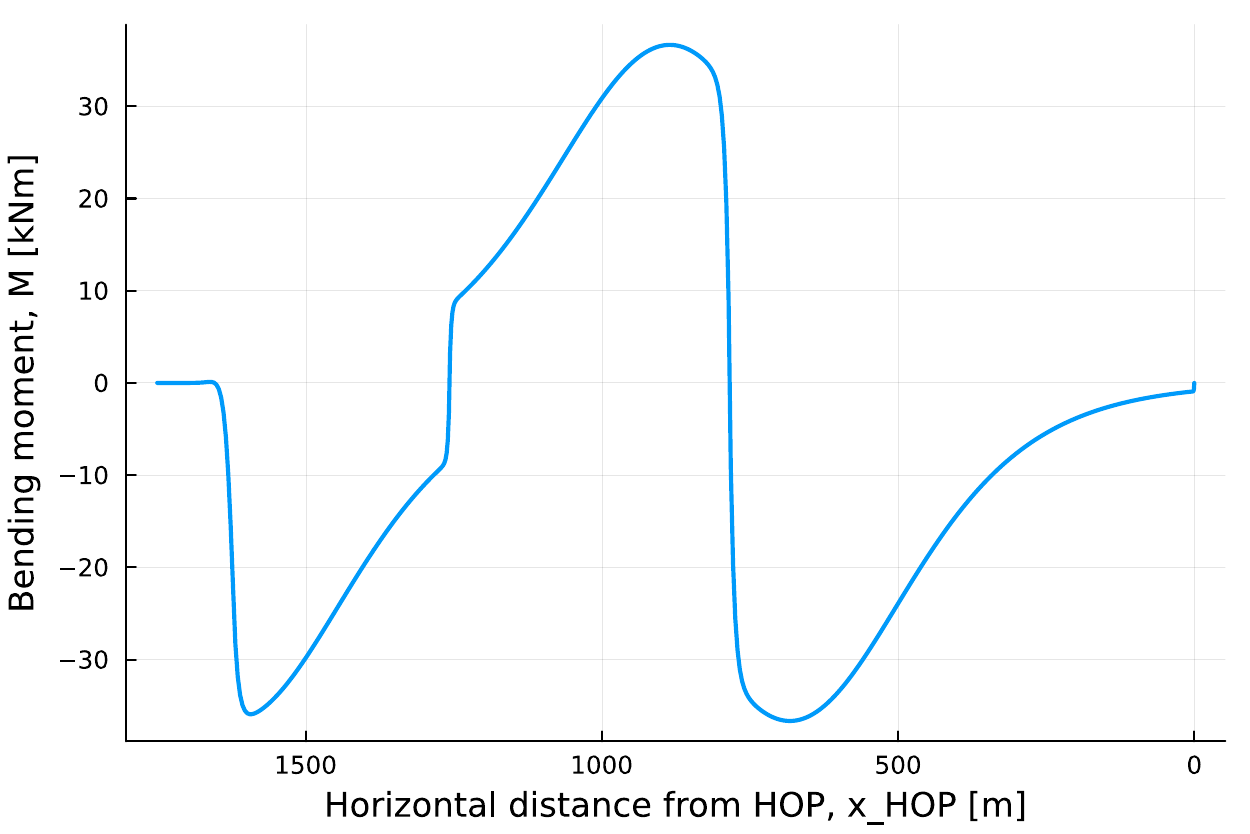}
    \caption{Computed bending-moment distribution for the baseline case. The maximum value is compared with the reference result in Table~\ref{tab:trapper_validation}.}
    \label{fig:trapper_moment}
\end{figure}

In this section, the implemented analysis approach is validated by comparing its predictions with the results reported by \citet{trapper2020feasible}. Figs.~\ref{fig:trapper_configuration}--\ref{fig:trapper_moment} present the computed equilibrium configuration, embedment in the seabed, axial tension, shear force, and bending moment, respectively. Figs.~\ref{fig:trapper_configuration}--\ref{fig:trapper_moment} show the configuration and response obtained by the authors, while the quantitative comparison of selected response quantities with the reference results is provided in Table~\ref{tab:trapper_validation}.

Table~\ref{tab:basic_parameters} reports the basic lazy-wave case of \citet{trapper2020feasible}. The validation grid used in the present code has $\Delta s=4$~m, $N=756$, and hence a total length $L=3024$~m.

\begin{table*}[htbp!]
\centering
\caption{Baseline parameters for the lazy-wave configuration.}
\label{tab:basic_parameters}
\renewcommand{\arraystretch}{1.15}
\begin{tabular}{p{0.51\textwidth}cc}
\toprule
\textbf{Parameter} & \textbf{Symbol} & \textbf{Value} \\
\midrule
Outer diameter [m] & $D$ & 0.2032 \\
Wall thickness [m] & $t$ & 0.0191 \\
Elastic modulus [N/m$^2$] & $E$ & $2.1\times10^{11}$ \\
Steel density [kg/m$^3$] & $\rho_s$ & 7850 \\
Seawater density [kg/m$^3$] & $\rho_w$ & 1030 \\
Normal drag coefficient & $C_n$ & 1.2 \\
Tangential drag coefficient & $C_\tau$ & 0.024 \\
Gravitational acceleration [m/s$^2$] & $g$ & 9.81 \\
Horizontal component of top tension [N] & $P_0$ & $0.20\times10^6$ \\
Water depth [m] & $h_w$ & 2000 \\
Upper-section length [m] & $L_u$ & 1800 \\
Buoyancy-section length [m] & $L_b$ & 600 \\
Reference uplift ratio & $r_{\mathrm{up}}$ & 2 \\
Seabed stiffness per unit line length [N/m$^2$] & $k_s$ & $4\times10^3$ \\
Surface-current velocity [m/s] & $v_0$ & 0 \\
Finite difference spacing [m] & $\Delta s$ & 4 \\
\bottomrule
\end{tabular}
\end{table*}

Table~\ref{tab:trapper_validation} summarizes the comparison of the maximum embedment, axial tension, and bending moment for the baseline case defined in Table~\ref{tab:basic_parameters}. The reference values are estimated from the convergence plots at $N=756$ in Fig.~8 of \citet{trapper2020feasible}. The relative differences are approximately $0.1\%$ or less for these three quantities. Because the reference values are read from plots rather than tabulated directly, these differences should be interpreted as approximate measures of agreement, not as precise estimates of numerical error.

The computed shear-force distribution in Fig.~\ref{fig:trapper_shear} exhibits localized peaks at the boundaries of the buoyancy section, where the distributed buoyancy load changes abruptly. At the lower boundary, $s=L-L_u-L_b$, the computed peak is $1.61\times10^{3}$~N, close to the upper end of the approximate range $(1.25\text{--}1.6)\times10^{3}$~N read from Fig.~8 of \citet{trapper2020feasible}. At the upper boundary, $s=L-L_u$, the computed peak has a magnitude of $4.59\times10^{3}$~N. Since this peak is not separately tabulated in the reference study, the quantitative shear comparison in Table~\ref{tab:trapper_validation} is restricted to the lower boundary. These results positively support the consistency of the present implementation with the results reported by \citet{trapper2020feasible}.

\begin{table*}[htbp!]
\centering
\caption{Comparison of maximum response quantities with
\citet{trapper2020feasible} for the baseline case ($\Delta s=4$~m, $N=756$).}
\label{tab:trapper_validation}
\renewcommand{\arraystretch}{1.15}
\begin{tabular}{p{0.30\textwidth}cccc}
\toprule
\textbf{Quantity} & \textbf{Symbol} & \textbf{This work} & \textbf{Trapper (2020)} & \textbf{Diff.\ [\%]} \\
\midrule
Maximum embedment [m]                       & $e_{\max}$    & $0.1884$                 & $\approx 0.1883$               & $0.05$ \\
Maximum axial tension [N]                   & $T_{\max}$    & $1.2695\times10^{6}$     & $\approx 1.2683\times10^{6}$   & $0.09$ \\
Maximum bending moment [N\,m]               & $M_{\max}$    & $3.666\times10^{4}$      & $\approx 3.6623\times10^{4}$   & $0.10$ \\
Shear peak at lower buoyancy boundary [N]   & $S_{peak}$           & $1.61\times10^{3}$       & $\approx (1.25\text{--}1.6)\times10^{3}$ & n/a \\
\bottomrule
\end{tabular}
\par\smallskip
\parbox{\textwidth}{\small Reference values are estimated from Fig.~8 of \citet{trapper2020feasible} at $N=756$. Relative differences are calculated as the absolute difference divided by the reference value and expressed as percentages. The shear comparison uses an approximate range read from the plot rather than a single reference value.}
\end{table*}

\section{Sensitivity analysis of the structural response}
\label{sec:sensitivity}

Gradient-based optimization requires derivatives of the objective and constraints with respect to the nodal uplift ratios. The objective in Eq.~\eqref{eq:objective_buoyancy} is explicit and has the analytical gradient $2\vect r$. The curvature and tension constraints depend on the design both explicitly and through the equilibrium angles, and their gradients are evaluated by a discrete adjoint method. Central finite difference sensitivities are used only as an independent verification tool.

\subsection{Discrete adjoint sensitivity analysis formulation}
\label{sec:adjoint_formulation}

Consider a scalar response $g=g(\vect \theta(\vect r),\vect r)$ and the reduced equilibrium equations
\begin{equation}
    \widetilde{\vect R}(\vect \theta,\vect r)=\vect 0,
    \label{eq:reduced_equilibrium_sensitivity}
\end{equation}
where $\vect \theta=[\theta_2,\ldots,\theta_N]^{\mathsf T}$ contains the independent angles. Direct differentiation with respect to a design variable $r_i$ gives
\begin{equation}
    \widetilde{\vect J}\frac{\mathrm d\vect \theta}{\mathrm d r_i}
    +\frac{\partial\widetilde{\vect R}}{\partial r_i}=\vect 0,
    \qquad
    \widetilde{\vect J}=\frac{\partial\widetilde{\vect R}}{\partial\vect \theta}.
    \label{eq:direct_sensitivity}
\end{equation}
The total response derivative is
\begin{equation}
    \frac{\mathrm dg}{\mathrm d r_i}
    =\frac{\partial g}{\partial r_i}
    +\frac{\partial g}{\partial\vect \theta}
      \frac{\mathrm d\vect \theta}{\mathrm d r_i}.
    \label{eq:total_derivative}
\end{equation}
Solving Eq.~\eqref{eq:direct_sensitivity} separately for every $r_i$ would require one linear system solution per design variable. Instead, define the adjoint vector by
\begin{equation}
    \widetilde{\vect J}^{\mathsf T}\vect\lambda_g
    =-\left(\frac{\partial g}{\partial\vect \theta}\right)^{\mathsf T}.
    \label{eq:adjoint_system}
\end{equation}
The complete gradient is then
\begin{equation}
    \frac{\mathrm dg}{\mathrm d\vect r}
    =\frac{\partial g}{\partial\vect r}
    +\vect\lambda_g^{\mathsf T}
      \frac{\partial\widetilde{\vect R}}{\partial\vect r}.
    \label{eq:adjoint_gradient}
\end{equation}
Thus, one adjoint system solve is required for each scalar response function that needs to be differentiated, independent of the number of (buoyancy) design variables.

An important implementation detail concerns the HOP condition in Eq.~\eqref{eq:hop_discrete_bc}. Although the full structural routine evaluates the residual and its Jacobian in terms of all nodal angles, the angle \(\theta_{N+1}\) is not an independent state variable because the zero moment boundary condition imposes \(\theta_{N+1}=\theta_N\). Consequently, a perturbation of the last independent angle \(\theta_N\) also produces the same perturbation in \(\theta_{N+1}\), that is,
\begin{equation}
\mathrm d\theta_{N+1}=\mathrm d\theta_N.
\end{equation}
For an equilibrium residual \(R_i\), the contributions associated with these two angles are therefore
\begin{equation}\label{eq:hop_jacobian_chain_rule}
\begin{split}
\mathrm dR_i
&=
\cdots+
\frac{\partial R_i}{\partial\theta_N}\mathrm d\theta_N
+
\frac{\partial R_i}{\partial\theta_{N+1}}\mathrm d\theta_{N+1}
\\
&=\cdots+
\left(
\frac{\partial R_i}{\partial\theta_N}
+
\frac{\partial R_i}{\partial\theta_{N+1}}
\right)\mathrm d\theta_N.
\end{split}
\end{equation}
Hence, after eliminating the dependent angle \(\theta_{N+1}\), the derivative of the reduced residual with respect to the last independent angle is
\begin{equation}
\frac{\mathrm d\widetilde R_i}{\mathrm d\theta_N}
=
\frac{\partial R_i}{\partial\theta_N}
+
\frac{\partial R_i}{\partial\theta_{N+1}}.
\label{eq:hop_jacobian_reduction}
\end{equation}
Let $\mathcal I=\{2,\ldots,N\}$ denote the interior index set. The reduced Jacobian is assembled by retaining the rows and columns associated with the indices in the set $\mathcal I$ and adding the entries of the full Jacobian $\theta_{N+1}$ column in these rows to the retained $\theta_N$ column. Using local row and column indices $i,j=1,\ldots,N-1$, its entries are
\begin{equation}
    \begin{gathered}
        \widetilde J_{ij}=
        \begin{cases}
            \displaystyle\frac{\partial R_{i+1}}{\partial\theta_{j+1}},
            & j=1,\ldots,N-2,\\[8pt]
            \displaystyle\frac{\partial R_{i+1}}{\partial\theta_N}
            +\frac{\partial R_{i+1}}{\partial\theta_{N+1}},
            & j=N-1,
        \end{cases}\\
        i=1,\ldots,N-1.
    \end{gathered}
    \label{eq:folded_jacobian}
\end{equation}
The first $N-2$ columns therefore retain the derivatives with respect to $\theta_2,\ldots,\theta_{N-1}$, while the final column accounts for the simultaneous variation of $\theta_N$ and $\theta_{N+1}$ imposed by the HOP condition. This preserves the contribution of the dependent angle after enforcing $\theta_{N+1}=\theta_N$. The same chain rule is applied to the full-angle derivative of a scalar response $g$: the derivatives with respect to $\theta_N$ and $\theta_{N+1}$ are added to form the last entry of the reduced response derivative used in the adjoint system.

For a design variable $r_i$ located at global node $i$, the direct effect of buoyancy on the equilibrium residual follows from Eq.~\eqref{eq:design_nodal_weight}. Holding the angles fixed gives
\begin{equation}
    \left.\frac{\partial R_j}{\partial r_i}\right|_{\vect\theta}
    =
    \begin{cases}
        -w_g\Delta s^2\cos\theta_j, & j\ge i,\\
        0, & j<i,
    \end{cases}
    \label{eq:residual_design_derivative}
\end{equation}
followed by restriction to the equilibrium rows $\mathcal I$. This matrix is common to the tension and curvature adjoint calculations.

\subsection{Curvature constraint}

The curvature aggregate is given by Eq.~\eqref{eq:kappa_p}, with
\begin{equation}
    \kappa_i=\frac{\theta_i-\theta_{i-1}}{\Delta s},
    \qquad i=2,\ldots,N+1.
    \label{eq:curvature_fd_sensitivity}
\end{equation}
Its derivative with respect to the segment curvature is
\begin{equation}
    \frac{\partial\kappa_p}{\partial\kappa_i}
    =\frac{|\kappa_i|^{p-1}\operatorname{sgn}(\kappa_i)}
    {\kappa_p^{p-1}},
    \label{eq:kappa_p_derivative}
\end{equation}
and
\begin{equation}
    \frac{\partial\kappa_i}{\partial\theta_k}
    =\frac{1}{\Delta s}\left(\delta_{k,i}-\delta_{k,i-1}\right).
    \label{eq:kappa_theta_derivative}
\end{equation}
The vector $\partial\kappa_p/\partial\vect\theta$ is assembled by the chain rule and folded similarly to Eq.~\eqref{eq:folded_jacobian}. Since the curvature aggregate has no explicit dependence on $\vect r$ at fixed angles, its gradient is
\begin{equation}\label{eq:kappa_adjoint_gradient}
\begin{split}
    &\frac{\mathrm d\kappa_p}{\mathrm d\vect r}
    =\vect\lambda_\kappa^{\mathsf T}
     \frac{\partial\widetilde{\vect R}}{\partial\vect r},\\
    &\widetilde{\vect J}^{\mathsf T}\vect\lambda_\kappa
    =-\left(\frac{\partial\kappa_p}{\partial\vect \theta}\right)^{\mathsf T}.
\end{split}
\end{equation}

\subsection{Tension constraint}
\label{sec:tension_sensitivity}

For the tension aggregate of Eq.~\eqref{eq:tension_p}, define
\begin{equation}
    \omega_j^T
    =\frac{|T_j|^{p-1}\operatorname{sgn}(T_j)}{T_p^{p-1}}.
    \label{eq:tension_p_weight}
\end{equation}
Then
\begin{equation} \label{eq:tension_chain_rule}
\begin{split}
    &\frac{\partial T_p}{\partial\theta_k}
    =\sum_{j=1}^{N+1}\omega_j^T
      \frac{\partial T_j}{\partial\theta_k},\\
    &\left.\frac{\partial T_p}{\partial r_i}\right|_{\vect\theta}
    =\sum_{j=1}^{N+1}\omega_j^T
      \left.\frac{\partial T_j}{\partial r_i}\right|_{\vect\theta}.
\end{split}
\end{equation}
The nodal tension depends on the HOP reaction and on the cumulative horizontal, vertical, and gravity loads. These terms are differentiated analytically in the implementation. Under the sign convention in which the recovered tensile force is positive, the explicit contribution of a buoyancy variable at node $i$ is
\begin{equation}
    \left.\frac{\partial T_j}{\partial r_i}\right|_{\vect\theta}
    =
    \begin{cases}
        -w_g\Delta s\sin\theta_j, & j < i,\\
        0, & j\ge i.
    \end{cases}
    \label{eq:tension_explicit_design_derivative}
\end{equation}
The state derivative $\partial T_i/\partial\theta_k$ includes the derivative of the HOP reaction and, when current is present, the derivatives of the drag forces. For example, the geometry and velocity contributions satisfy
\begin{equation}\label{eq:current_state_derivatives}
\begin{split}
    &\frac{\partial y_i}{\partial\theta_k}
    =
    \begin{cases}
        -\Delta s\cos\theta_k, & i\le k\le N,\\
        0, & \text{otherwise},
    \end{cases}\\
   &\frac{\partial v_i}{\partial\theta_k}
    =-\frac{2v_0y_i}{h_w^2}
      \frac{\partial y_i}{\partial\theta_k},
\end{split}
\end{equation}
with the remaining drag derivatives obtained directly from Eqs.~\eqref{eq:normal_drag}--\eqref{eq:global_drag}. The complete tension gradient is finally
\begin{equation}\label{eq:tension_adjoint_gradient}
\begin{split}
    &\frac{\mathrm dT_p}{\mathrm d\vect r}
    =\left.\frac{\partial T_p}{\partial\vect r}\right|_{\vect \theta}
     +\vect\lambda_T^{\mathsf T}
      \frac{\partial\widetilde{\vect R}}{\partial\vect r}, \\
    &\widetilde{\vect J}^{\mathsf T}\vect\lambda_T
    =-\left(\frac{\partial T_p}{\partial\vect \theta}\right)^{\mathsf T}.
\end{split}
\end{equation}

\subsection{Finite difference verification of the analytical gradients}

\begin{figure}[htbp!]
    \centering
    \includegraphics[width=\linewidth]{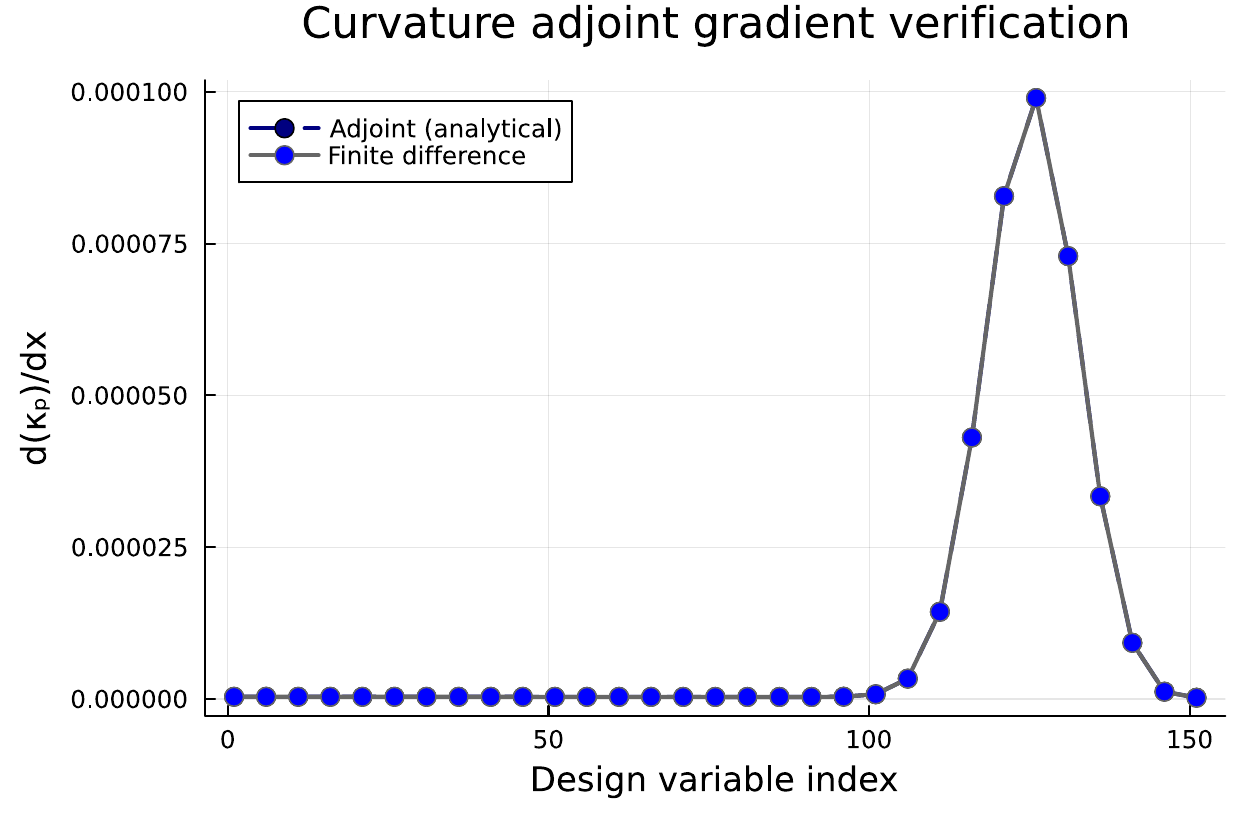}
    \caption{Comparison of adjoint and central finite difference derivatives of the $p$-norm curvature aggregate at the 31 sampled design-variable indices ($p=50$, $h=10^{-5}$).}
    \label{fig:curvature_gradient_check}
\end{figure}

\begin{figure}[htbp!]
    \centering
    \includegraphics[width=\linewidth]{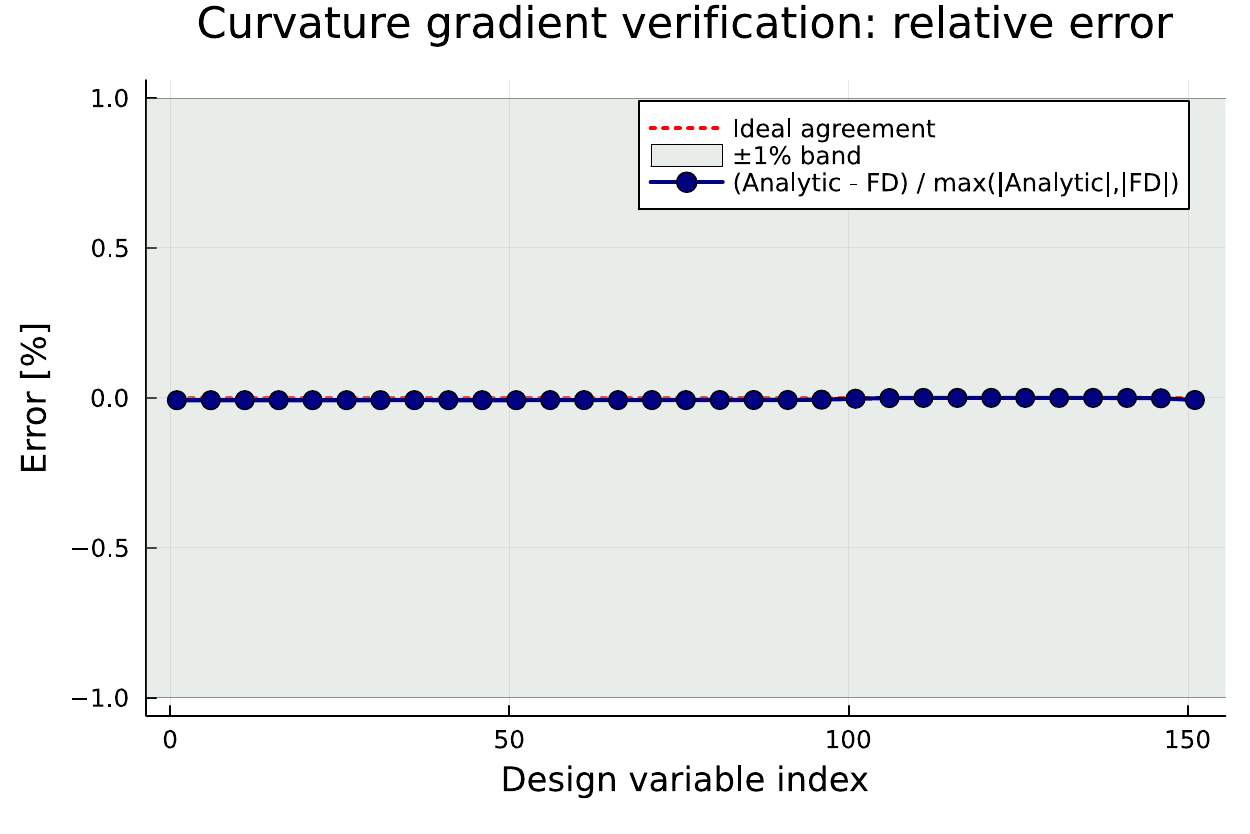}
    \caption{Signed relative discrepancies between adjoint and finite difference curvature derivatives, normalized by the larger absolute derivative and expressed as percentages ($p=50$, $h=10^{-5}$).}
    \label{fig:curvature_gradient_error}
\end{figure}

\begin{figure}[htbp!]
    \centering
    \includegraphics[width=\linewidth]{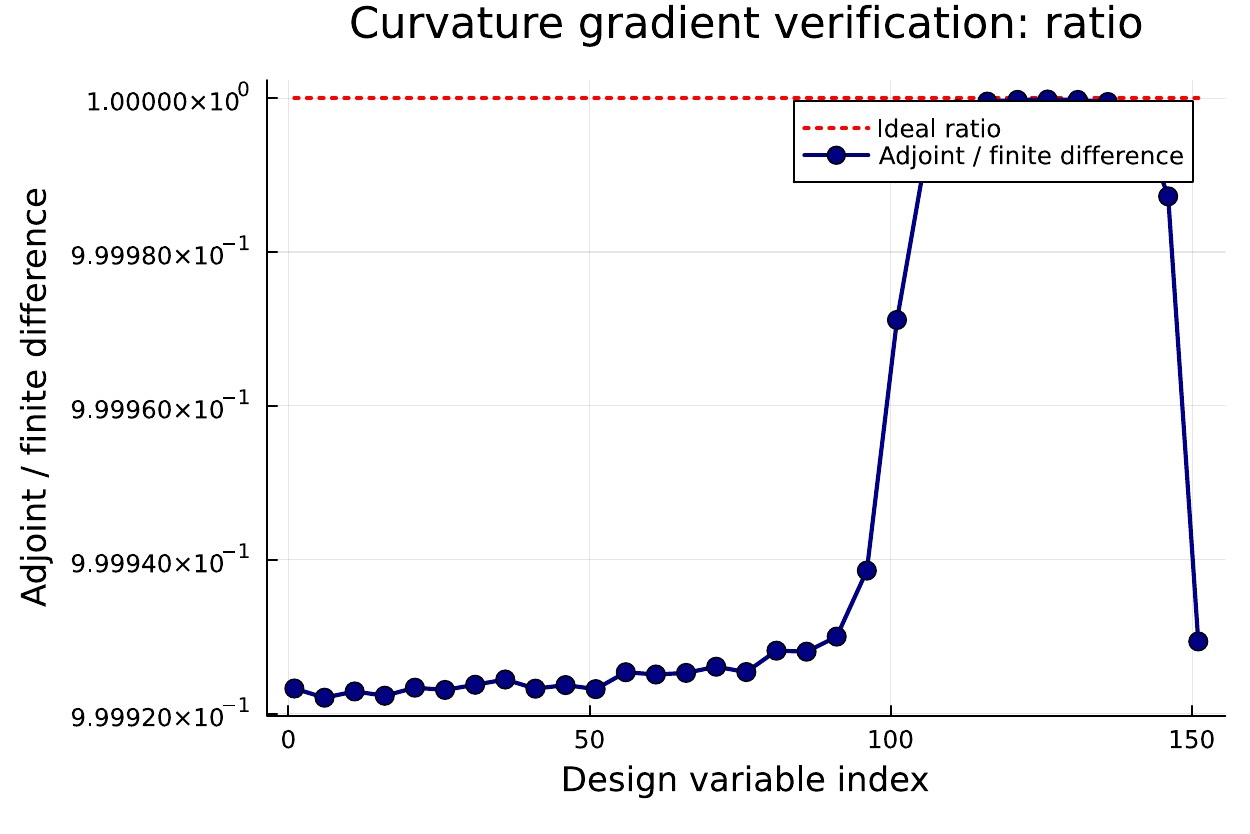}
    \caption{Ratio of adjoint to finite difference curvature derivatives. A ratio of unity indicates exact agreement ($p=50$, $h=10^{-5}$).}
    \label{fig:curvature_gradient_ratio}
\end{figure}

\begin{figure}[htbp!]
    \centering
    \includegraphics[width=\linewidth]{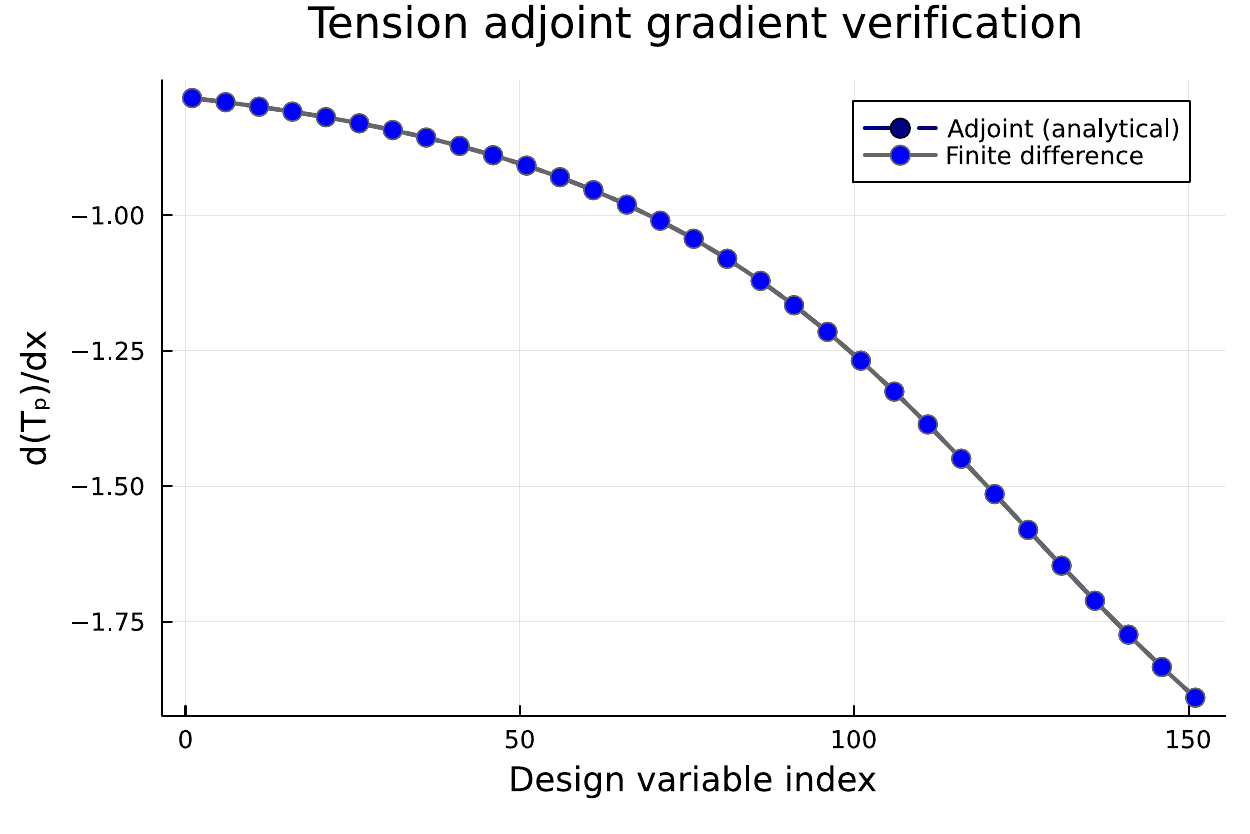}
    \caption{Comparison of adjoint and central finite difference derivatives of the $p$-norm tension aggregate at the 31 sampled design-variable indices ($p=50$, $h=10^{-5}$).}
    \label{fig:tension_gradient_check}
\end{figure}

\begin{figure}[htbp!]
    \centering
    \includegraphics[width=\linewidth]{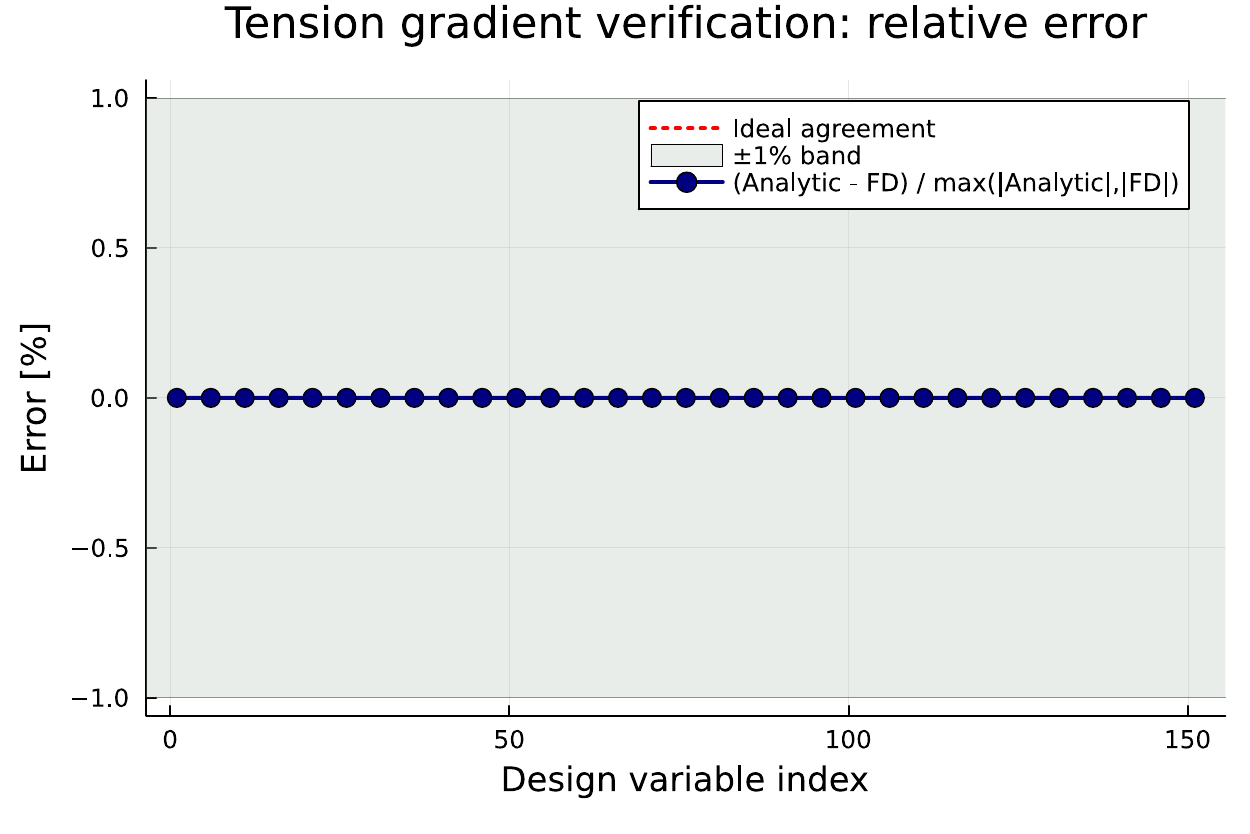}
    \caption{Signed relative discrepancies between adjoint and finite difference tension derivatives, normalized by the larger absolute derivative and expressed as percentages ($p=50$, $h=10^{-5}$).}
    \label{fig:tension_gradient_error}
\end{figure}

\begin{figure}[htbp!]
    \centering
    \includegraphics[width=\linewidth]{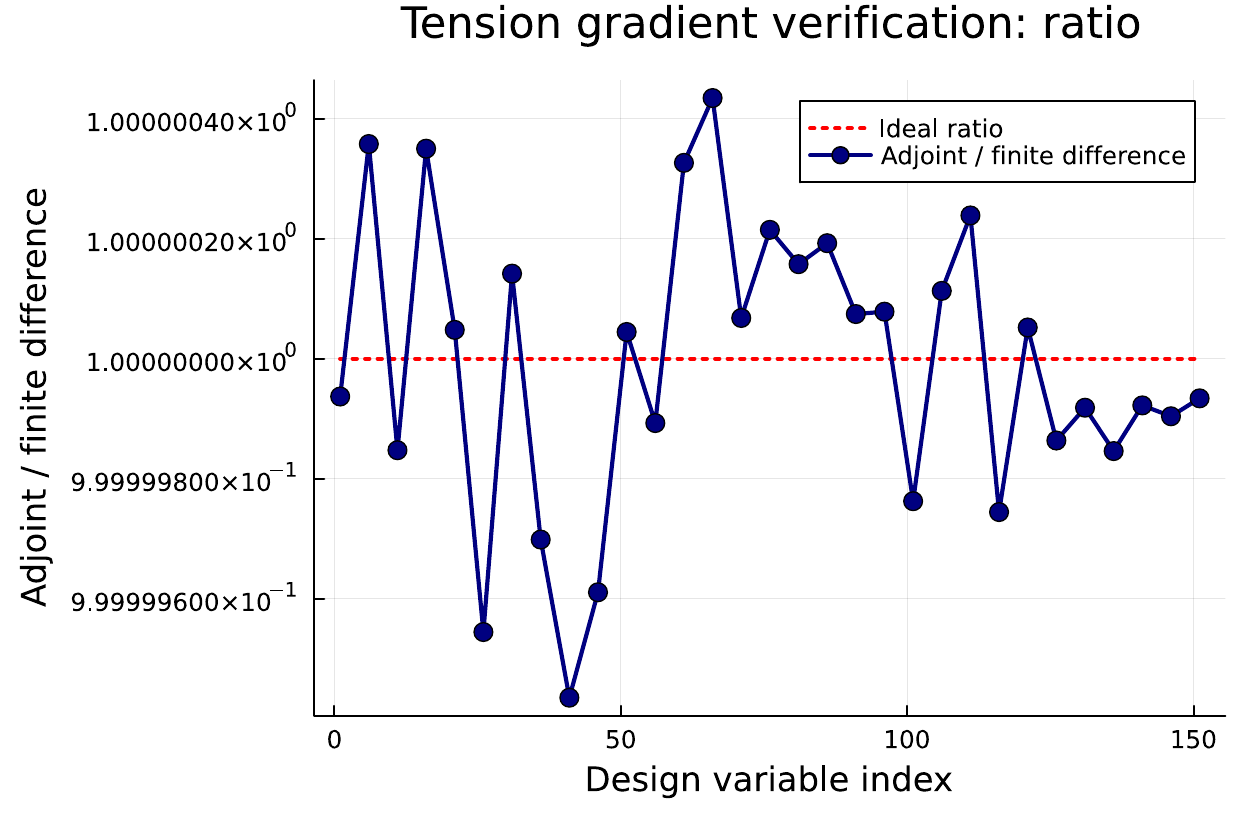}
    \caption{Ratio of adjoint to finite difference tension derivatives. A ratio of unity indicates exact agreement ($p=50$, $h=10^{-5}$).}
    \label{fig:tension_gradient_ratio}
\end{figure}

This section presents the numerical verification procedure for the analytical gradients computed using the adjoint variable method described in Secs.~\ref{sec:adjoint_formulation}--\ref{sec:tension_sensitivity}. Specifically, the gradients of the $p$-norm curvature and tension constraints are compared with approximate gradients obtained using central finite differences. For a scalar response $q$ and the $i$th design variable, the central finite difference approximation is given by Eq.~\eqref{eq:central_difference_check}:
\begin{equation}
    \left.\frac{\mathrm dq}{\mathrm dr_i}\right|_{\mathrm{FD}}
    \approx
    \frac{q(\vect r+h\vect e_i)-q(\vect r-h\vect e_i)}{2h}.
    \label{eq:central_difference_check}
\end{equation}
Here, $\vect e_i\in\mathbb R^{n_d}$ is a vector with its $i$th entry equal to one and all other entries equal to zero. Thus, $\vect r\pm h\vect e_i$ perturbs only the design variable $r_i$ by $\pm h$, leaving all other design variables unchanged. 
Each perturbed evaluation requires a complete nonlinear equilibrium solution, so central differences require approximately $2n_d$ response analyses for a full gradient, where $n_d$ denotes the number of buoyancy design variables. In contrast, after one equilibrium solution the adjoint approach requires only one additional linear solve per function differentiated. Finite differences are therefore used only for validation purposes rather than inside the optimization loop. A perturbation size of $h=10^{-5}$ is used for the finite difference verification shown in Figs.~\ref{fig:curvature_gradient_check}--\ref{fig:tension_gradient_ratio}. 
The verification case contains $n_d=151$ buoyancy design variables, with a uniform uplift ratio $r_i=2$, $\Delta s=4$~m, $N=756$, and zero current. The $p$-norm exponent is set to $p=50$ for both the curvature and tension aggregates, matching the value used in the optimization examples. A subset of 31 variables, with indices $i=1,6,\ldots,151$, is checked using central finite differences. The two perturbations for each selected variable require 62 additional second stage equilibrium solves, with the total line length held fixed. 

Figs.~\ref{fig:curvature_gradient_check}--\ref{fig:tension_gradient_ratio} compare the analytical and finite difference derivatives at the sampled indices, together with their relative errors and ratios. Since the uplift ratios are dimensionless, the curvature derivatives $\mathrm d\kappa_p/\mathrm dr_i$ have units of m$^{-1}$, and the tension derivatives $\mathrm dT_p/\mathrm dr_i$ have units of force. 
In the implementation, and therefore in Figs.~\ref{fig:tension_gradient_check}--\ref{fig:tension_gradient_ratio}, forces are expressed in kN. For $h=10^{-5}$, the maximum relative discrepancies are $0.00779\%$ for the curvature gradient and $0.0000565\%$ for the tension gradient. 
The close agreement supports the accuracy of the implemented adjoint sensitivities for the tested design and perturbation size.

The same verification of the sensitivities is also evaluated using $h=10^{-6}$. Table~\ref{tab:gradient_check_p50} reports the maximum relative discrepancies between the adjoint and central finite difference gradients for two perturbation sizes. The maxima are taken over the sampled variables for each constraint gradient. Both checks show excellent agreement, with maximum relative discrepancies below $0.01\%$. Increasing the perturbation size from $10^{-6}$ to $10^{-5}$ slightly reduces the maximum relative discrepancy for the curvature gradient and reduces that for the tension gradient by a factor of approximately 12.

\begin{table}[htbp!]
    \centering
    \caption{Maximum relative discrepancies between adjoint and central finite difference constraint gradients for $p=50$, evaluated over 31 sampled design variables.}
    \label{tab:gradient_check_p50}
    \small
    \setlength{\tabcolsep}{2pt}
    \begin{tabular}{@{}ccc@{}}
        \toprule
        Perturbation & \multicolumn{2}{c}{Maximum relative discrepancy [\%]} \\
        size $h$ & Curvature gradient & Tension gradient \\
        \midrule
        $10^{-6}$ & $0.00929$ & $0.000669$ \\
        $10^{-5}$ & $0.00779$ & $0.0000565$ \\
        \bottomrule
    \end{tabular}
\end{table}

\section{Numerical examples}
\label{sec:examples}

This section presents numerical applications of the proposed optimization approach to the design of buoyancy distributions for offshore lazy-wave risers. All examples are based on the optimization problem defined in Eq.~\eqref{eq:optimization_problem}, but employ different design parametrizations. In general, the physical nodal uplift ratios are expressed as a function $\vect r=\vect r(\vect x)$ of the optimization variables $\vect x$. The selected parametrization determines the number and meaning of the optimization variables and the degree to which the resulting buoyancy distribution can be implemented using a limited set of module sizes. The following three parametrizations are considered:
\begin{itemize}
\item In Sec.~\ref{sub:param1}, every nodal uplift ratio is an independent optimization variable. This continuous parametrization provides the greatest design freedom but may produce many distinct uplift ratio values.
\item In Sec.~\ref{sub:param2}, each design node is assigned in advance to a prescribed size-group. The optimization variables are the uplift ratios of the groups, and the physical nodal uplift ratios are obtained based on the fixed group assignments.
\item In Sec.~\ref{sub:param3}, both the uplift ratios of the prescribed number of size-groups and their assignment to the design nodes are optimized. This parametrization allows us to optimize the available module sizes and their spatial distribution simultaneously.
\end{itemize}

In addition to the different design parametrizations, the numerical examples consider different prescribed values of the curvature limit $\kappa_{\mathrm{lim}}$ and tension limit $T_{\mathrm{lim}}$, defined in Eqs.~\eqref{eq:max_curvature_constraint} and~\eqref{eq:max_tension_constraint}, respectively. In this way it is possible to assess the influence of the structural performance requirements on the optimized buoyancy distribution. 
Unless otherwise stated, the structural properties, environmental conditions, loading parameters, and numerical discretization are those reported in Table~\ref{tab:basic_parameters}. In all optimization analyses, the exponent of the $p$-norm aggregates in Eqs.~\eqref{eq:kappa_p} and~\eqref{eq:tension_p} is set to $p=50$.

The structural analysis and adjoint sensitivity procedures described in Secs.~\ref{sec:response} and~\ref{sec:sensitivity} were implemented by the authors in the Julia programming language \citep{Julia-2017}. The design optimization problems were solved using \texttt{Ipopt.jl} \citep{wachter2006implementation}, interfaced through the \texttt{Optimization.jl} package \citep{vaibhav_kumar_dixit_2023_7738525}.

The nonconvex problems are solved using Ipopt and analytical gradients. The reported optimized solutions significantly improve the initial design guesses, while satisfying the prescribed constraints within the numerical tolerances. No claim of global optimality of the solutions obtained is made.

\subsection{Continuous design parametrization}
\label{sub:param1}

This section presents the results obtained using a continuous design parametrization in which the nodal uplift ratios $r_i$ are treated as independent design variables. The optimization algorithm may select each uplift ratio freely within its prescribed lower and upper bounds, thereby allowing the buoyancy distribution to vary at every design node. This parametrization provides the greatest design freedom and can yield a highly efficient design solution. However, the resulting continuously varying uplift ratios may be impractical to realize using commercially available buoyancy modules. Rounding or grouping the optimized values into a limited set of module sizes would therefore be required for practical implementation and could reduce the performance of the optimized design or lead to constraint violations. 
These continuous designs show how the optimized uplift ratio varies along the buoyant section and provide a baseline for comparison with designs obtained using a limited number of module sizes, which are reported in the following sections. 

For this parametrization, the following lower and upper bounds are imposed on all nodal uplift ratios:
\begin{equation}
    0\le r_i\le 3,
    \qquad i=1,\ldots,n_d.
    \label{eq:continuous_design_bounds}
\end{equation}
The optimization is started from a uniform initial design in which all nodal uplift ratios are set to $r_i=2$, $i=1,\ldots,n_d$. In the configuration plots, the reference curve is the unbuoyed catenary ($r_i=0$), rather than the optimization starting design.

\begin{figure}[htbp!]
    \centering
    \includegraphics[width=\linewidth]{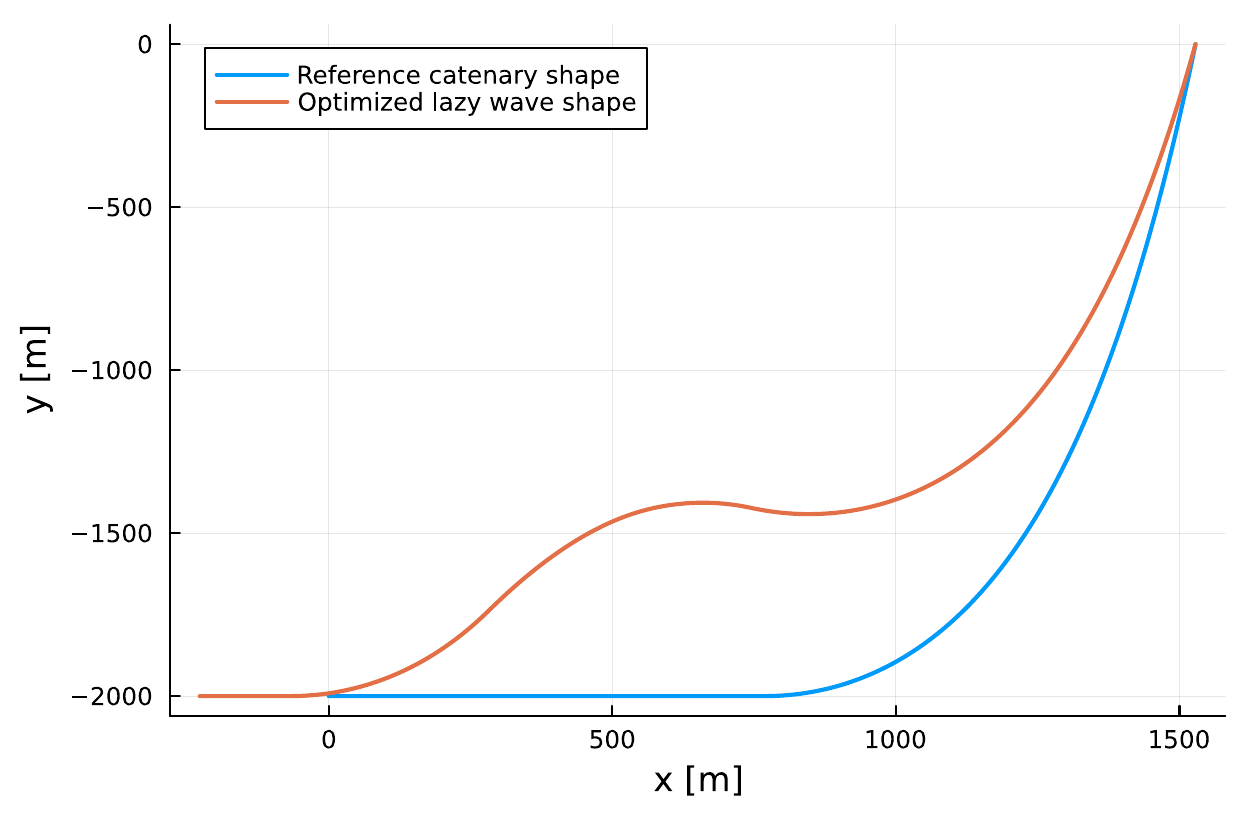}
    \caption{Reference and optimized riser configurations for the results of Sec.~\ref{sub:param1}, with $T_{\mathrm{lim}}=0.75T_{\max}^{\mathrm{cat}}$ and $\kappa_{\mathrm{lim}}=1.20\kappa_{\max}^{\mathrm{cat}}$.}
    \label{fig:continuous_075_120_shape}
\end{figure}

\begin{figure}[htbp!]
    \centering
    \includegraphics[width=\linewidth]{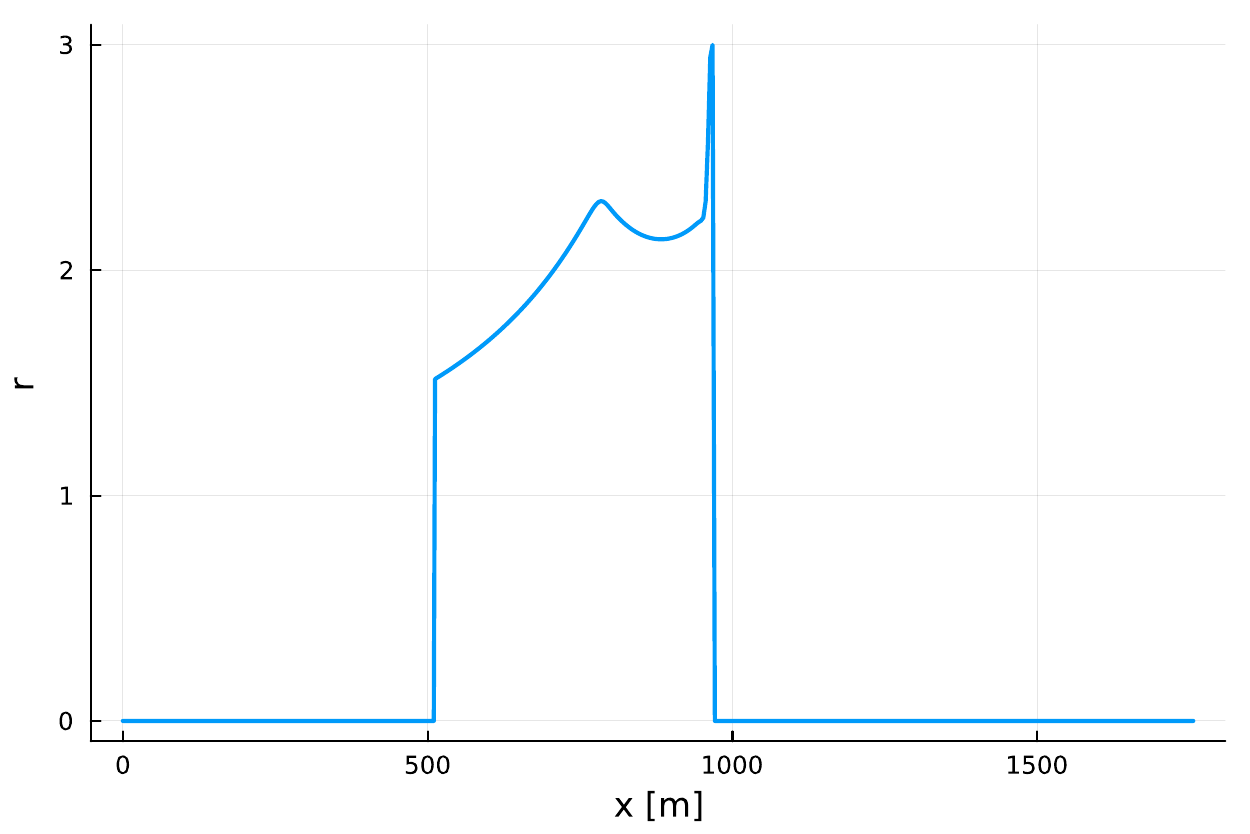}
    \caption{Optimized buoyancy distribution for the results of Sec.~\ref{sub:param1}, with $T_{\mathrm{lim}}=0.75T_{\max}^{\mathrm{cat}}$ and $\kappa_{\mathrm{lim}}=1.20\kappa_{\max}^{\mathrm{cat}}$.}
    \label{fig:continuous_075_120_buoyancy}
\end{figure}

\begin{figure}[htbp!]
    \centering
    \includegraphics[width=\linewidth]{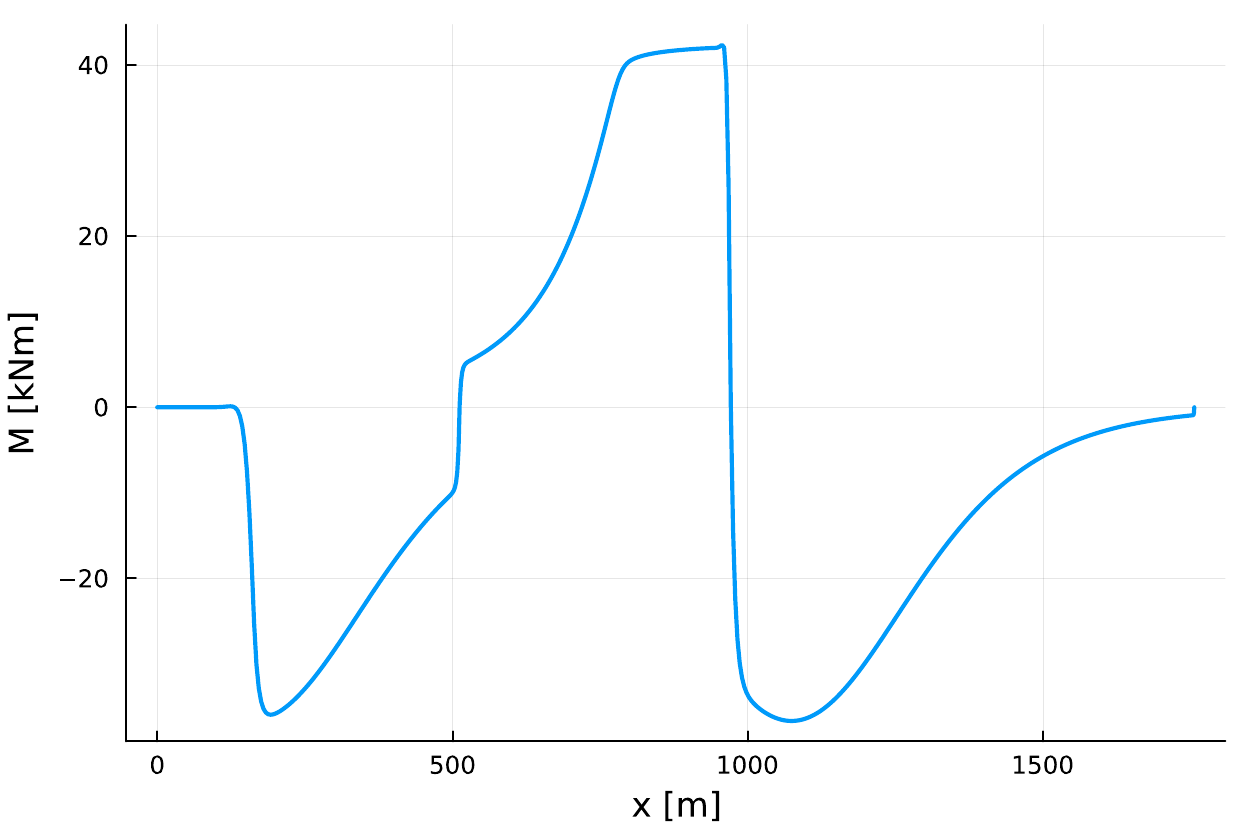}
    \caption{Bending moment response for the optimized design of Sec.~\ref{sub:param1} with $T_{\mathrm{lim}}=0.75T_{\max}^{\mathrm{cat}}$ and $\kappa_{\mathrm{lim}}=1.20\kappa_{\max}^{\mathrm{cat}}$.}
    \label{fig:continuous_075_120_moment}
\end{figure}

\begin{figure}[htbp!]
    \centering
    \includegraphics[width=\linewidth]{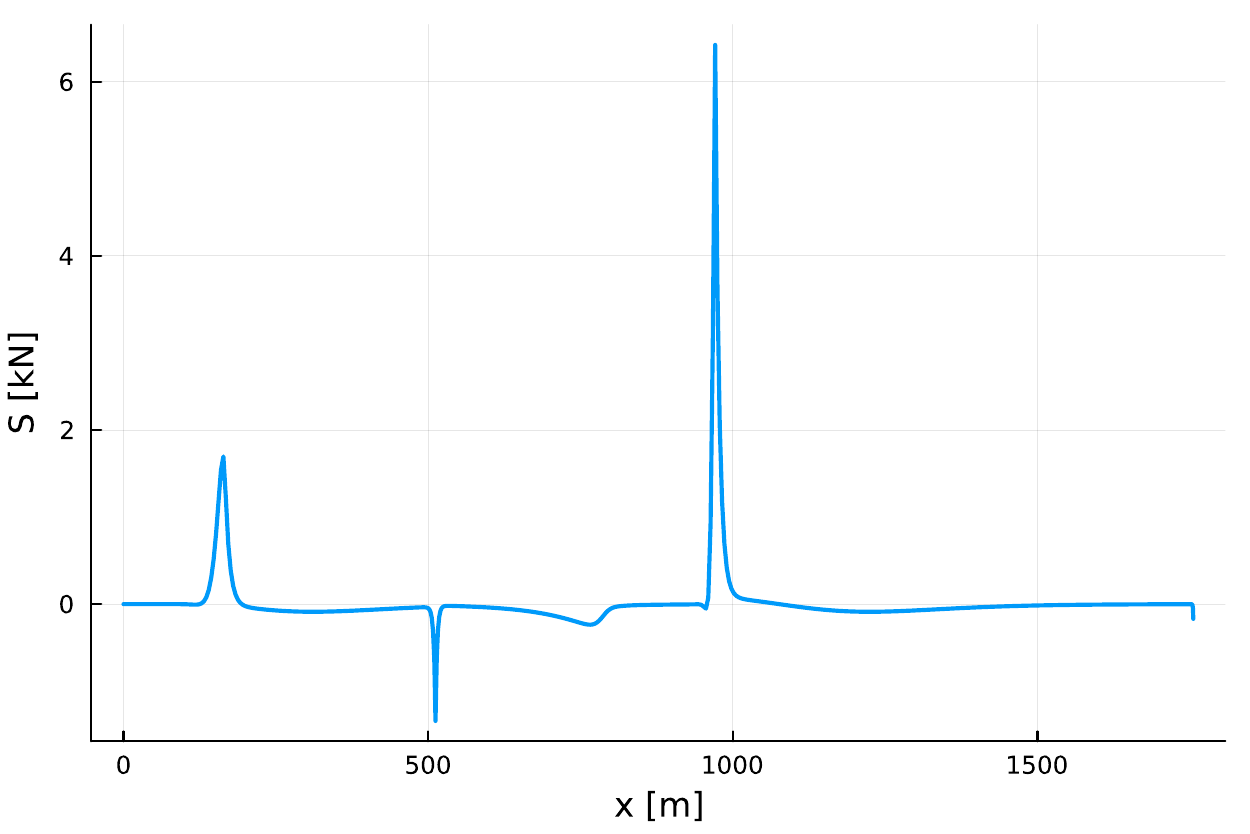}
    \caption{Shear response for the optimized design of Sec.~\ref{sub:param1} with $T_{\mathrm{lim}}=0.75T_{\max}^{\mathrm{cat}}$ and $\kappa_{\mathrm{lim}}=1.20\kappa_{\max}^{\mathrm{cat}}$.}
    \label{fig:continuous_075_120_shear}
\end{figure}

\begin{figure}[htbp!]
    \centering
    \includegraphics[width=\linewidth]{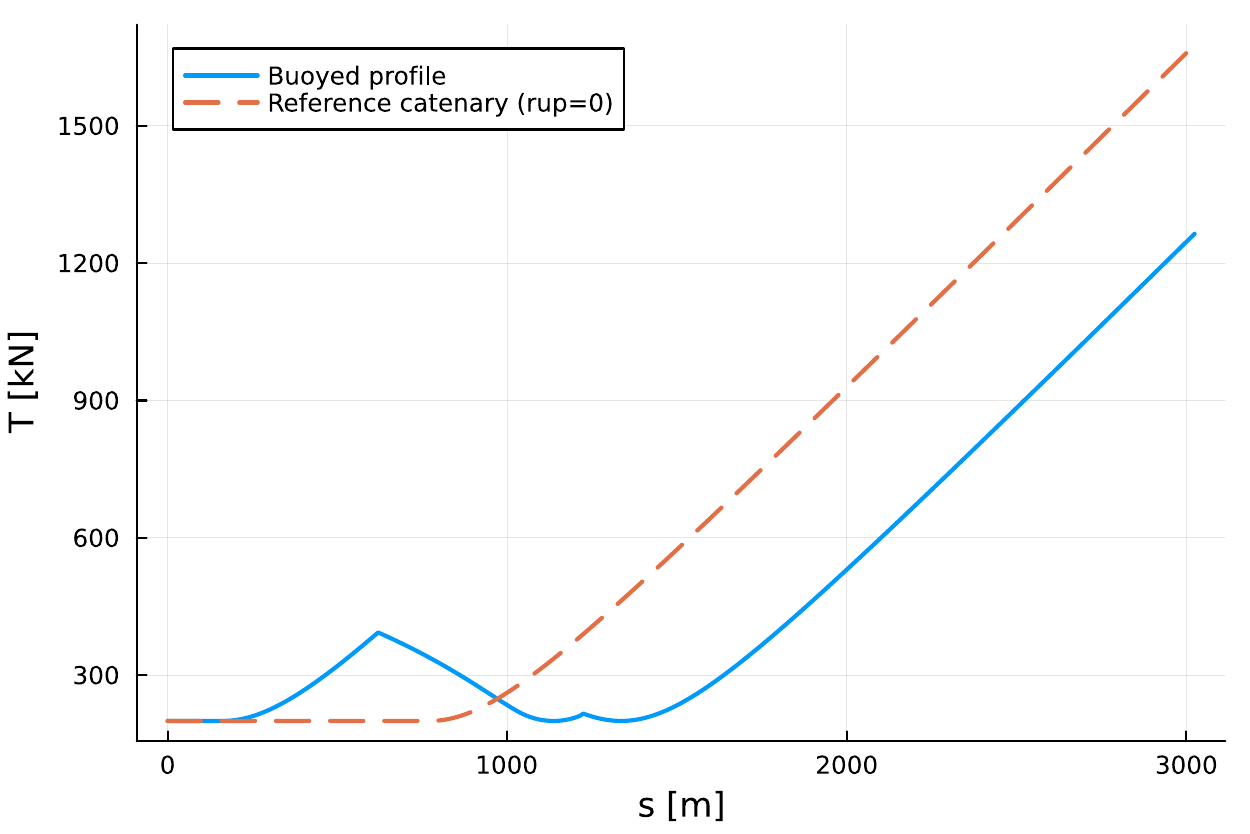}
    \caption{Tension response for the optimized design of Sec.~\ref{sub:param1} with $T_{\mathrm{lim}}=0.75T_{\max}^{\mathrm{cat}}$ and $\kappa_{\mathrm{lim}}=1.20\kappa_{\max}^{\mathrm{cat}}$.}
    \label{fig:continuous_075_120_tension}
\end{figure}

The tension and curvature limits are defined relative to the response of the corresponding free hanging catenary configuration, obtained for the same riser without buoyancy modules. Denoting by $T_{\max}^{\mathrm{cat}}$ and $\kappa_{\max}^{\mathrm{cat}}$ the $p$-norm aggregates of the effective tension and curvature of this reference configuration, computed as in Eqs.~\eqref{eq:kappa_p} and~\eqref{eq:tension_p} with the same exponent $p$, the limits are prescribed as
\begin{equation}
    T_{\mathrm{lim}}=0.75T_{\max}^{\mathrm{cat}},
    \qquad
    \kappa_{\mathrm{lim}}=1.20\kappa_{\max}^{\mathrm{cat}}.
    \label{eq:continuous_design_constraint_limits}
\end{equation}
The tension requirement therefore seeks a $25\%$ reduction relative to the unbuoyed catenary. By contrast, the curvature limit permits an increase of up to $20\%$, because the introduction of buoyancy modules produces the additional hogbend and sagbend curvature characteristic of a lazy-wave profile. The curvature constraint limits this increase and prevents the optimized buoyancy distribution from generating excessive local bending.  
The optimization results for this constraint case are presented in Figs.~\ref{fig:continuous_075_120_shape}--\ref{fig:continuous_075_120_tension}. In particular, these figures show the optimized lazy-wave configuration, the corresponding buoyancy distribution, the bending moment and shear responses, and, finally, the effective tension response.

\begin{figure}[htbp!]
    \centering
    \includegraphics[width=\linewidth]{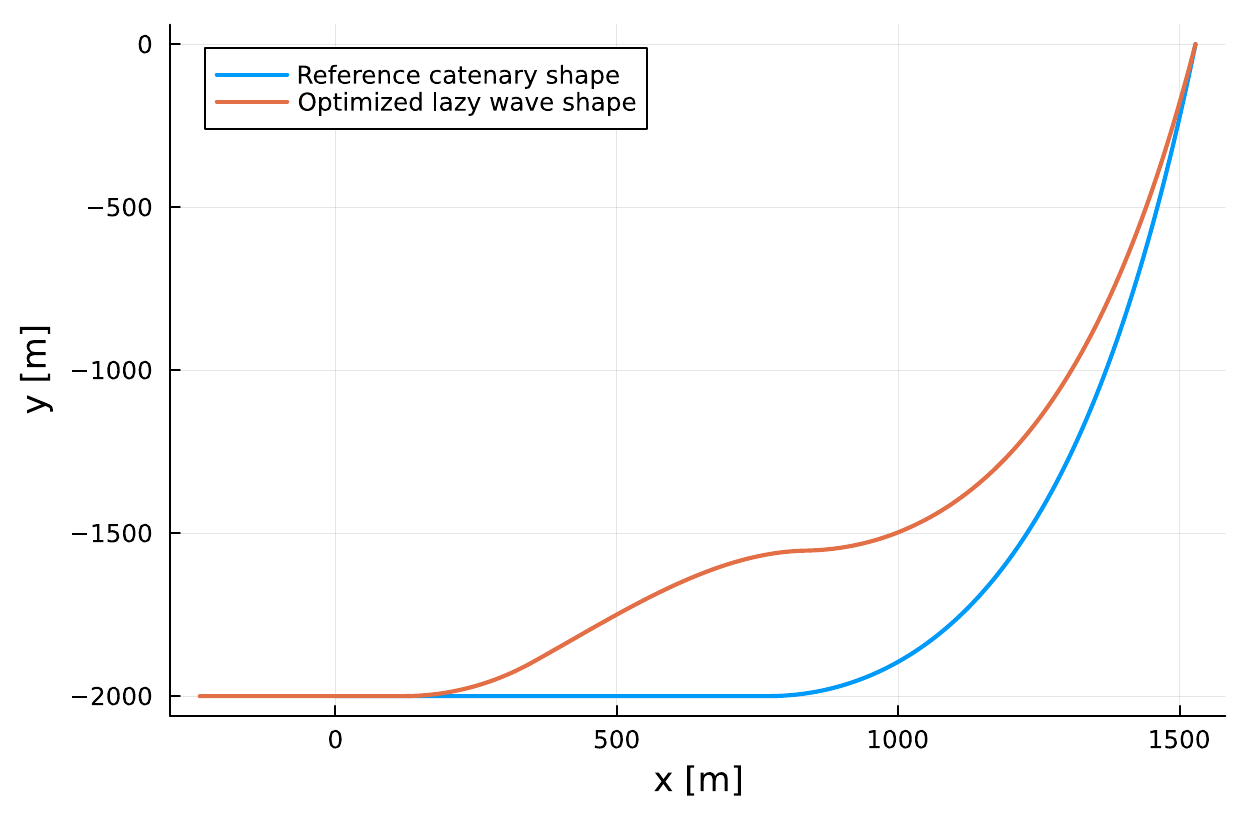}
    \caption{Reference and optimized riser configurations for the results of Sec.~\ref{sub:param1}, with $T_{\mathrm{lim}}=0.80T_{\max}^{\mathrm{cat}}$ and $\kappa_{\mathrm{lim}}=1.10\kappa_{\max}^{\mathrm{cat}}$.}
    \label{fig:continuous_080_110_shape}
\end{figure}

\begin{figure}[htbp!]
    \centering
    \includegraphics[width=\linewidth]{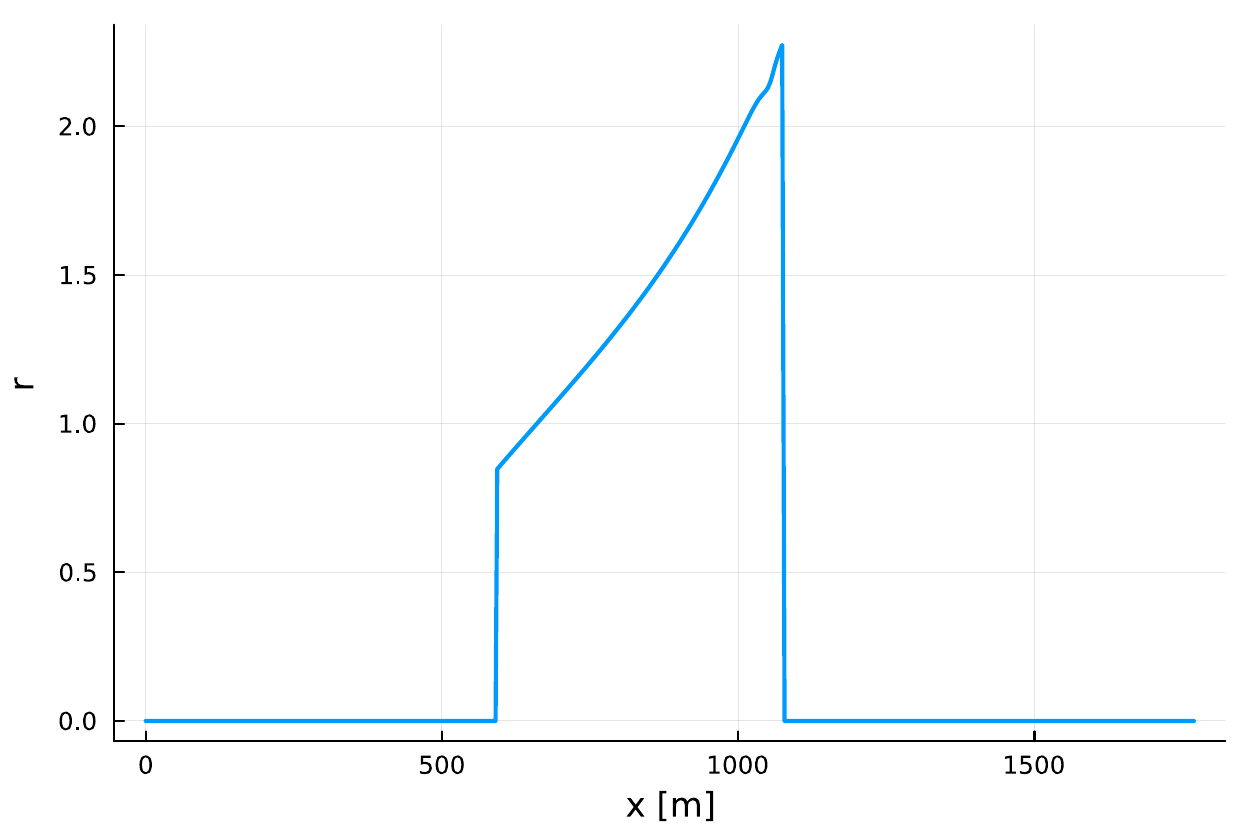}
    \caption{Optimized buoyancy distribution for the results of Sec.~\ref{sub:param1}, with $T_{\mathrm{lim}}=0.80T_{\max}^{\mathrm{cat}}$ and $\kappa_{\mathrm{lim}}=1.10\kappa_{\max}^{\mathrm{cat}}$.}
    \label{fig:continuous_080_110_buoyancy}
\end{figure}

\begin{figure}[htbp!]
    \centering
    \includegraphics[width=\linewidth]{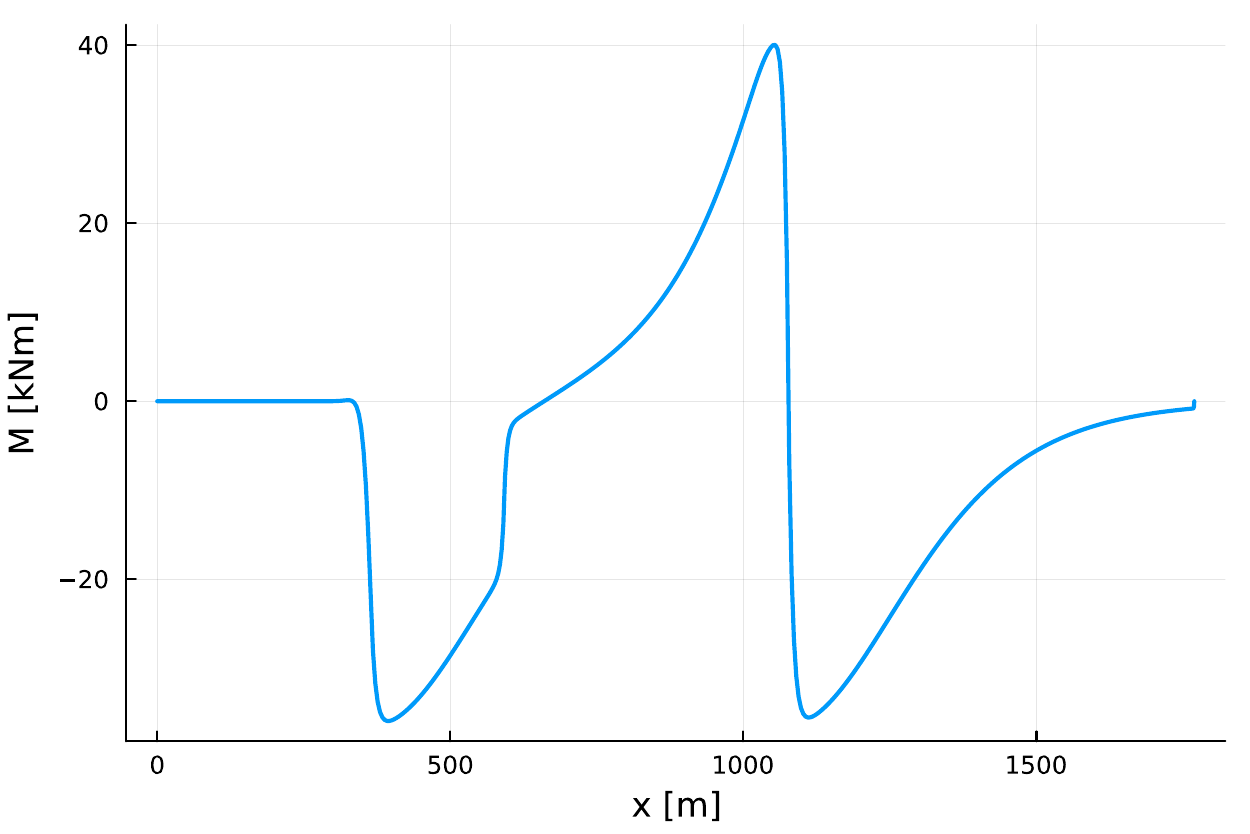}
    \caption{Bending moment response for the optimized design of Sec.~\ref{sub:param1} with $T_{\mathrm{lim}}=0.80T_{\max}^{\mathrm{cat}}$ and $\kappa_{\mathrm{lim}}=1.10\kappa_{\max}^{\mathrm{cat}}$.}
    \label{fig:continuous_080_110_moment}
\end{figure}

\begin{figure}[htbp!]
    \centering
    \includegraphics[width=\linewidth]{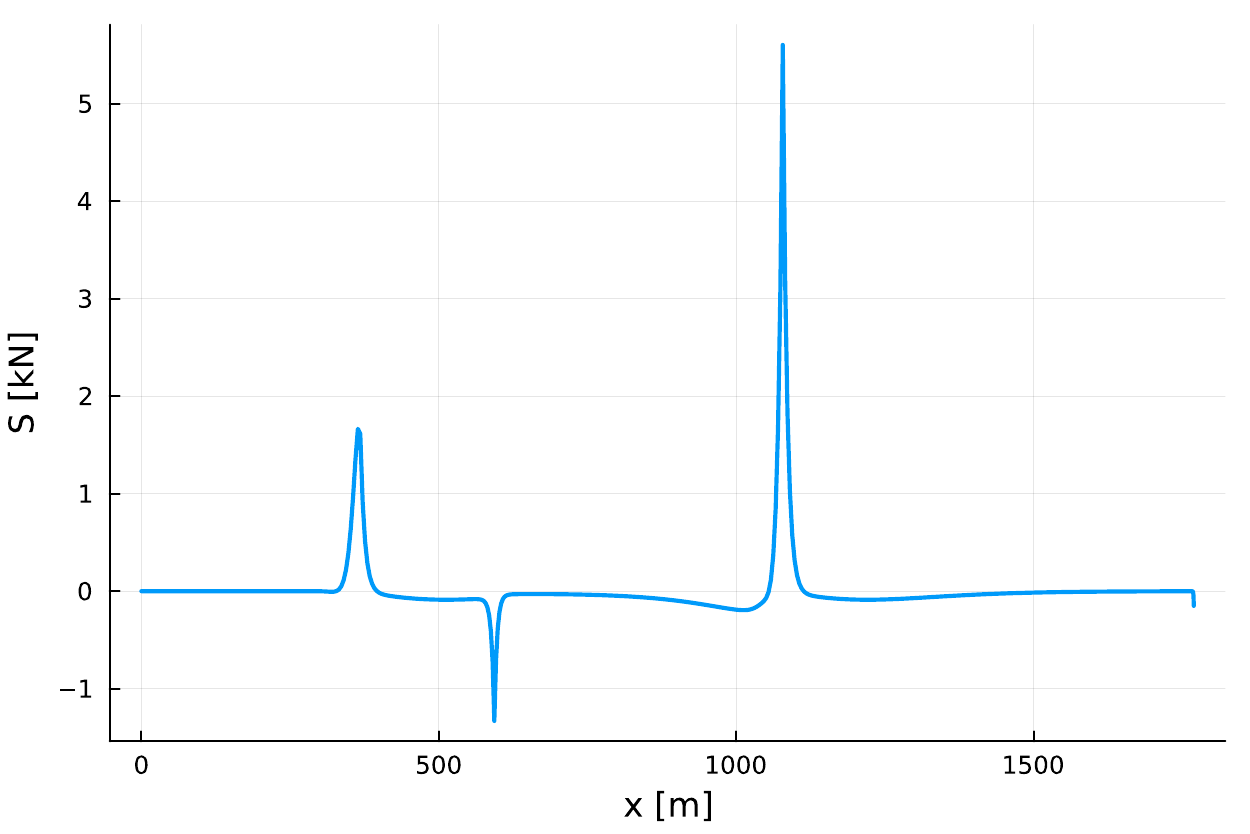}
    \caption{Shear response for the optimized design of Sec.~\ref{sub:param1} with $T_{\mathrm{lim}}=0.80T_{\max}^{\mathrm{cat}}$ and $\kappa_{\mathrm{lim}}=1.10\kappa_{\max}^{\mathrm{cat}}$.}
    \label{fig:continuous_080_110_shear}
\end{figure}

\begin{figure}[htbp!]
    \centering
    \includegraphics[width=\linewidth]{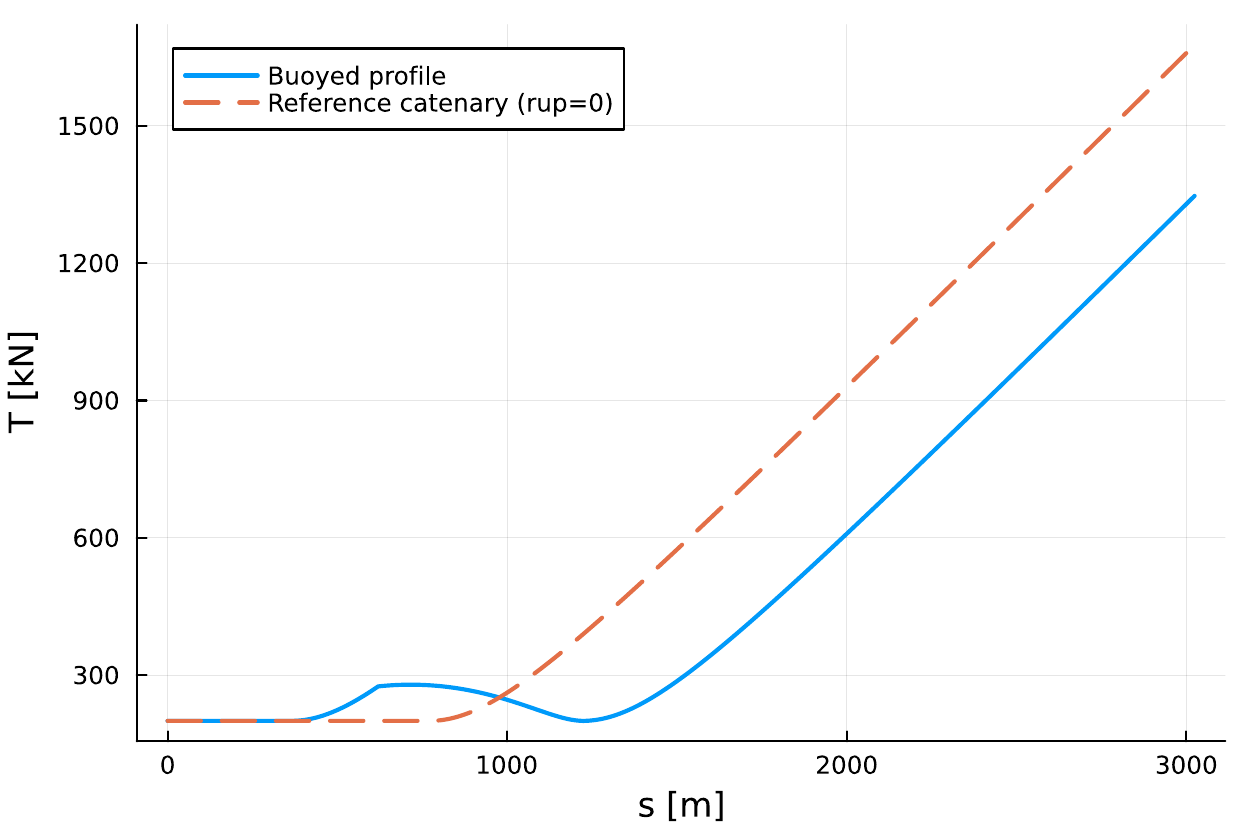}
    \caption{Tension response for the optimized design of Sec.~\ref{sub:param1} with $T_{\mathrm{lim}}=0.80T_{\max}^{\mathrm{cat}}$ and $\kappa_{\mathrm{lim}}=1.10\kappa_{\max}^{\mathrm{cat}}$.}
    \label{fig:continuous_080_110_tension}
\end{figure}

A second constraint case is also considered, with
\begin{equation}
    T_{\mathrm{lim}}=0.80T_{\max}^{\mathrm{cat}},
    \qquad
    \kappa_{\mathrm{lim}}=1.10\kappa_{\max}^{\mathrm{cat}}.
    \label{eq:continuous_design_constraint_limits_second}
\end{equation}
This case relaxes the required tension reduction while imposing a more restrictive curvature limit.  
The optimization results for this second constraint case are presented in Figs.~\ref{fig:continuous_080_110_shape}--\ref{fig:continuous_080_110_tension}.

The two constraint cases lead to markedly different optimized uplift ratio distributions. In the case of the second set of constraints given in Eq.~\eqref{eq:continuous_design_constraint_limits_second}, the distribution associated with the configuration in Fig.~\ref{fig:continuous_080_110_shape}, shown in Fig.~\ref{fig:continuous_080_110_buoyancy}, varies smoothly and follows an approximately linear, mildly parabolic trend along the buoyancy section. By contrast, with the first set of constraints given in Eq.~\eqref{eq:continuous_design_constraint_limits}, the obtained buoyancy distribution associated with Fig.~\ref{fig:continuous_075_120_shape}, shown in Fig.~\ref{fig:continuous_075_120_buoyancy}, is more strongly nonuniform, with an increase followed by a pronounced decrease in uplift ratio. The optimized lazy-wave profile for the $0.80/1.10$ case is therefore smoother, consistent with its more restrictive curvature limit.

The tension responses in Figs.~\ref{fig:continuous_075_120_tension} and~\ref{fig:continuous_080_110_tension} confirm the reduction in peak effective tension relative to the reference catenary profile. In terms of the $p$-norm aggregate of Eq.~\eqref{eq:tension_p}, the tension is limited to $75\%$ of the corresponding catenary value in the first case and $80\%$ in the second case, in accordance with the respective prescribed tension limits.

The computational effort recorded by IPOPT for the two continuous- design cases is summarized in Table~\ref{tab:continuous_optimization_cost}.
\begin{table}[htbp!]
\centering
\caption{IPOPT iterations, reported solver times, and final objective values for the continuous design cases of Sec.~\ref{sub:param1}.}
\label{tab:continuous_optimization_cost}
\small
\setlength{\tabcolsep}{2pt}
\begin{tabular}{ccccc}
\toprule
$(\alpha_T,\alpha_\kappa)$ & \begin{tabular}[c]{@{}c@{}}Design\\ variables\end{tabular} & Iterations & \begin{tabular}[c]{@{}c@{}}IPOPT\\ time (s)\end{tabular} & \begin{tabular}[c]{@{}c@{}}Final\\ objective\end{tabular} \\
\midrule
$(0.75,\,1.20)$ & 151 & 66 & 1303.094 & 582.101 \\
$(0.80,\,1.10)$ & 151 & 10 & 77.305 & 327.463 \\
\bottomrule
\end{tabular}
\end{table}
The $0.75/1.20$ case required substantially greater solver effort, with 66 iterations and a reported IPOPT time approximately 16.9 times that of the $0.80/1.10$ case. These results show that the computational effort differed markedly between the two cases. Because both the tension and curvature limits change, the difference cannot be attributed to either limit alone.

\begin{figure*}[t]
    \centering
    \begin{minipage}{0.49\textwidth}
        \centering
        \includegraphics[width=\linewidth]{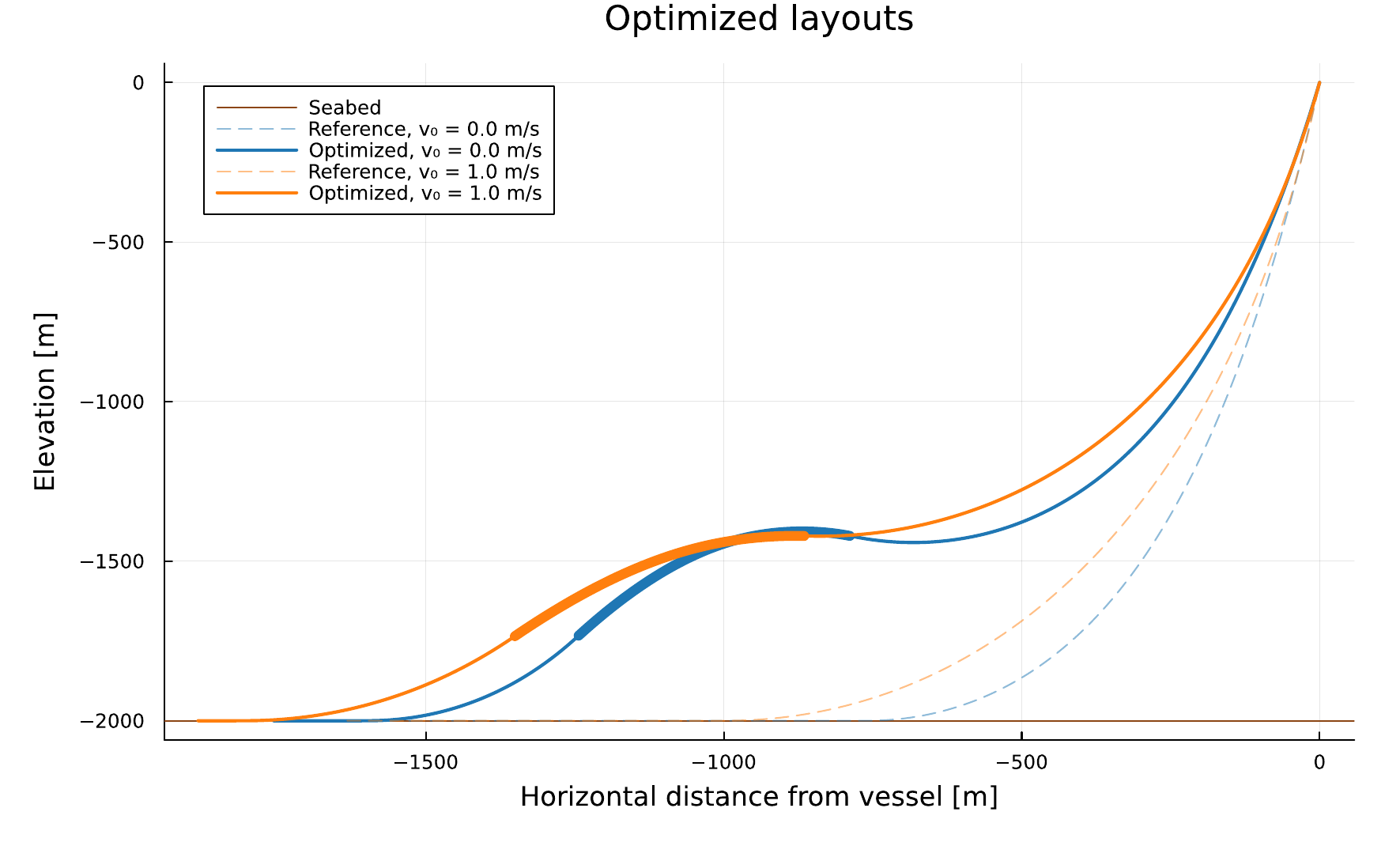}
        \par\small (a) Layouts
    \end{minipage}\hfill
    \begin{minipage}{0.49\textwidth}
        \centering
        \includegraphics[width=\linewidth]{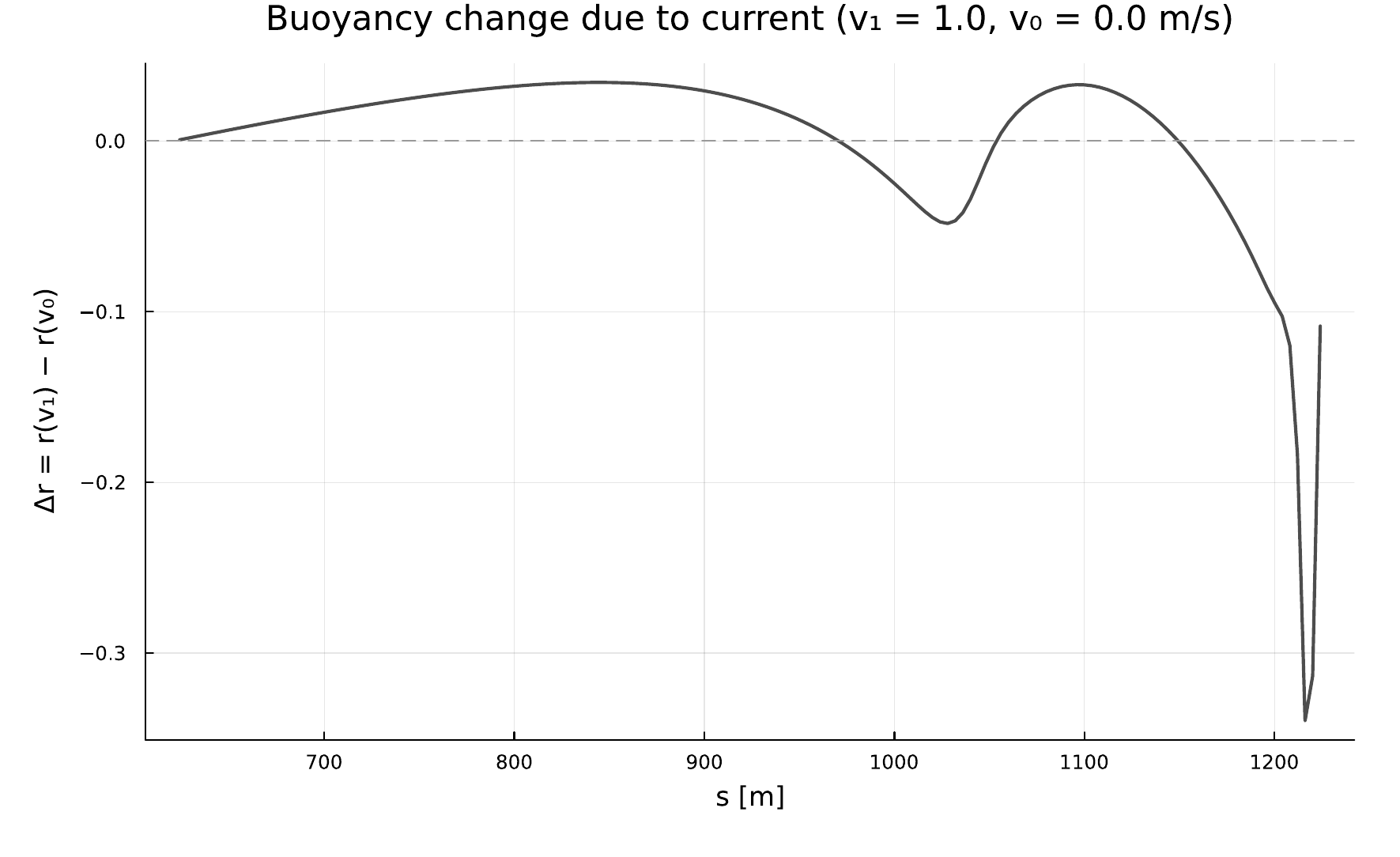}
        \par\small (b) Uplift-ratio difference
    \end{minipage}
    \par\medskip
    \begin{minipage}{0.49\textwidth}
        \centering
        \includegraphics[width=\linewidth]{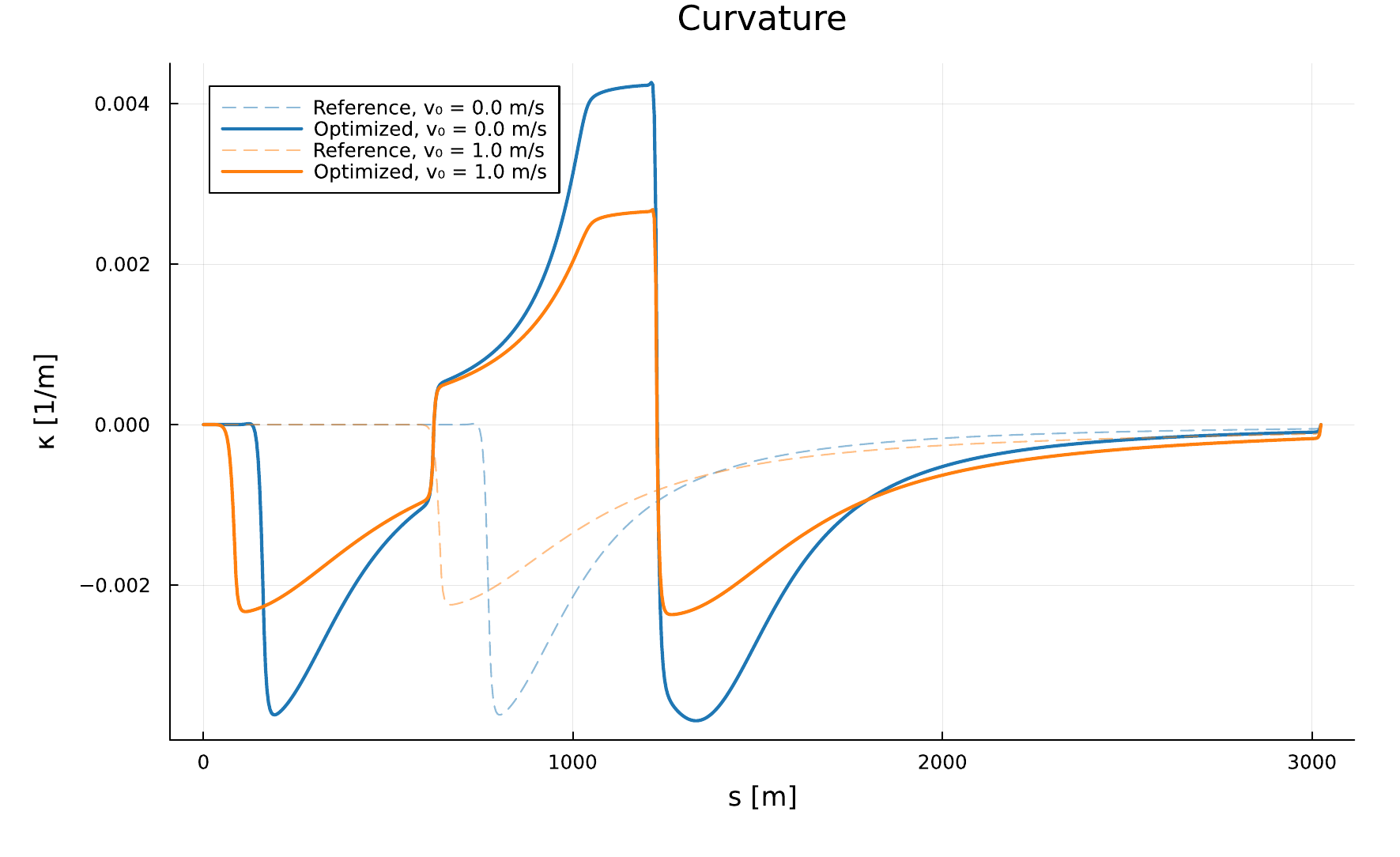}
        \par\small (c) Curvature
    \end{minipage}\hfill
    \begin{minipage}{0.49\textwidth}
        \centering
        \includegraphics[width=\linewidth]{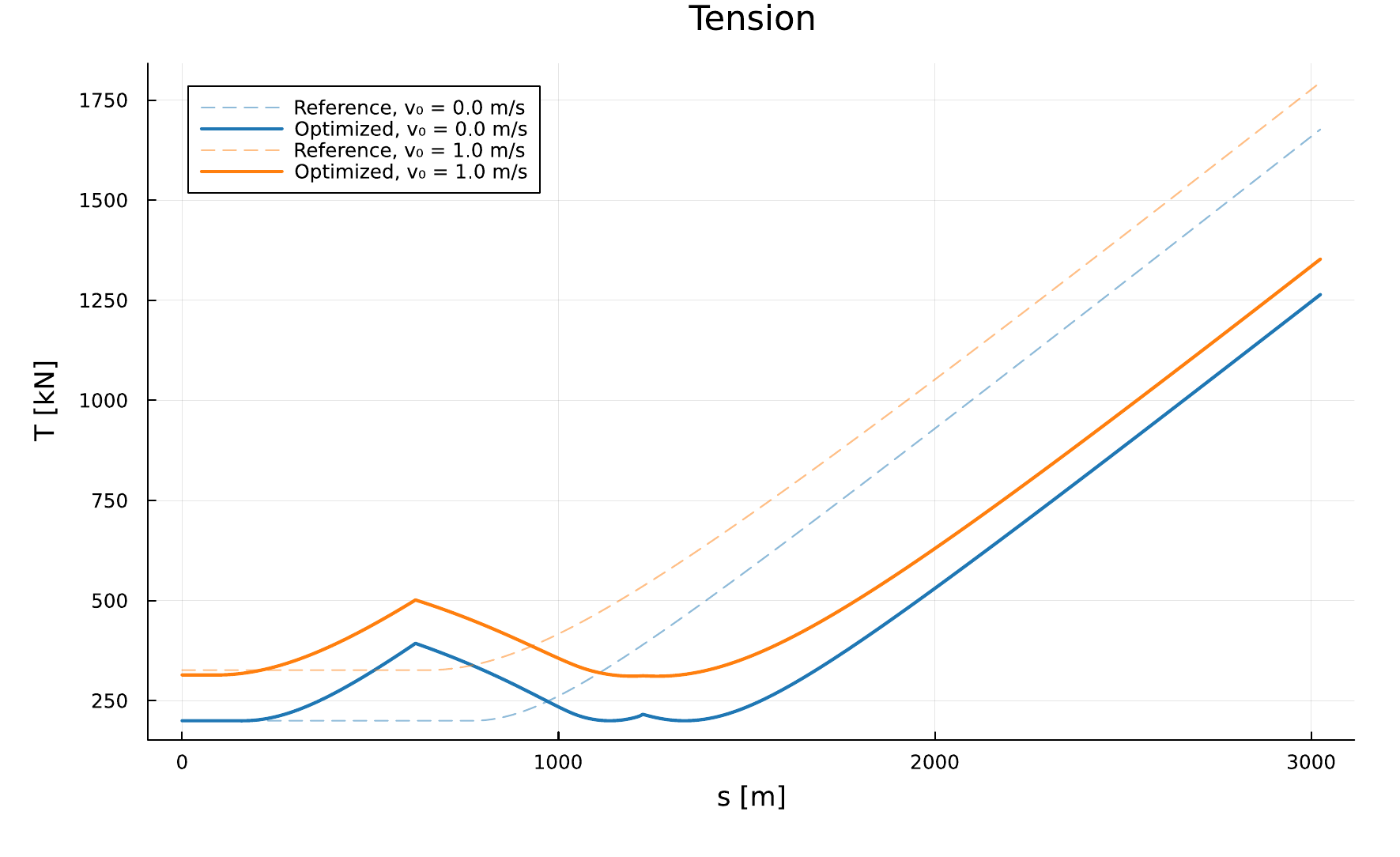}
        \par\small (d) Effective tension
    \end{minipage}
    \caption{Effect of current on the continuous designs for the $0.75/1.20$ constraint case of Sec.~\ref{sub:param1}: (a) reference and optimized layouts aligned at the HOP; (b) $\Delta r=r_{v_0=1}-r_{v_0=0}$; (c) curvature; and (d) effective tension. Each design and reference catenary is evaluated under its corresponding surface current velocity, $v_0=0$ or $1$~m/s.}
    \label{fig:continuous_current_effect}
\end{figure*}

An additional optimization was performed for the $0.75/1.20$ constraint case to examine the effect of current on the final design, using the profile in Eq.~\eqref{eq:current_profile} with surface velocity $v_0=1$~m/s. The remaining model and optimization settings were unchanged. The limits were recalculated from the unbuoyed reference catenary subject to current: $T_{\mathrm{lim}}$ increases from $1321.00$ to $1415.22$~kN, whereas $\kappa_{\mathrm{lim}}$ decreases from $4.5228\times10^{-3}$ to $2.8327\times10^{-3}$~m$^{-1}$. Figure~\ref{fig:continuous_current_effect} compares the optimized layouts, uplift ratio differences, and responses, with each design evaluated under its corresponding current.

The current case retains the lazy-wave configuration but increases the horizontal HOP--TDP distance from $1591.8$ to $1794.4$~m. The uplift redistribution is modest over most of the buoyant span, with the largest absolute change, $|\Delta r|=0.340$, near its upper boundary. The quadratic objective changes only from $582.101$ to $580.566$ (a $0.26\%$ reduction). The curvature peaks decrease, while the HOP tension increases from $1264.2$ to $1352.6$~kN. These results illustrate how the approach effectively adapts the optimized buoyancy distribution to the current loading to satisfy the response constraints.

\subsection{Optimization of buoyancy size-groups}
\label{sub:param2}

This section presents the results obtained using a grouped design parametrization. Each design node is assigned in advance to one of a prescribed number of buoyancy size-groups, and all nodes belonging to the same group share a common uplift ratio. The nodal group assignments are fixed during the optimization, and only the uplift ratio associated with each group is treated as a design variable. Consequently, the number of optimization variables is equal to the number of size-groups rather than the number of physical design nodes. This parametrization produces a buoyancy distribution with a limited number of distinct module sizes and is therefore more directly compatible with practical manufacturing and installation requirements than the fully continuous parametrization of Sec.~\ref{sub:param1}.

For these grouped design examples, the curvature and tension limits are set as specified in Eq.~\eqref{eq:continuous_design_constraint_limits}. We consider three, five, and six size-groups. In each case, the buoyancy nodes are partitioned, in order of increasing arc length, into contiguous groups whose numbers of nodes differ by at most one. The uplift ratio associated with each group is treated as an optimization variable and optimized subject to the stated constraints, starting from all group uplift ratios equal to 2, that is, from the same uniform initial design used in Sec.~\ref{sub:param1}.

Let the $n_d$ buoyancy nodes be indexed by $i=1,\ldots,n_d$, and let the prescribed size-groups be indexed by $k=1,\ldots,n_{\mathrm{sg}}$. Their fixed assignments are represented by the binary membership matrix
\begingroup\footnotesize
\begin{equation}
\begin{split}
    & B_{ik}=
    \begin{cases}
        1, & \text{if buoyancy node $i$ belongs to size-group $k$}\\
        0, & \text{otherwise}
    \end{cases}\\
    & \sum_{k=1}^{n_{\mathrm{sg}}}B_{ik}=1.
\end{split}
    \label{eq:size_group_membership}
\end{equation}
\endgroup
Thus, every design node belongs to exactly one size-group. The optimization variables are the size-group uplift ratios
\begin{equation}
    \vect r^{\mathrm{sg}}
    =\begin{bmatrix}
        r^{\mathrm{sg}}_1 & \cdots & r^{\mathrm{sg}}_{n_{\mathrm{sg}}}
      \end{bmatrix}^{\mathsf T},
    \qquad
    \underline r\le r^{\mathrm{sg}}_k\le\overline r,
    \label{eq:size_group_design_vector}
\end{equation}
and the physical nodal uplift ratios are obtained through the linear mapping
\begin{equation}
    \vect r=\vect B\vect r^{\mathrm{sg}}.
    \label{eq:size_group_mapping}
\end{equation}
The objective in Eq.~\eqref{eq:objective_buoyancy} can then be written directly in terms of the group variables as
\begin{equation}
    f\!\left(\vect r^{\mathrm{sg}}\right)
    =\left(\vect r^{\mathrm{sg}}\right)^{\mathsf T} \vect B^{\mathsf T}
     \vect B\vect r^{\mathrm{sg}}
\end{equation}
with gradient
\begin{equation}
    \nabla_{\vect r^{\mathrm{sg}}}f
    =2\vect B^{\mathsf T}\vect B\vect r^{\mathrm{sg}}.
    \label{eq:size_group_objective_gradient}
\end{equation}
More generally, with this parametrization of the design, for any scalar response or constraint $g$ the nodal gradient obtained from the adjoint analysis (see Sec.~\ref{sec:sensitivity}) is mapped to the optimization variables by means of the chain rule:
\begin{equation}
    \frac{\mathrm dg}{\mathrm d\vect r^{\mathrm{sg}}}
    =\vect B ^{\mathsf T} \frac{\mathrm dg}{\mathrm d\vect r}.
    \label{eq:size_group_gradient_mapping}
\end{equation}
The structural and adjoint equations therefore remain formulated in terms of the physical nodal vector $\vect r$. Only the final gradient is changed according to the actual design parametrization considered.

The design optimization results are reported in the following figures. Figures~\ref{fig:size_groups_3_buoyancy} and~\ref{fig:size_groups_3_ratios} show the optimized buoyancy distribution and the uplift ratios of the first design case with three size-groups available. Similarly, the  results for the case of five size-groups are shown in Figs.~\ref{fig:size_groups_5_buoyancy} and~\ref{fig:size_groups_5_ratios}, while Figs.~\ref{fig:size_groups_6_buoyancy} and~\ref{fig:size_groups_6_ratios} present the results for six size-groups. 
For the three size-group design case, the uplift ratios increase as one moves along the riser from the seabed toward the floating station (Figs.~\eqref{fig:size_groups_3_buoyancy} and \eqref{fig:size_groups_3_ratios}). With five size-groups (Figs.~\eqref{fig:size_groups_5_buoyancy} and \eqref{fig:size_groups_5_ratios}), the optimized discrete distribution begins to capture the increase and then decrease pattern in uplift ratio observed in the continuous design case (see Fig.~\ref{fig:continuous_075_120_buoyancy}). This pattern is even more clearly expressed in the six size-group uplift ratios case, which is shown in Fig.~\ref{fig:size_groups_6_buoyancy}.

\begin{figure}[htbp!]
    \centering
    \includegraphics[width=\linewidth]{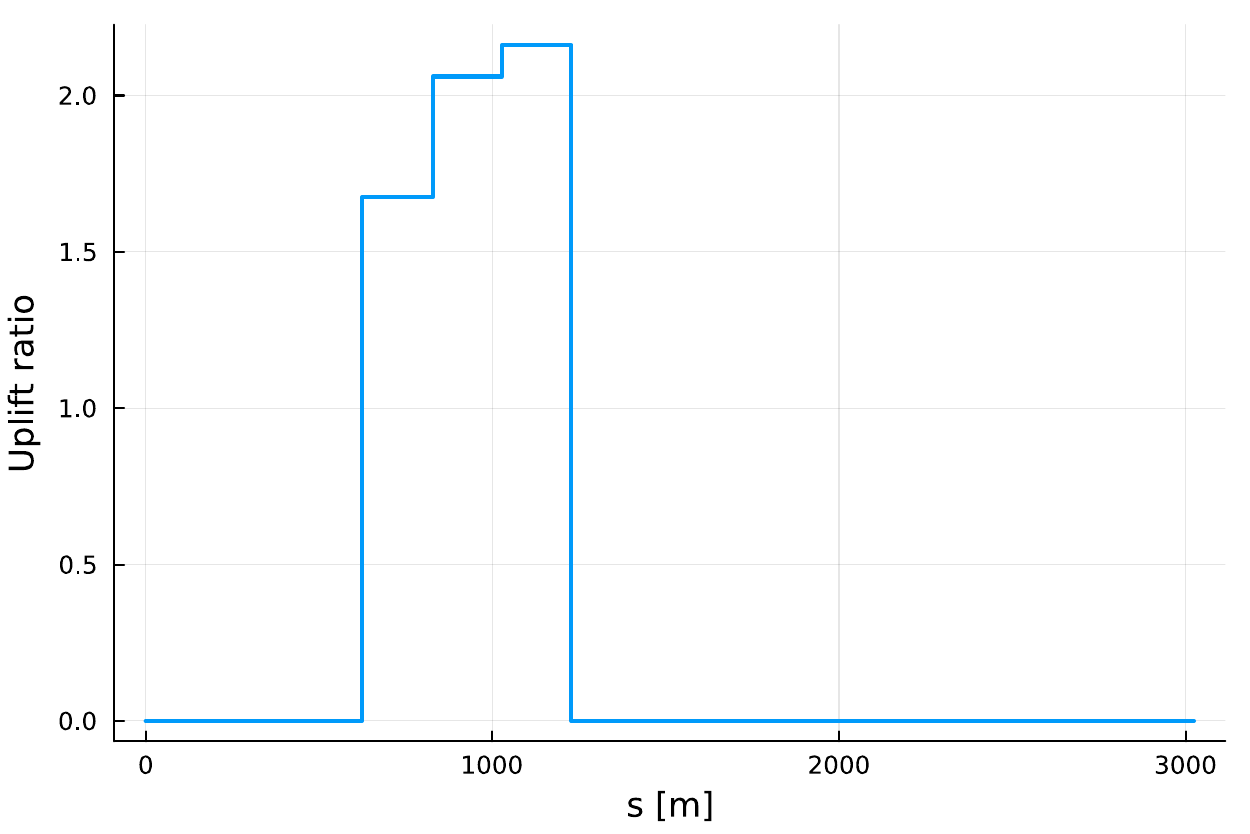}
    \caption{Optimized buoyancy distribution obtained in Sec.~\ref{sub:param2} using three size-groups.}
    \label{fig:size_groups_3_buoyancy}
\end{figure}

\begin{figure}[htbp!]
    \centering
    \includegraphics[width=\linewidth]{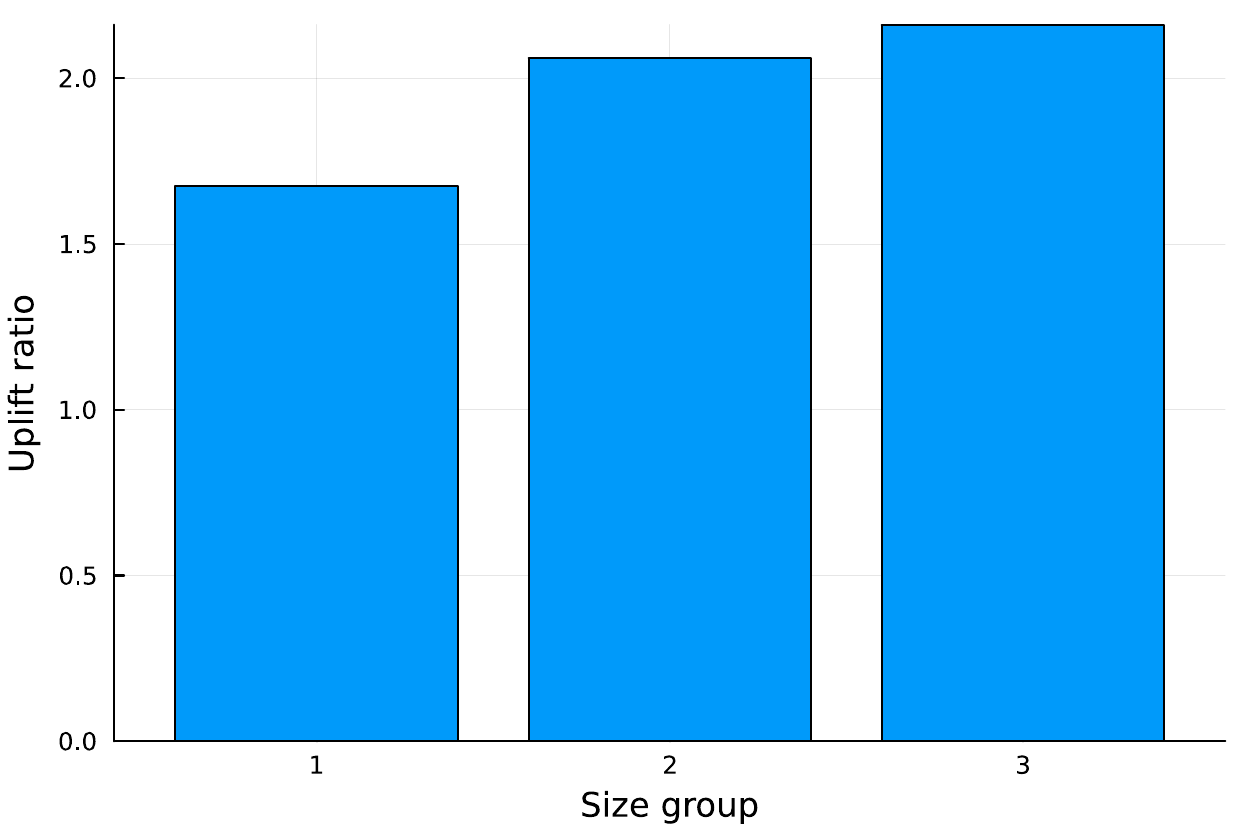}
    \caption{Optimized size-group uplift ratios obtained in Sec.~\ref{sub:param2} using three size-groups.}
    \label{fig:size_groups_3_ratios}
\end{figure}

\begin{figure}[htbp!]
    \centering
    \includegraphics[width=\linewidth]{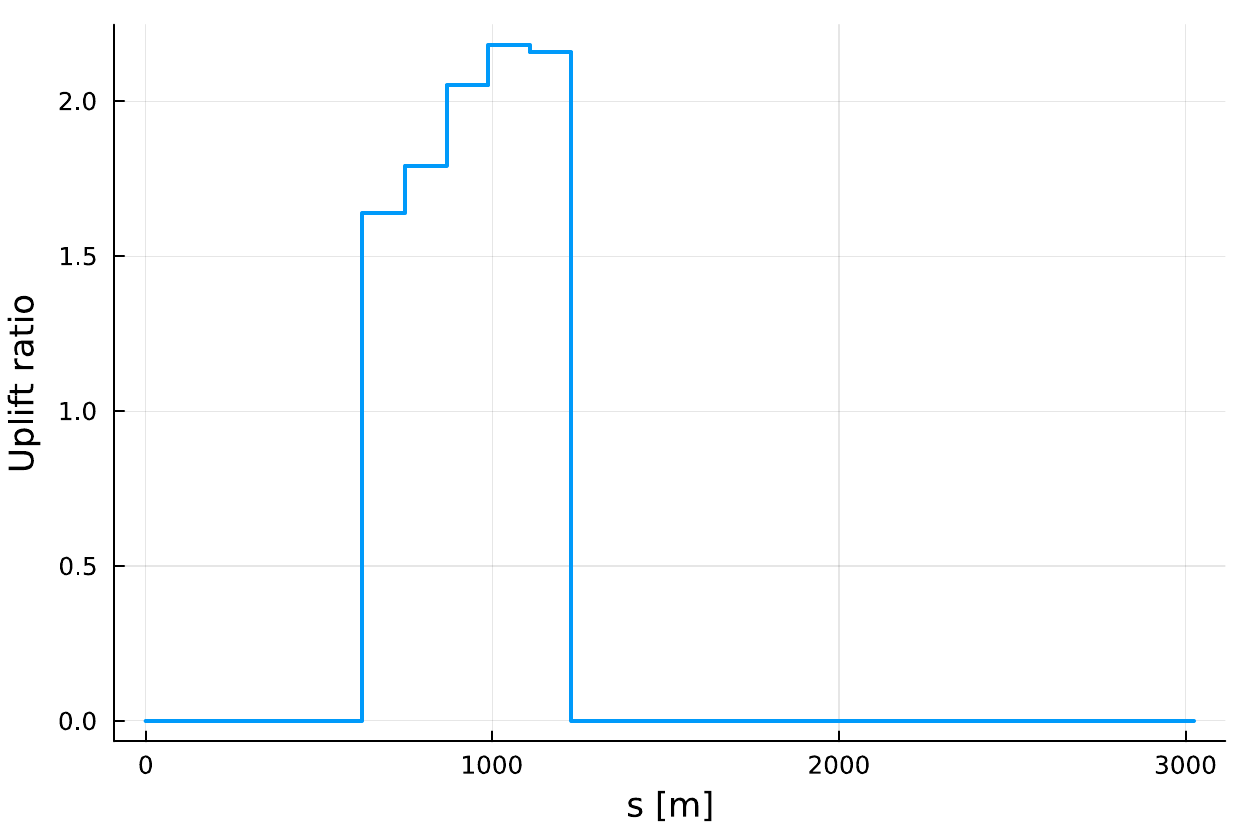}
    \caption{Optimized buoyancy distribution obtained in Sec.~\ref{sub:param2} using five size-groups.}
    \label{fig:size_groups_5_buoyancy}
\end{figure}

\begin{figure}[htbp!]
    \centering
    \includegraphics[width=\linewidth]{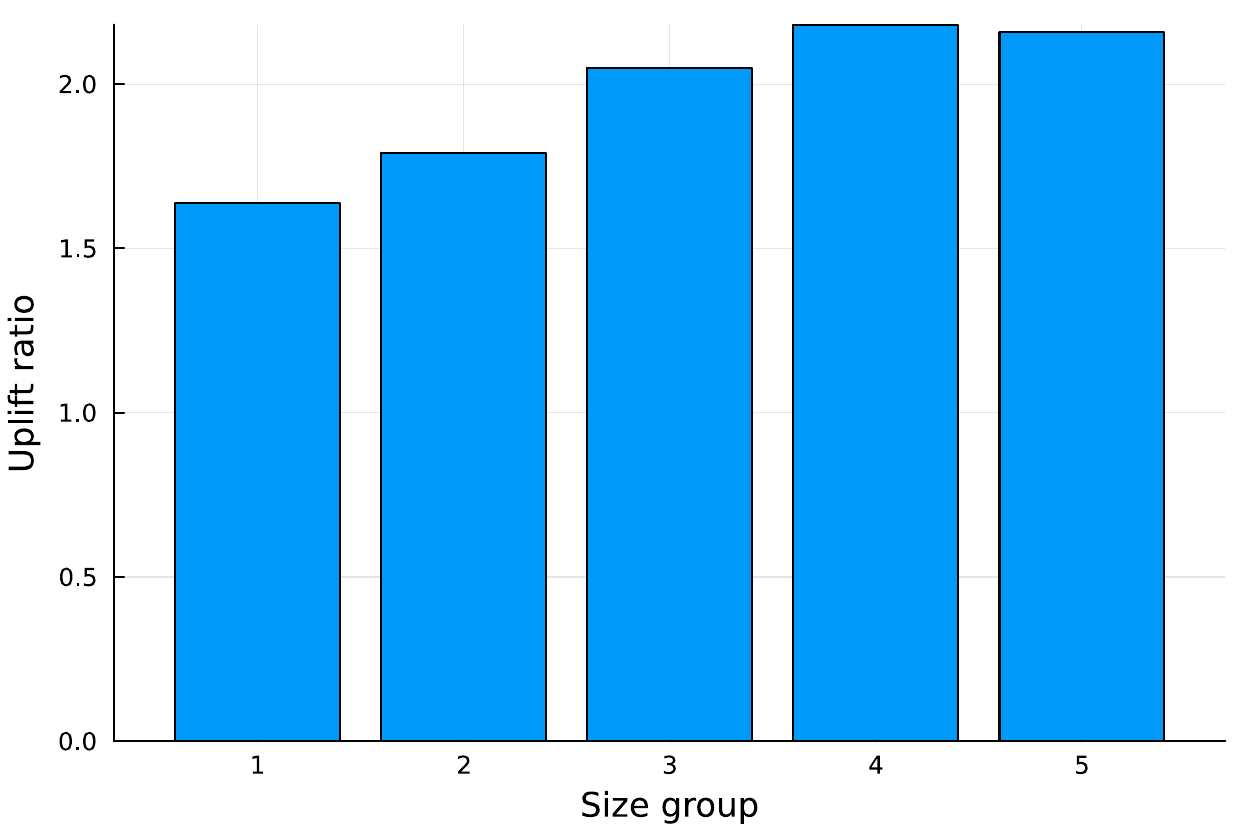}
    \caption{Optimized size-group uplift ratios obtained in Sec.~\ref{sub:param2} using five size-groups.}
    \label{fig:size_groups_5_ratios}
\end{figure}

\begin{figure}[htbp!]
    \centering
    \includegraphics[width=\linewidth]{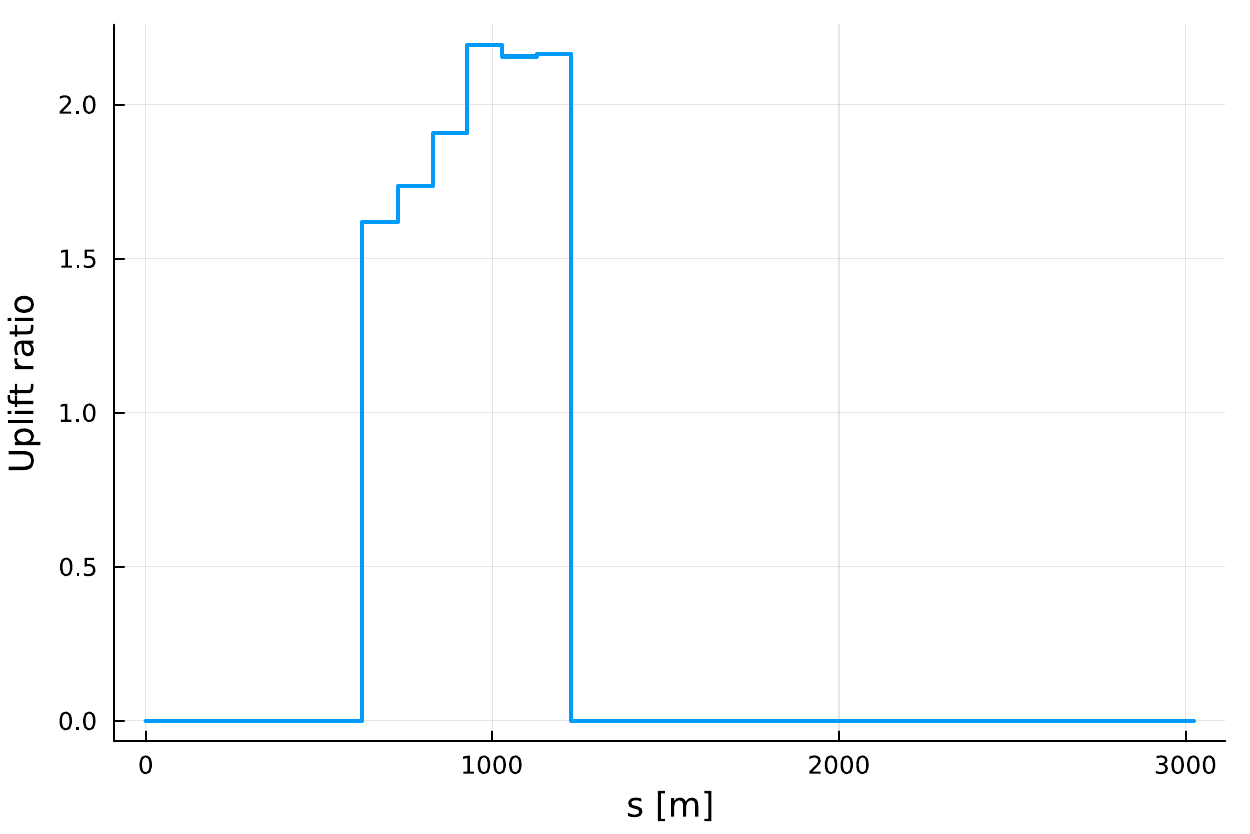}
    \caption{Optimized buoyancy distribution obtained in Sec.~\ref{sub:param2} using six size-groups.}
    \label{fig:size_groups_6_buoyancy}
\end{figure}

Figs.~\ref{fig:size_groups_6_shape}--\ref{fig:size_groups_6_tension} present the structural results for the six-size-group design. Fig.~\ref{fig:size_groups_6_shape} compares the unbuoyed reference catenary with the optimized lazy-wave layout. Fig.~\ref{fig:size_groups_6_moment} shows the bending moment distribution, and Fig.~\ref{fig:size_groups_6_tension} shows the effective tension along the riser. In the tension plot, the dashed curve is the catenary tension response corresponding to the blue catenary profile in Fig.~\ref{fig:size_groups_6_shape}. Its peak value provides the reference used to define the upper bound on tension in the optimization of the lazy-wave configuration.

\begin{figure}[htbp!]
    \centering
    \includegraphics[width=\linewidth]{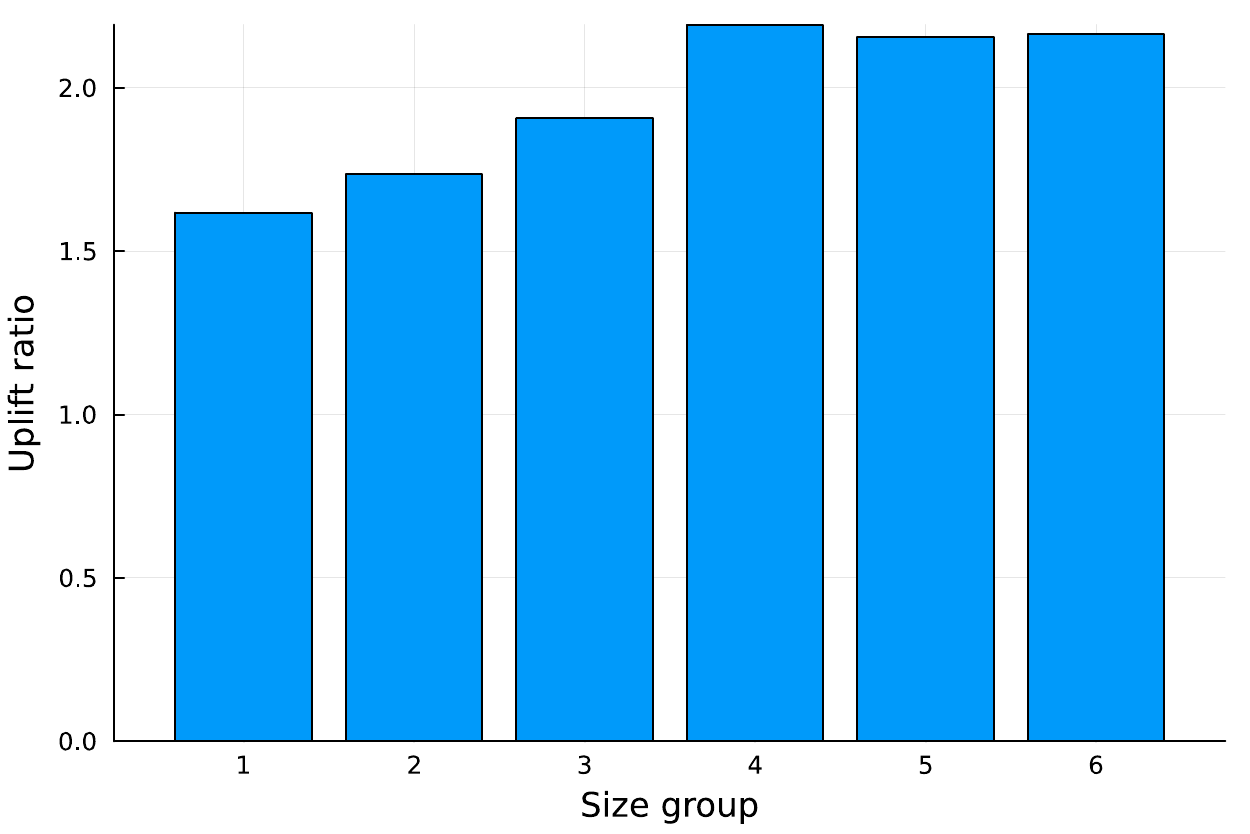}
    \caption{Optimized size-group uplift ratios obtained in Sec.~\ref{sub:param2} using six size-groups.}
    \label{fig:size_groups_6_ratios}
\end{figure}

\begin{figure}[htbp!]
    \centering
    \includegraphics[width=\linewidth]{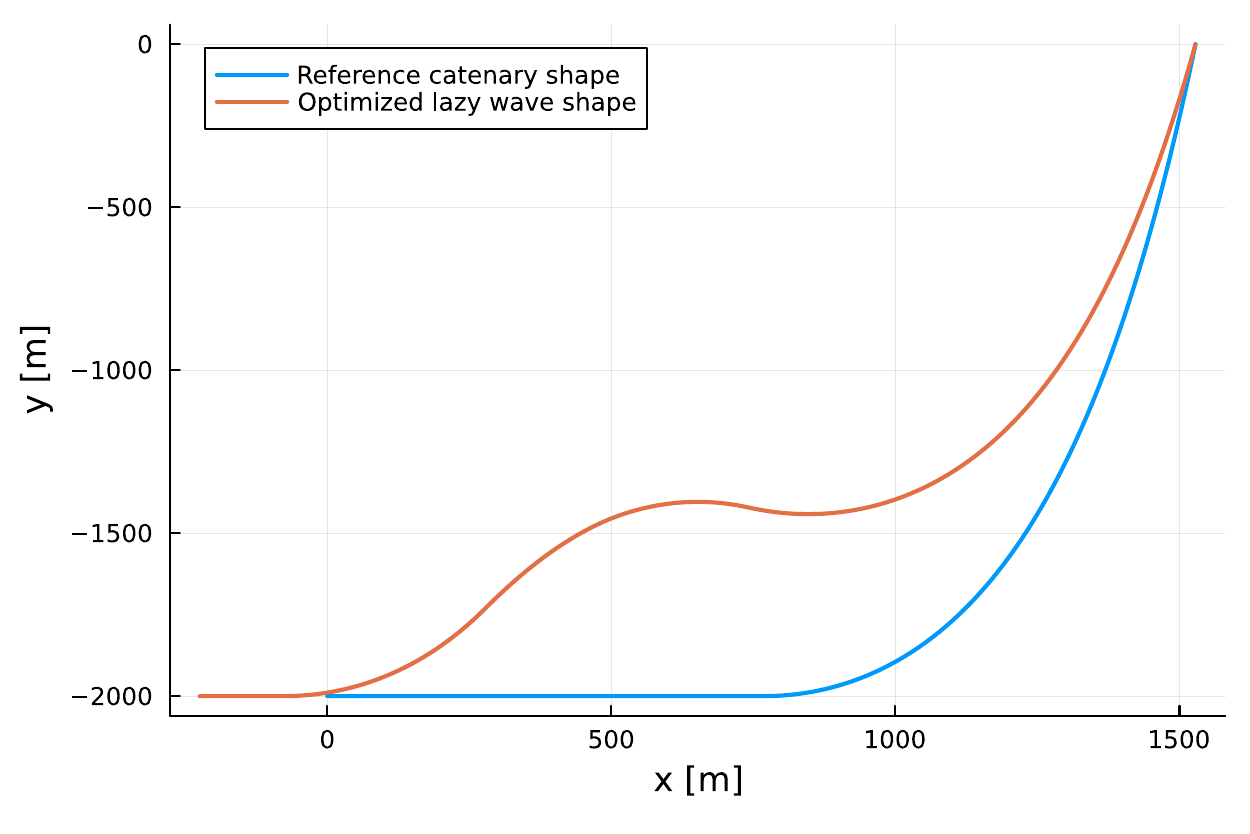}
    \caption{Reference and optimized riser configurations for the six-size-group design in Sec.~\ref{sub:param2}.}
    \label{fig:size_groups_6_shape}
\end{figure}

\begin{figure}[htbp!]
    \centering
    \includegraphics[width=\linewidth]{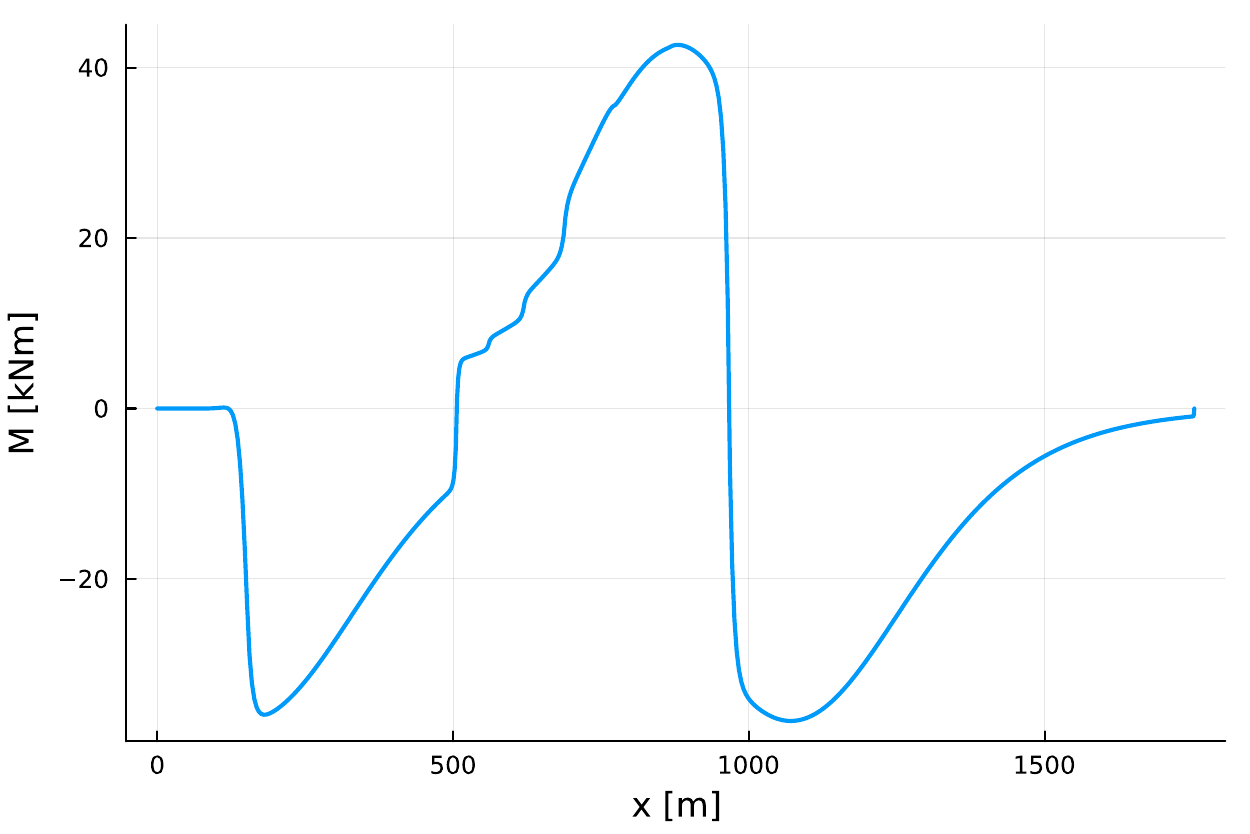}
    \caption{Bending-moment response for the optimized six-size-group design in Sec.~\ref{sub:param2}.}
    \label{fig:size_groups_6_moment}
\end{figure}

\begin{figure}[htbp!]
    \centering
    \includegraphics[width=\linewidth]{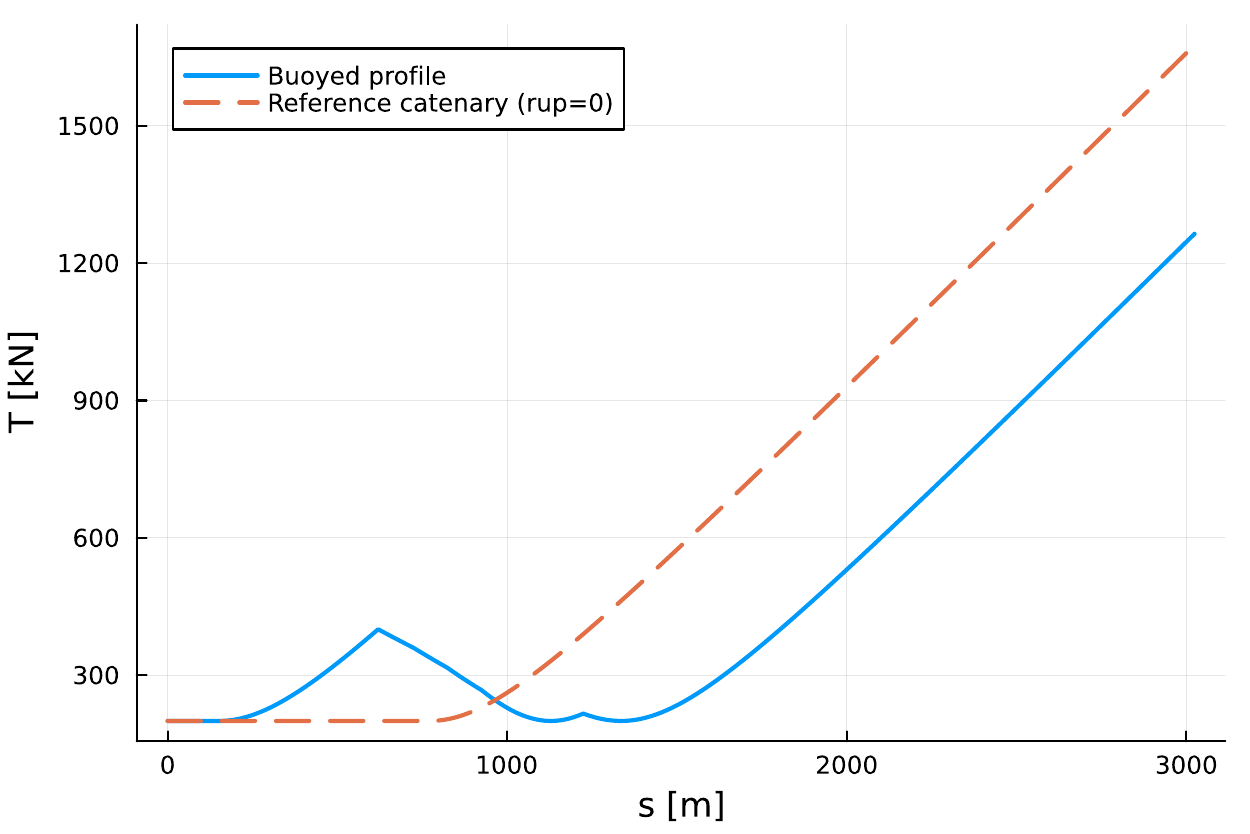}
    \caption{Effective-tension response for the optimized six-size-group design in Sec.~\ref{sub:param2}.}
    \label{fig:size_groups_6_tension}
\end{figure}

The computational effort required by the three size-grouped design optimization cases is summarized in Table~\ref{tab:size_group_optimization_cost}. The table reports the IPOPT iteration count and total solver time recorded in the corresponding output logs.
\begin{table}[htbp!]
    \centering
    \caption{IPOPT iteration counts, reported solver times, and final objective values for the optimization design cases of Sec.~\ref{sub:param2}.}
    \label{tab:size_group_optimization_cost}
    \small
    \setlength{\tabcolsep}{2pt}
    \begin{tabular}{cccc}
        \toprule
        \begin{tabular}[c]{@{}c@{}}Size-group\\ count\end{tabular} & Iterations & \begin{tabular}[c]{@{}c@{}}IPOPT\\ time (s)\end{tabular} & \begin{tabular}[c]{@{}c@{}}Final\\ objective\end{tabular} \\
        \midrule
        3 & 10 & 88.082 & 589.223 \\
        5 & 19 & 175.328 & 588.077 \\
        6 & 18 & 151.894 & 587.935 \\
        \bottomrule
    \end{tabular}
\end{table}
The three-, five-, and six-group formulations yield different buoyancy distributions with closely comparable objective values while satisfying the prescribed constraints. These examples illustrate the ability to accommodate different restrictions on the number of module sizes with little change in the achieved objective value.

The five size-group case required the greatest computational effort, with nearly twice the solver time of the three size-group case. Although the six size-group formulation introduces one additional design variable relative to the five size-group case, it converged in slightly fewer iterations and required less solver time. Thus, the computational cost increases overall relative to the three size-group case but is not strictly monotonic in the number of groups. 


\subsection{Simultaneous optimization of buoyancy size-group selection and buoyancy sizes}
\label{sub:param3}
In this section we discuss the results obtained with the last formulation proposed in this paper. This formulation is inspired by the approaches for discrete material optimization (DMO) discussed by \citet{Stegmann2005discrete,lund2005structural}. 
This formulation optimizes both the assignment of buoyant nodes to size-groups and the uplift ratio associated with each group. 
Let the $n_d$ buoyancy nodes be indexed by $i=1,\ldots,n_d$, and let $n_g$ be the prescribed number of size-groups. The assignment matrix is $\vect X=[x_{ij}]\in\mathbb{R}^{n_d\times n_g}$, where $x_{ij}\in[0,1]$ is the weight assigning buoyant node $i$ to group $j$. The group uplift ratios are collected in $\vect r^g=[r^g_1,\ldots,r^g_{n_g}]^{\mathsf T}$, and the vector with the optimization variables is $\vect z=[\vect X[:];\vect r^g]\footnote{Here, $[:]$ denotes vectorization: it stacks all matrix entries into a single vector.}\in\mathbb{R}^{n_d n_g+n_g}$.

Each buoyant node has a mechanical uplift ratio, which determines its structural response, and a cost uplift ratio, which enters the quadratic buoyancy objective:
\begin{equation}
    \begin{aligned}
        r_i^{\mathrm{mech}} &= \sum_{j=1}^{n_g}\varphi_{p_{\mathrm{DMO}}}(x_{ij})r_j^g,\\
        r_i^{\mathrm{cost}} &= \sum_{j=1}^{n_g}x_{ij}r_j^g,
        \qquad i=1,\ldots,n_d.
    \end{aligned}
    \label{eq:dmo_parametrization}
\end{equation}

The entries of $\vect X$ are continuous design variables bounded between zero and one, but only the endpoint values have a direct physical interpretation: $x_{ij}=1$ assigns node $i$ to group $j$, whereas $x_{ij}=0$ indicates that node $i$ is not assigned to group $j$. To encourage the optimizer to converge toward binary assignments, we use an interpolation scheme that discourages intermediate values.
The RAMP \citep{stolpe2001alternative} penalization used here is defined as:
\begin{equation}
    \varphi_{p_{\mathrm{DMO}}}(x)
    =\frac{x}{1+p_{\mathrm{DMO}}(1-x)},
    \qquad p_{\mathrm{DMO}}\ge0.
    \label{eq:dmo_penalty}
\end{equation}
It satisfies $\varphi_{p_{\mathrm{DMO}}}(0)=0$ and $\varphi_{p_{\mathrm{DMO}}}(1)=1$. For $p_{\mathrm{DMO}}>0$, it also satisfies $\varphi_{p_{\mathrm{DMO}}}(x)<x$ for $x\in(0,1)$, while retaining a nonzero slope at zero, $\varphi'_{p_{\mathrm{DMO}}}(0)=1/(1+p_{\mathrm{DMO}})$. Thus $r_i^{\mathrm{mech}}\le r_i^{\mathrm{cost}}$. 
The inequality is strict for fractional assignments, while binary assignments give equal values. At $p_{\mathrm{DMO}}=0$, the penalty vanishes and $r_i^{\mathrm{mech}}=r_i^{\mathrm{cost}}$. 
For $p_{\mathrm{DMO}}>0$, fractional assignments enter the quadratic objective through $r_i^{\mathrm{cost}}$ while contributing a smaller uplift $r_i^{\mathrm{mech}}$ to the equilibrium equations. The RAMP penalization is applied only to the mechanical interpolation, while the cost interpolation remains unpenalized. A fractional assignment costs according to its unpenalized uplift contribution, but delivers a reduced mechanical uplift. This mismatch is the penalization mechanism. Only $0$ or $1$ values of $x_{ij}$ avoid this loss of mechanical uplift relative to the cost attributed to them, encouraging the optimization algorithm to converge towards discrete selections within the continuous relaxation $x_{ij}\in[0,1]$ of the original binary definition of the variables. 

The DMO formulation of the problem reads as follows: 
\begingroup\footnotesize
\begin{equation}
    \begin{aligned}
        \minimize_{\vect X,\vect r^g}\quad
            & \left(\vect r^{cost}\right)^{\mathsf T}\vect r^{cost}\\
        \text{subject to}\quad
            & \kappa_p\left(\vect\theta(\vect r^{\mathrm{mech}}),\vect r^{\mathrm{mech}}\right)\le \kappa_{\mathrm{lim}},\\
            & T_p\left(\vect\theta(\vect r^{\mathrm{mech}}),\vect r^{\mathrm{mech}}\right)\le T_{\mathrm{lim}},\\
            & \sum_{j=1}^{n_g}x_{ij}\le1,
                \qquad i=1,\ldots,n_d,\\
            & r^g_{j+1}-r^g_j\ge\delta,
                \qquad j=1,\ldots,n_g-1,\\
            & x_{ij}\ge0,
                \qquad i=1,\ldots,n_d,\quad j=1,\ldots,n_g,\\
            & r_{\min}\le r^g_j\le r_{\max},
                \qquad j=1,\ldots,n_g,\\
            & \vect R\left(\vect\theta(\vect r^{\mathrm{mech}}),\vect r^{\mathrm{mech}}\right) = \vect 0.
    \end{aligned}
    \label{eq:dmo_optimization_problem}
\end{equation}
\endgroup
Here $\vect R=\vect 0$ is the discretized equilibrium system solved for the nodal angles $\vect\theta$ given the mechanical uplift. As a reminder, the response p-norm constraints have limits that are scaled from the catenary reference without buoyancy:
\begin{equation}
    \kappa_{\mathrm{lim}}=\alpha_\kappa \kappa_{\max}^{\mathrm{cat}} ,
    \qquad
    T_{\mathrm{lim}}=\alpha_T T_{\max}^{\mathrm{cat}} ,
    \label{eq:dmo_limits}
\end{equation}
where $(\alpha_T,\alpha_\kappa)=(0.75,1.20)$ or $(0.80,1.10)$ for the two cases considered below. 
The linear constraints 
\begin{equation}\label{eq:dmo_order} r^{g}_{j+1}-r^{g}_{j}\ge\delta, \end{equation} 
with $\delta>0$, prevent groups from collapsing onto the same size, which would make the assignment non unique and reduce the number of distinct module types. They do not, however, force every group to be used.

The row-sum constraint $\sum_{j=1}^{n_g}x_{ij}\le1$ caps the total assignment weight at each node and admits an all-zero row, which represents a node without a buoyancy module. Together with $x_{ij}\ge0$, it also implies $x_{ij}\le1$. Because $\vect X$ is continuous, this constraint does not by itself enforce a single group selection per node, and weight may be split among several groups. Such fractional assignments are discouraged by the RAMP penalization in \eqref{eq:dmo_penalty}, as discussed previously. In the examples, $n_g=5$, $\delta=0.05$, $r_{\min}=0$, and $r_{\max}=3$. 
Additionally, a continuation strategy is used in which $p_{\mathrm{DMO}}$ is increased from 1 to 5 in two subsequent optimization stages.
The gradients of the nonlinear curvature and tension constraints are obtained following the same approach as in Sec.~\ref{sub:param2}: the nodal sensitivities computed with the adjoint analysis of Sec.~\ref{sec:sensitivity} are mapped to the design variables by the chain rule, as in \eqref{eq:size_group_gradient_mapping}. 

As with the previous formulations, problem \eqref{eq:dmo_optimization_problem} is nonconvex, and its solution depends strongly on the initial design. In our numerical experiments, starting from a uniform uplift ratio tended to drive all nodes toward a single size-group. Better results were obtained by initializing the starting point from the optimized continuous design of Sec.~\ref{sub:param1}, which provides one uplift ratio $\bar r_i$ per buoyant node. The initial group uplift ratios $\vect r^{g,(0)}$ are taken as the $k$-means centers ($k=n_g$) of $\{\bar r_i\}$, sorted in ascending order to satisfy \eqref{eq:dmo_order}. Each node is then assigned a weight $w=0.9$ on the group whose uplift ratio is closest to $\bar r_i$, and $(1-w)/(n_g-1)$ on each of the remaining groups, so that every row of $\vect X^{(0)}$ sums to one. The starting point is then defined by $\vect z^{(0)}=[\vect X^{(0)}[:];\vect r^{g,(0)}]$.


\begin{figure}[htbp!]
    \centering
    \includegraphics[width=\linewidth]{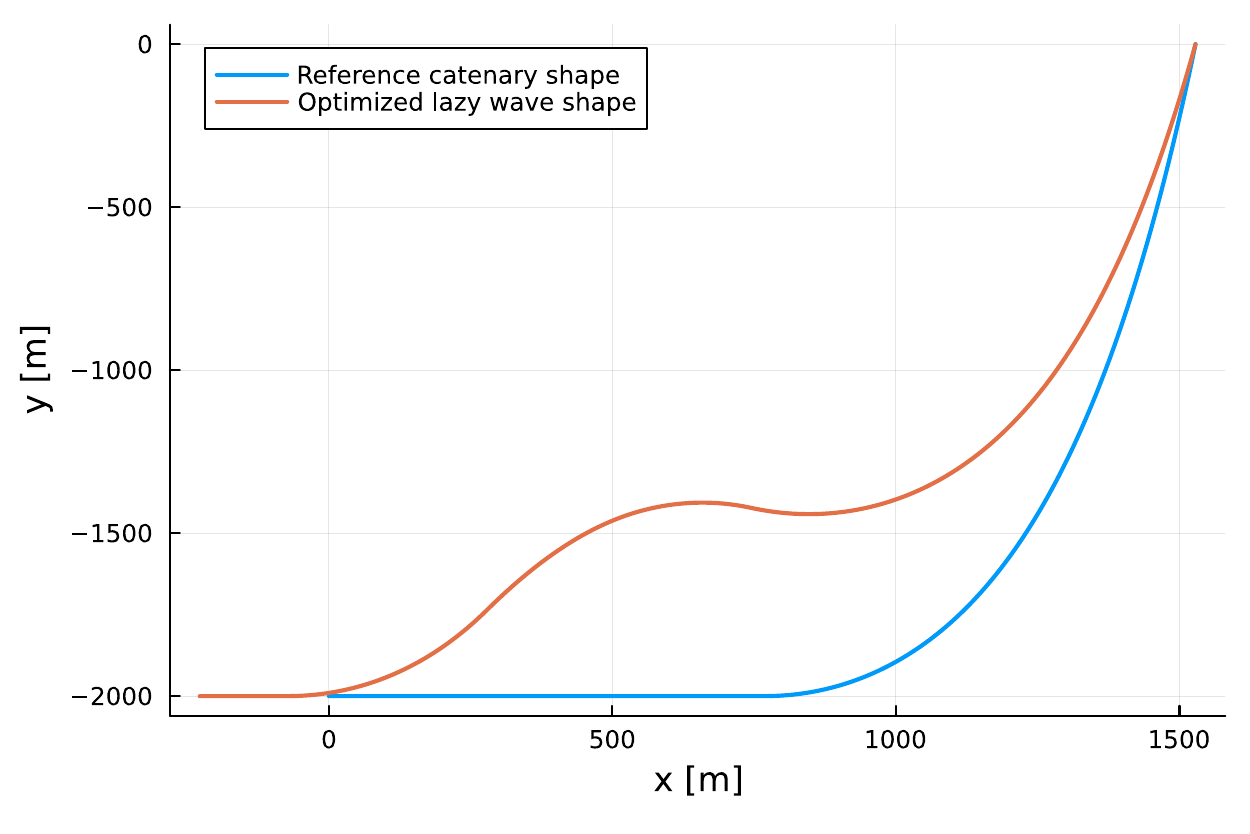}
    \caption{Reference and optimized riser layouts for the results of Sec.~\ref{sub:param3}, with $T_{\mathrm{lim}}=0.75T_{\max}^{\mathrm{cat}}$ and $\kappa_{\mathrm{lim}}=1.20\kappa_{\max}^{\mathrm{cat}}$.}
    \label{fig:dmo075_layout}
\end{figure}

\begin{figure}[htbp!]
    \centering
    \includegraphics[width=\linewidth]{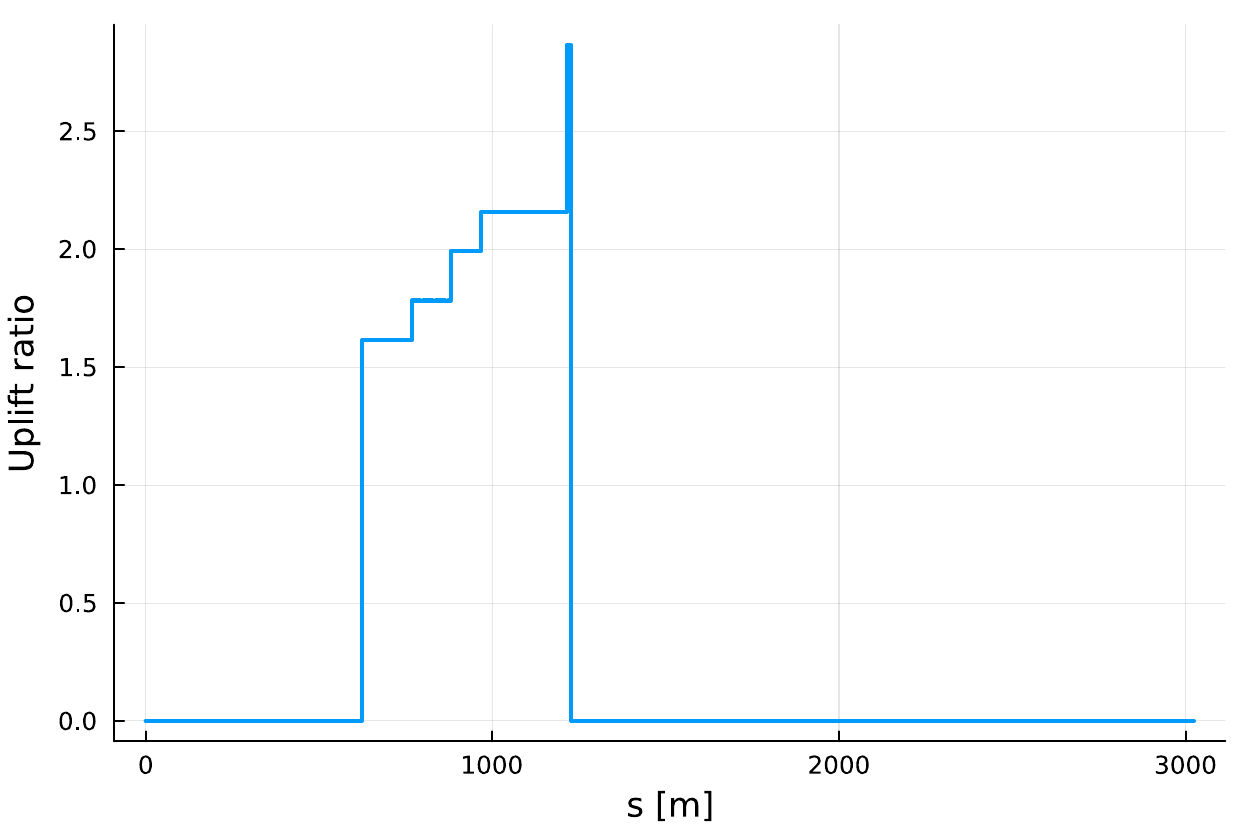}
    \caption{Optimized buoyancy distribution for the results of Sec.~\ref{sub:param3}, with $T_{\mathrm{lim}}=0.75T_{\max}^{\mathrm{cat}}$ and $\kappa_{\mathrm{lim}}=1.20\kappa_{\max}^{\mathrm{cat}}$.}
    \label{fig:dmo075_buoyancy}
\end{figure}

\begin{figure}[htbp!]
    \centering
    \includegraphics[width=\linewidth]{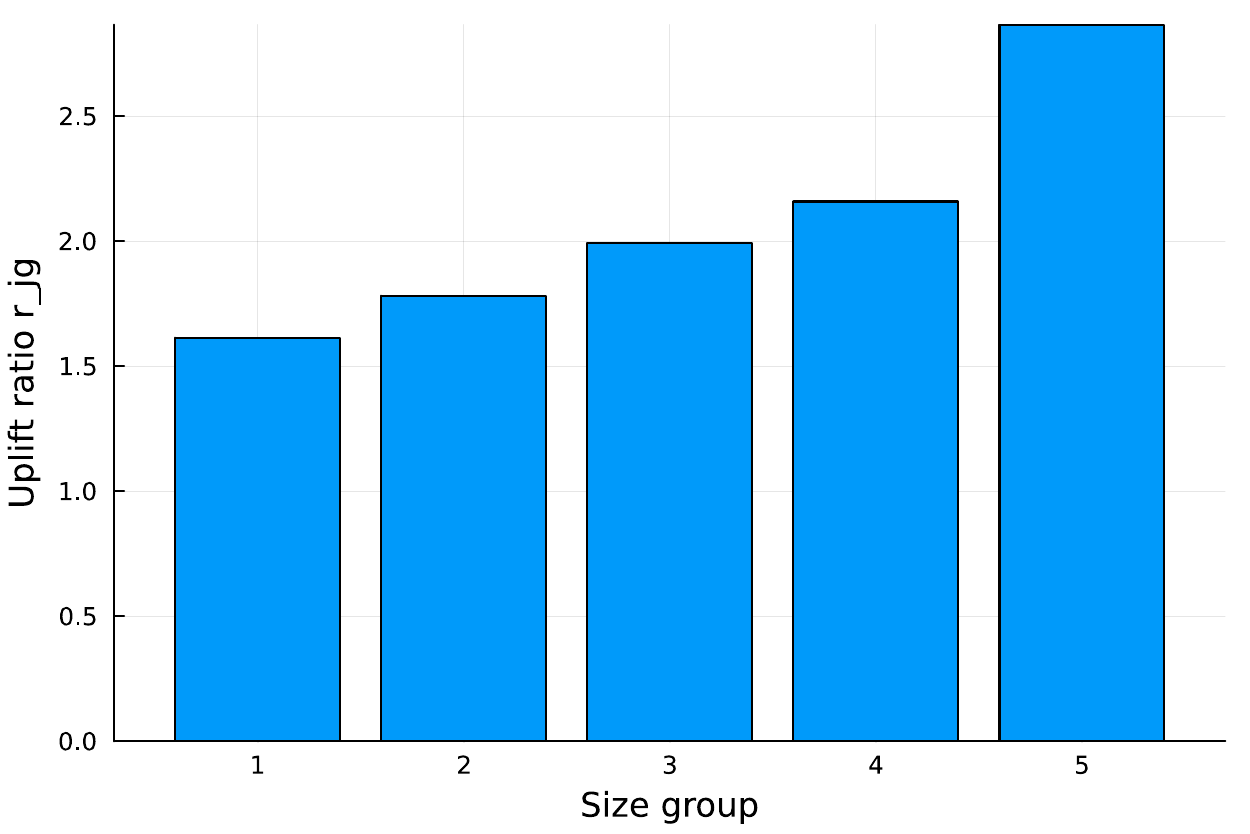}
    \caption{Optimized uplift ratios of the five buoyancy size-groups for the results of Sec.~\ref{sub:param3}, with $T_{\mathrm{lim}}=0.75T_{\max}^{\mathrm{cat}}$ and $\kappa_{\mathrm{lim}}=1.20\kappa_{\max}^{\mathrm{cat}}$.}
    \label{fig:dmo075_groups}
\end{figure}

\begin{figure}[htbp!]
    \centering
    \includegraphics[width=\linewidth]{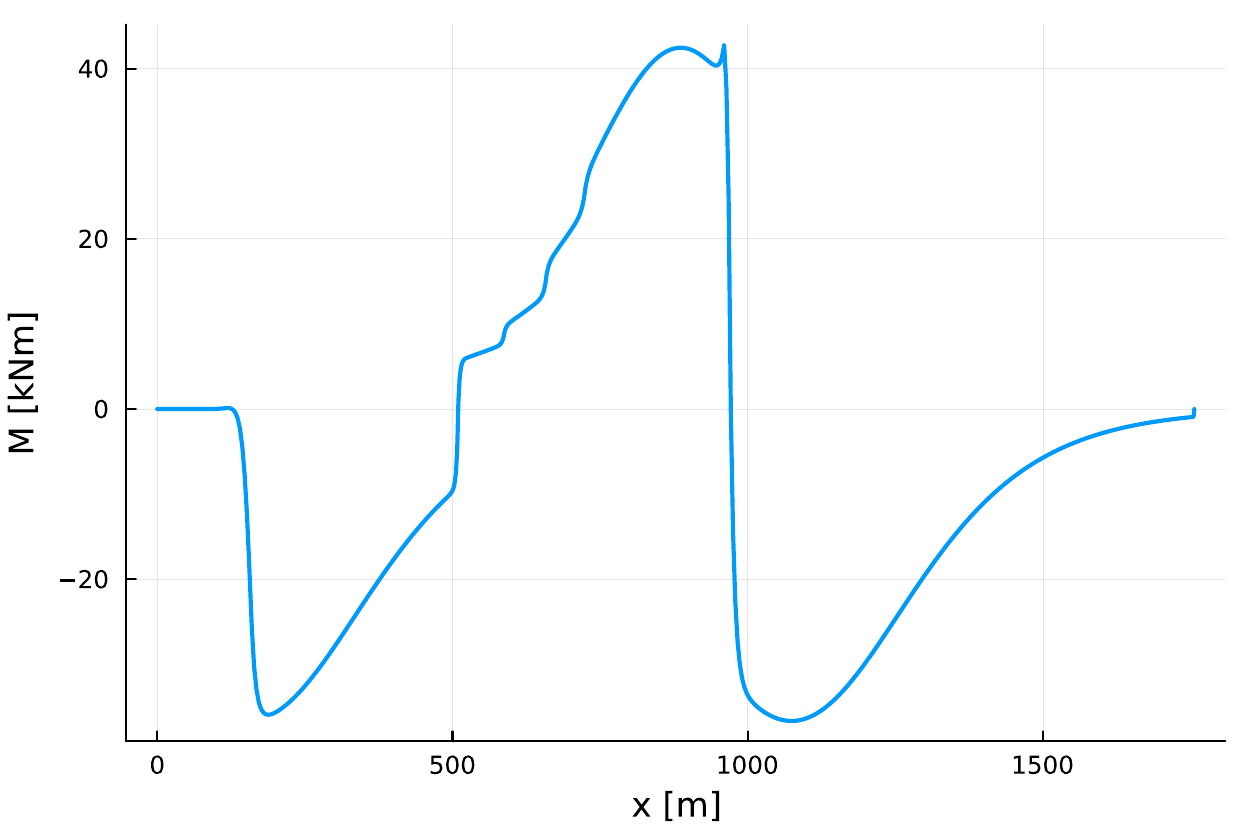}
    \caption{Bending moment response for the optimized five-size-group design of Sec.~\ref{sub:param3} with $T_{\mathrm{lim}}=0.75T_{\max}^{\mathrm{cat}}$ and $\kappa_{\mathrm{lim}}=1.20\kappa_{\max}^{\mathrm{cat}}$.}
    \label{fig:dmo075_moment}
\end{figure}

\begin{figure}[htbp!]
    \centering
    \includegraphics[width=\linewidth]{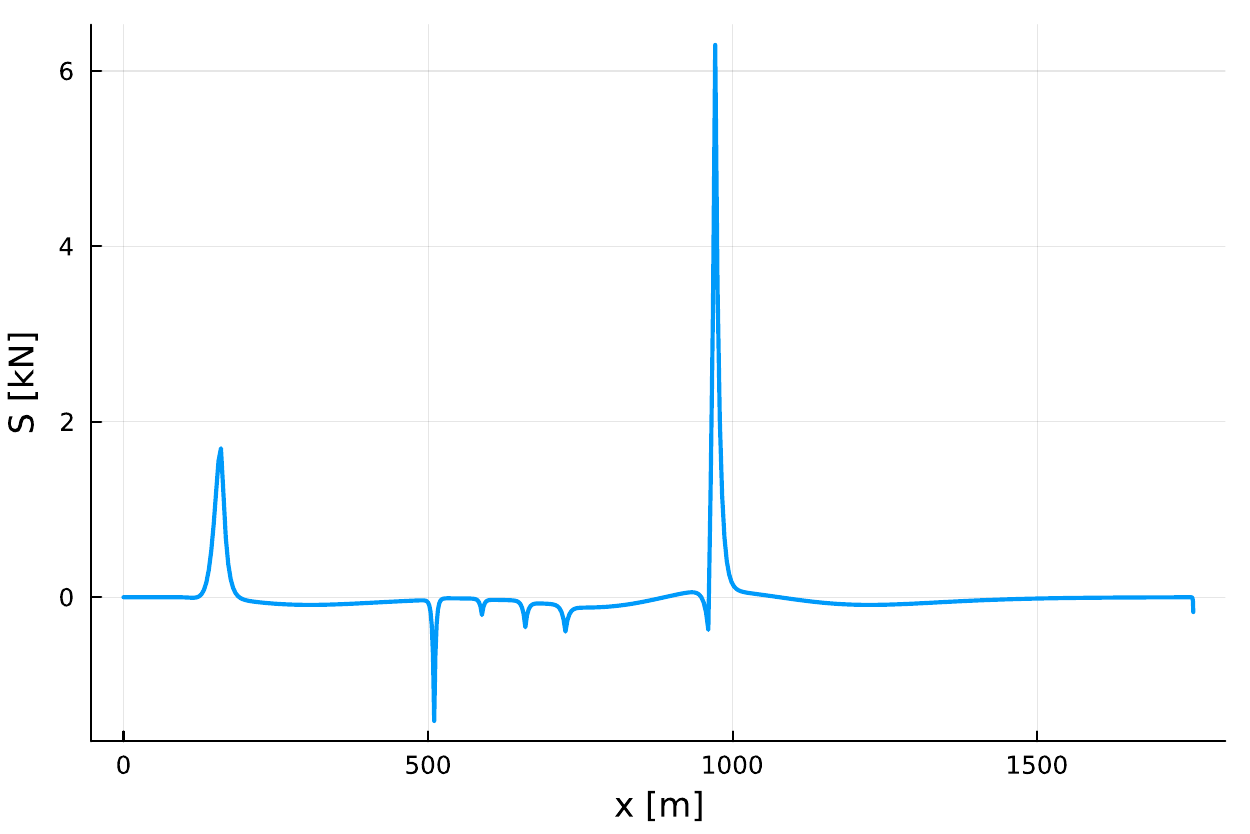}
    \caption{Shear force response for the optimized five-size-group design of Sec.~\ref{sub:param3} with $T_{\mathrm{lim}}=0.75T_{\max}^{\mathrm{cat}}$ and $\kappa_{\mathrm{lim}}=1.20\kappa_{\max}^{\mathrm{cat}}$.}
    \label{fig:dmo075_shear}
\end{figure}

\begin{figure}[htbp!]
    \centering
    \includegraphics[width=\linewidth]{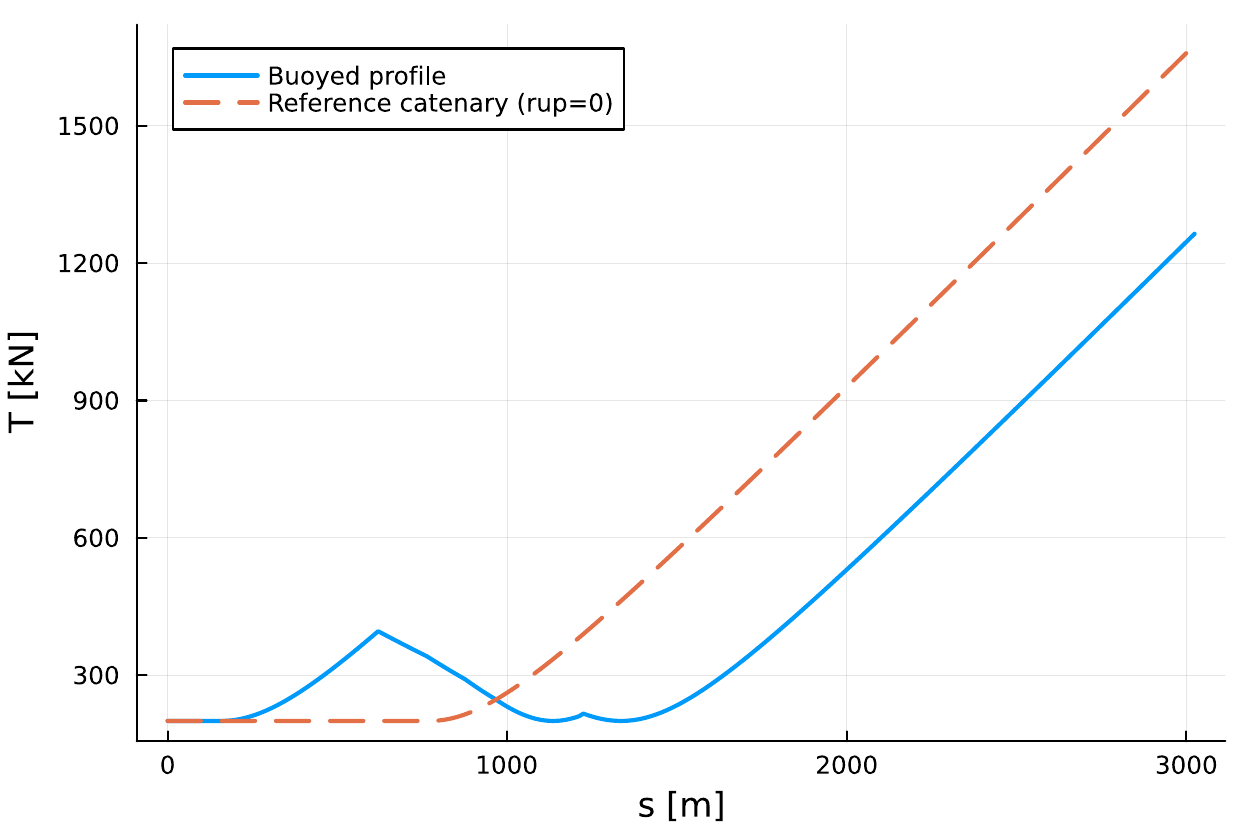}
    \caption{Effective tension response for the optimized five-size-group design of Sec.~\ref{sub:param3} with $T_{\mathrm{lim}}=0.75T_{\max}^{\mathrm{cat}}$ and $\kappa_{\mathrm{lim}}=1.20\kappa_{\max}^{\mathrm{cat}}$.}
    \label{fig:dmo075_tension}
\end{figure}

The DMO design formulation discussed in this section was applied with five buoyancy size-groups for the two combinations of tension and curvature limits introduced in \eqref{eq:dmo_limits}. First, for $T_{\mathrm{lim}}=0.75T_{\max}^{\mathrm{cat}}$ and $\kappa_{\mathrm{lim}}=1.20\kappa_{\max}^{\mathrm{cat}}$, the optimized group uplift ratios were $r^g_j=1.615$, $1.782$, $1.993$, $2.159$, and $2.866$, in ascending group order. The optimized layout, buoyancy distribution, and group uplift ratios are shown in Figs.~\ref{fig:dmo075_layout}--\ref{fig:dmo075_groups}. The associated bending moment, shear force, and effective tension are shown in Figs.~\ref{fig:dmo075_moment}--\ref{fig:dmo075_tension}.

\begin{figure}[htbp!]
    \centering
    \includegraphics[width=\linewidth]{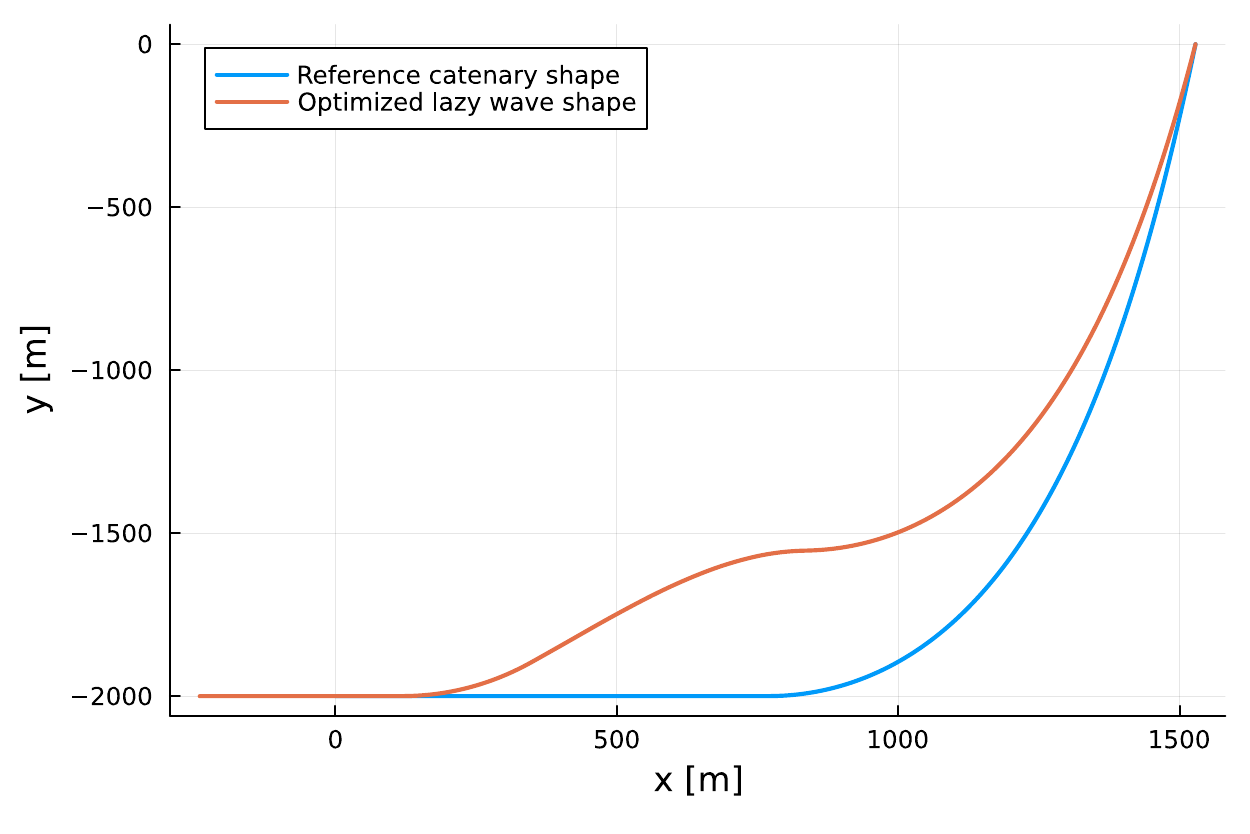}
    \caption{Reference and optimized riser layouts for the results of Sec.~\ref{sub:param3}, with $T_{\mathrm{lim}}=0.80T_{\max}^{\mathrm{cat}}$ and $\kappa_{\mathrm{lim}}=1.10\kappa_{\max}^{\mathrm{cat}}$.}
    \label{fig:dmo080_layout}
\end{figure}

\begin{figure}[htbp!]
    \centering
    \includegraphics[width=\linewidth]{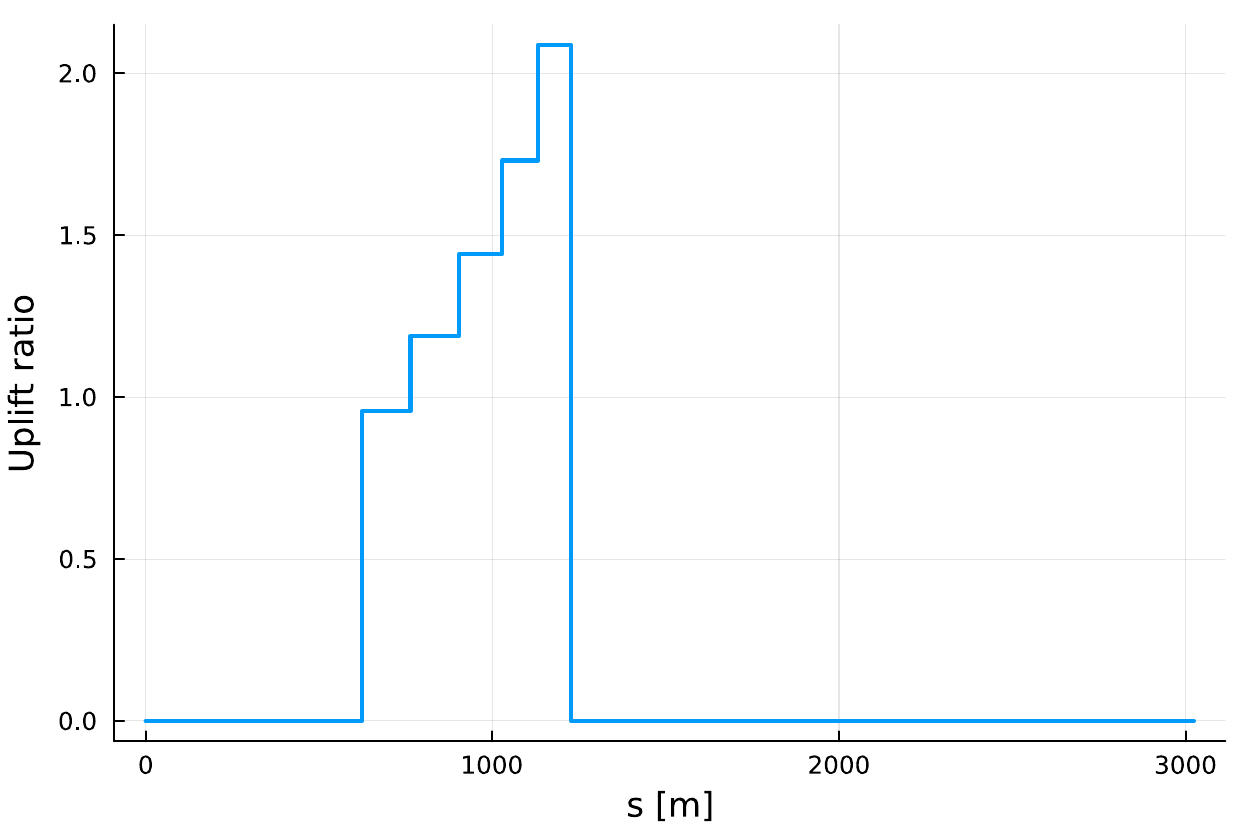}
    \caption{Optimized buoyancy distribution for the results of Sec.~\ref{sub:param3}, with $T_{\mathrm{lim}}=0.80T_{\max}^{\mathrm{cat}}$ and $\kappa_{\mathrm{lim}}=1.10\kappa_{\max}^{\mathrm{cat}}$.}
    \label{fig:dmo080_buoyancy}
\end{figure}

\begin{figure}[htbp!]
    \centering
    \includegraphics[width=\linewidth]{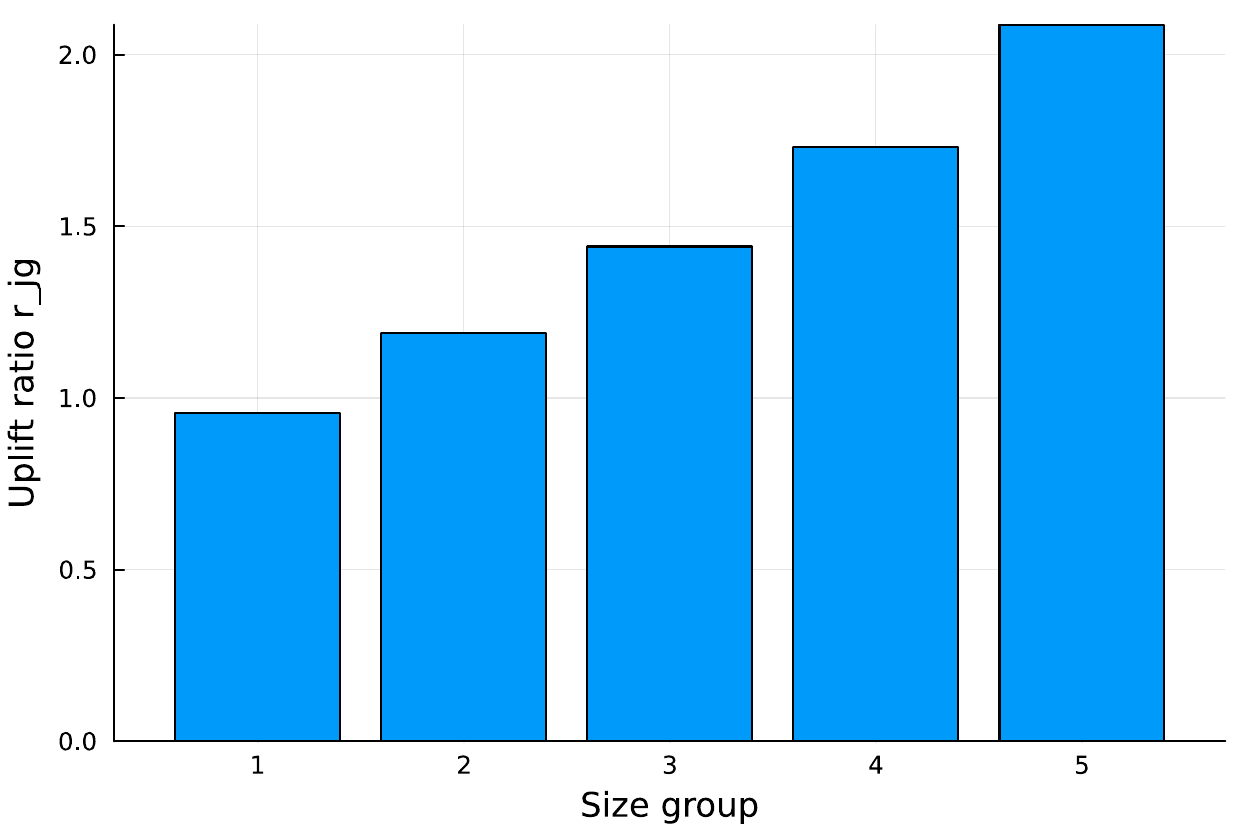}
    \caption{Optimized uplift ratios of the five buoyancy size-groups for the results of Sec.~\ref{sub:param3}, with $T_{\mathrm{lim}}=0.80T_{\max}^{\mathrm{cat}}$ and $\kappa_{\mathrm{lim}}=1.10\kappa_{\max}^{\mathrm{cat}}$.}
    \label{fig:dmo080_groups}
\end{figure}

\begin{figure}[htbp!]
    \centering
    \includegraphics[width=\linewidth]{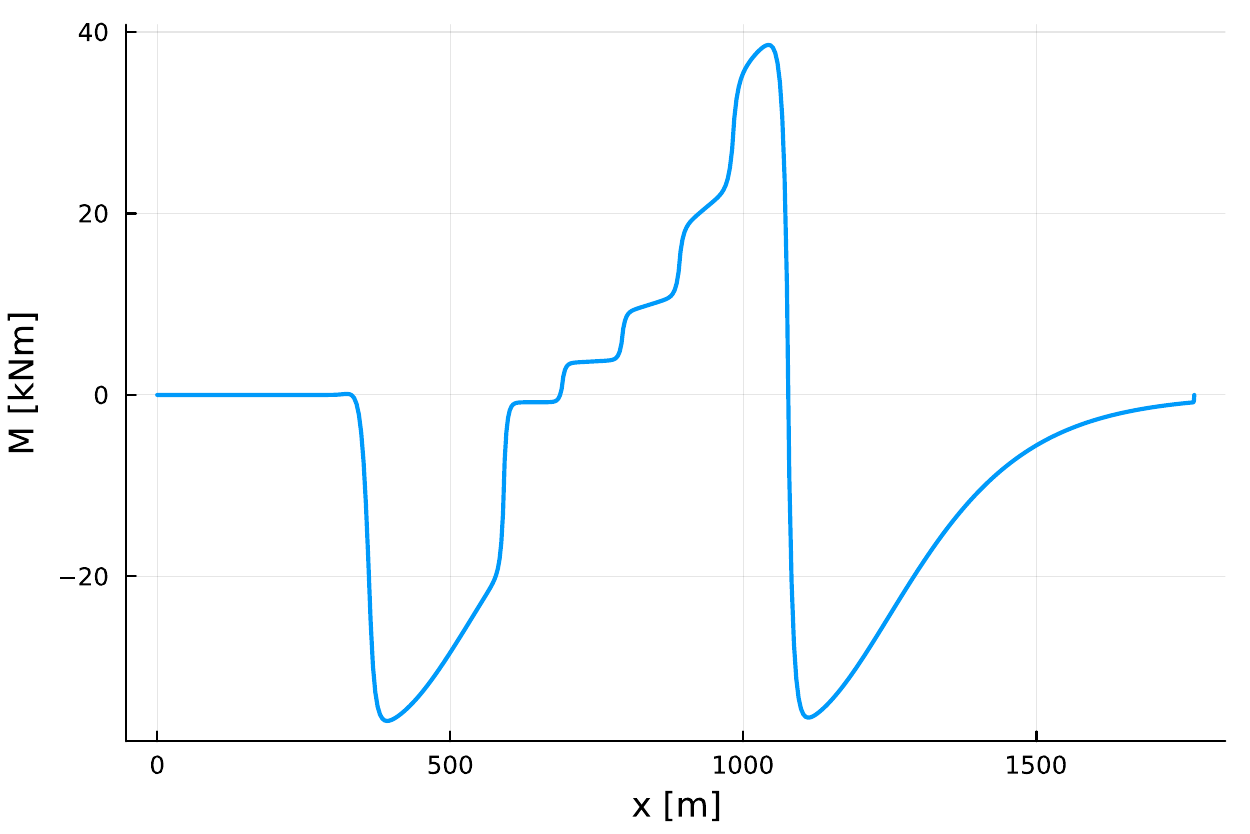}
    \caption{Bending moment response for the optimized five-size-group design of Sec.~\ref{sub:param3} with $T_{\mathrm{lim}}=0.80T_{\max}^{\mathrm{cat}}$ and $\kappa_{\mathrm{lim}}=1.10\kappa_{\max}^{\mathrm{cat}}$.}
    \label{fig:dmo080_moment}
\end{figure}

\begin{figure}[htbp!]
    \centering
    \includegraphics[width=\linewidth]{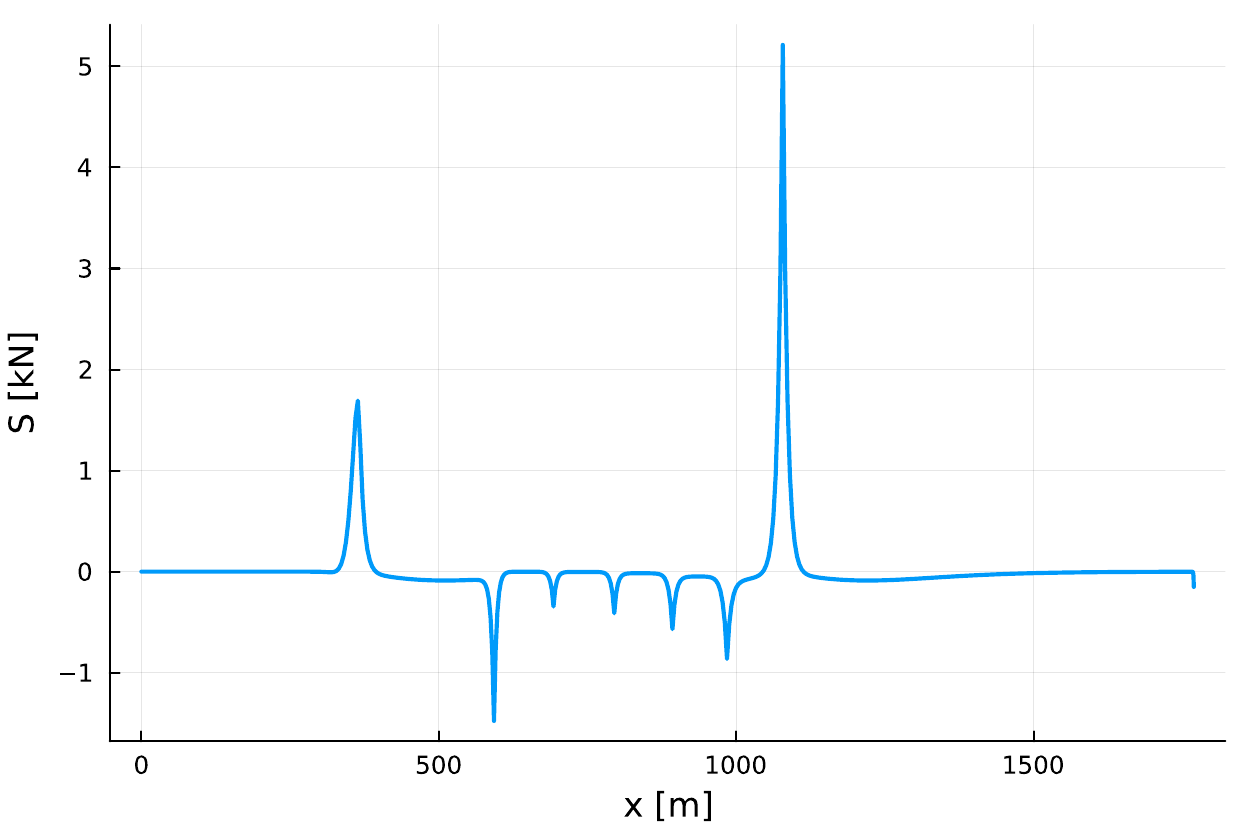}
    \caption{Shear force response for the optimized five-size-group design of Sec.~\ref{sub:param3} with $T_{\mathrm{lim}}=0.80T_{\max}^{\mathrm{cat}}$ and $\kappa_{\mathrm{lim}}=1.10\kappa_{\max}^{\mathrm{cat}}$.}
    \label{fig:dmo080_shear}
\end{figure}

\begin{figure}[htbp!]
    \centering
    \includegraphics[width=\linewidth]{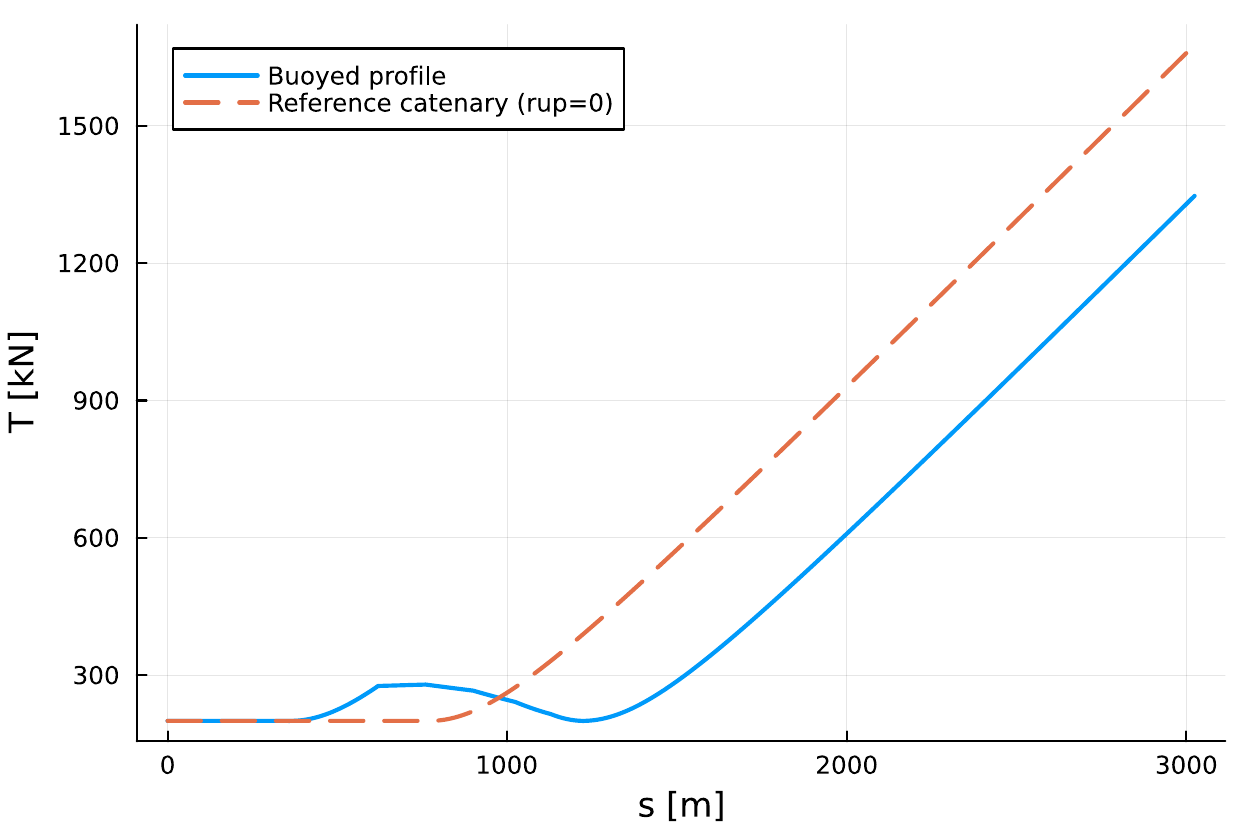}
    \caption{Effective tension response for the optimized five-size-group design of Sec.~\ref{sub:param3} with $T_{\mathrm{lim}}=0.80T_{\max}^{\mathrm{cat}}$ and $\kappa_{\mathrm{lim}}=1.10\kappa_{\max}^{\mathrm{cat}}$.}
    \label{fig:dmo080_tension}
\end{figure}

The second case uses $T_{\mathrm{lim}}=0.80 T_{\max}^{\mathrm{cat}}$ and $\kappa_{\mathrm{lim}}=1.10 \kappa_{\max}^{\mathrm{cat}}$. Its optimized uplift ratios are $r^g_j=0.957$, $1.189$, $1.442$, $1.731$, and $2.088$. The corresponding layout, buoyancy distribution, and group uplift ratios are shown separately in Figs.~\ref{fig:dmo080_layout}--\ref{fig:dmo080_groups}, with the bending moment, shear force, and effective tension responses in Figs.~\ref{fig:dmo080_moment}--\ref{fig:dmo080_tension}. Relative to the first case, the selected group uplift ratios are lower and more evenly spaced. 

Overall, the results of these two cases show that the DMO design formulation, combined with the gradient-based approach proposed in this paper, successfully identifies practical buoyancy distributions. Only the number of distinct size-groups is prescribed. Neither the number of nodes assigned to each group nor the uplift ratio that mechanically characterizes each group is restricted in advance, and both are determined by the optimization. 
The DMO designs also retain the main features of the corresponding continuous designs of Sec.~\ref{sub:param1}. For $T_{\mathrm{lim}}=0.75T_{\max}^{\mathrm{cat}}$ and $\kappa_{\mathrm{lim}}=1.20\kappa_{\max}^{\mathrm{cat}}$, the buoyancy distribution in Fig.~\ref{fig:dmo075_buoyancy} follows a similar pattern to that in Fig.~\ref{fig:continuous_075_120_buoyancy}: a smooth increase in uplift along the buoyancy section, followed by a sharp peak. Similarly, for $T_{\mathrm{lim}}=0.80T_{\max}^{\mathrm{cat}}$ and $\kappa_{\mathrm{lim}}=1.10\kappa_{\max}^{\mathrm{cat}}$, the distribution in Fig.~\ref{fig:dmo080_buoyancy} reproduces the approximately parabolic increase of buoyancy along the buoyancy section observed in Fig.~\ref{fig:continuous_080_110_buoyancy}. In both cases, the DMO distributions are a stepwise, jagged approximation of the corresponding continuous distributions, clearly because each node can take only one of the $n_g$ group uplift ratios.

As for the previous formulations, Table~\ref{tab:dmo_optimization_cost} reports the computational effort required by IPOPT to solve the two optimization problems discussed in this section. Compared with the continuous and size-group formulations (Tables~\ref{tab:continuous_optimization_cost} and~\ref{tab:size_group_optimization_cost}), the computational time increases. This is due to the combinatorial nature of the group assignment and sizing problem, the larger number of design variables (760, compared with 151 for the continuous formulation), and the two-stage continuation on $p_{\mathrm{DMO}}$, which is needed for smoother convergence. Nevertheless, the final objective values of the DMO designs are close to those of the continuous designs in Table~\ref{tab:continuous_optimization_cost}, where each node of the buoyancy section has an independent uplift ratio: 583.762 versus 582.101 for $(\alpha_T,\alpha_\kappa)=(0.75,1.20)$, and 328.488 versus 327.463 for $(\alpha_T,\alpha_\kappa)=(0.80,1.10)$, corresponding to differences of about 0.3\%.

\begin{table}[htbp!]
    \centering
    \caption{IPOPT iterations, reported solver times, and final objective values for the DMO design cases of Sec.~\ref{sub:param3}. Iterations and times are summed over the two continuation stages ($p_{\mathrm{DMO}}=1$ and $p_{\mathrm{DMO}}=5$).}
    \label{tab:dmo_optimization_cost}
    \small
\setlength{\tabcolsep}{2pt}
    \begin{tabular}{ccccc}
\toprule
$(\alpha_T,\alpha_\kappa)$ & \begin{tabular}[c]{@{}c@{}}Design\\ variables\end{tabular} & Iterations & \begin{tabular}[c]{@{}c@{}}IPOPT\\ time (s)\end{tabular} & \begin{tabular}[c]{@{}c@{}}Final\\ objective\end{tabular} \\
\midrule
$(0.75,\,1.20)$ & 760 & 97 & 3075.671 & 583.762 \\
$(0.80,\,1.10)$ & 760 & 55 & 1362.463 & 328.488 \\
\bottomrule
\end{tabular}
\end{table}
\section{Conclusions}
\label{sec:end}
In this paper, we developed and discussed a novel gradient-based optimization approach for the design of lazy-wave offshore riser configurations. The design task consists of the spatial distribution and sizing of buoyancy modules. The objective was to develop a computationally efficient approach suitable for the conceptual stages of engineering design, where numerous alternative configurations must be explored and assessed. The design optimization problem was formulated to minimize a quadratic measure of the nodal uplift ratios while satisfying structural constraints on effective tension and curvature under a prescribed horizontal hang-off force and possibly current drag loads.

The main contribution of this work is a novel framework that integrates nonlinear equilibrium analysis and discrete adjoint sensitivities for gradient-based optimization of buoyancy distributions in lazy-wave risers. Once the equilibrium configuration is known, one linear adjoint solve per response constraint yields its gradient with respect to all nodal uplift ratios, regardless of the number of design variables. The curvature and effective tension gradients agreed with central finite difference estimates to within 0.01\% across the design variables checked. The same response gradients support all three design parametrizations considered.

A further contribution is the development and investigation of different design parameterizations, ranging from a continuous distribution of independently optimized nodal buoyancy values to more practical formulations involving a limited number of buoyancy module sizes. In particular, a formulation inspired by structural topology optimization and discrete material optimization was introduced, allowing the available buoyancy module sizes and their assignment along the riser to be optimized simultaneously. These parameterizations provide different levels of design freedom and practical applicability, enabling the consideration of manufacturing and installation requirements already during the optimization process.

The numerical examples demonstrated the capability of the proposed framework to identify improved lazy-wave configurations while satisfying prescribed structural performance requirements. The results also highlighted the influence of the selected tension and curvature limits on the optimal buoyancy distribution and the trade off between design flexibility and practical restrictions on available module sizes. 
Overall, the proposed methodology offers a promising and computationally efficient tool for the design optimization of lazy-wave offshore systems, particularly during the early conceptual and intermediate design phases. 

Although the present work focuses on the static response of offshore risers, the general framework is also applicable to the optimization of lazy-wave power cables for floating offshore wind turbines. The optimized configurations can subsequently be assessed using more sophisticated numerical models, including three dimensional and dynamic analyses, during the detailed design and verification stages.

\section*{Conflict of interest}
\noindent The authors declare that they have no conflict of interest.

\section*{Acknowledgments}
\noindent Both authors gratefully acknowledge the fruitful discussions with members of the Multidisciplinary Design Optimization (MDO) research group at the Technion. 
NP also acknowledges the financial support from the Israel Science Foundation (ISF) under grant no. 304/24.

\section*{Replication of results}
\noindent The numerical and algorithmic details are provided
in sufficient detail in the text to be implemented. 

\bibliography{mybibfile}

\end{document}